 \documentclass[twocolumn,english,showpacs,preprintnumbers,amsmath,amssymb,floatfix]{revtex4}

\usepackage[T1]{fontenc}
\usepackage[latin9]{inputenc}
\usepackage{color}
\usepackage{array}
\usepackage{amstext}
\usepackage{graphicx}
\usepackage{esint}
\usepackage{rotating}
\usepackage{amsmath}
\usepackage{slashed}
\usepackage[font=small,labelfont=bf]{caption}
\usepackage[colorlinks = true,
linkcolor = blue,
urlcolor  = blue,
citecolor = blue,
anchorcolor = blue]{hyperref}

\makeatletter

\@ifundefined{textcolor}{}
{%
 \definecolor{BLACK}{gray}{0}
 \definecolor{WHITE}{gray}{1}
 \definecolor{RED}{rgb}{1,0,0}
 \definecolor{GREEN}{rgb}{0,1,0}
 \definecolor{BLUE}{rgb}{0,0,1}
 \definecolor{CYAN}{cmyk}{1,0,0,0}
 \definecolor{MAGENTA}{cmyk}{0,1,0,0}
 \definecolor{YELLOW}{cmyk}{0,0,1,0}
 }

\@ifundefined{definecolor}
 {\usepackage{color}}{}
\@ifundefined{definecolor}
 {\usepackage{color}}{}
\makeatother
\usepackage{babel}

\begin{document}


\title{Hybrid star within the framework of a lowest-order constraint variational method }

\author{S. Khanmohamadi$^{a}$}
\email{s.khanmohamadi@ut.ac.ir}
\author{H. R. Moshfegh$^{a}$  }
\email{hmoshfegh@ut.ac.ir}
\affiliation{
	$^{(a)}$Department of Physics, University of Tehran, P.O.Box 14395-547, Tehran, Iran
}
\author {S.~Atashbar Tehrani$^{b}$}
\email{Atashbar@ipm.ir}
\affiliation {
	$^{(b)}$School of Particles and Accelerators, Institute for Research in Fundamental Sciences (IPM), P.O.Box
	19395-5531, Tehran, Iran}
\date{\today}

%
%
\begin{abstract}\label{abstract}
The hadron-quark phase transition in the core of heavy neutron star (NS) has been studied. For the hadronic sector, we have used the lowest-order constraint variational method by employing $ AV_{18} $,  $  AV_{14}$,  $  UV_{14}$, and   Reid $  68 $ two-body nucleon-nucleon forces supplemented by the phenomenological Urbana-type three-body force. We have adopted the MIT bag model as well as three-flavor version of the Nambu- Jona-Lasinio (NJL) model to describe the quark phase. The equation of state (EOS) of a hybrid star (HS) is presented by combining two EOS of the hadronic sector and quark sector of a star, which are derived from independent models or theories. The hadron-quark transition is constructed by considering a sharp phase transition, i.e., Maxwell construction. The structure of the HS is calculated and reported by solving Tolman-Oppenheimer-Volkoff equations.
Finally the radii and tidal deformability of purely NS and HS for the mass of $ 1.4M_{\odot} $ is computed and new constraints on these quantities are checked.
 The maximum mass of HS is found more than $  2 M_{\odot} $ for both the NJL and MIT bag models. However, the maximum mass of $  1.796 M_{\odot} $ ($  1.896 M_{\odot} $) was the best result that would be calculated for a stable HS with the pure quark core within the MIT (NJL) model.

All the hybrid EOS fulfill the constraints on radii and tidal deformability extracted from the binary GW170817 for HSs.
A comprehensive analysis on the structure of purely NS and HS and also compactness, tidal Love number, and tidal deformability for the star with the mass of 1.4 $ M_{\odot} $ has been conducted for various EOS of the hadron sector and several parameter sets of the quark EOS. The results achieved in this study are in good concurrence with the other calculations reported on this subject.

\end{abstract}


\maketitle


%
\section{Introduction}\label{Introduction}
One of the issues in the context of the compact stars is the probable appearance of the quark degrees of freedom in the interior of the heavy neutron stars (NSs) ~\cite{Glendenning:2001pe,Witten:1984rs,Baym:1985tn}. The question of whether or not quark matter exists in the core of NSs has newly received interest ~\cite{Hempel:2009vp,Burgio:2002sn,Burgio:2015zka}, by the discovery of two massive neutron stars~\cite{Demorest:2010bx,Antoniadis:2013pzd,Lynch:2012vv,vanKerkwijk:2010mt,Fonseca:2016tux}. Microscopic calculations demonstrate that in heavy NSs ($M\approx2 M_{\odot}$) the density of the core reaches to around $ 1 $ fm$ ^{-3} $, and at such high densities, the Fermi energy level of particles increases enough to produce various exotic particles, and the appearance of the quark phase (in addition to the baryonic phase) is not unexpected~\cite{Maieron:2004af}. In fact in densities above nuclear saturation density( $  \rho\gg\rho_{0}=0.16 $ fm$ ^{-3} $) some other exotic particles may exist in the interior of a NS in addition to nucleons and leptons such as hyperons and  $\pi$ and $k $ condensation. In higher densities, nuclear matter may experience phase transition to a deconfined quark plasma of $ u$, $ d$, and $s $ quarks. However, the appearance of hyperons in beta stable matter would strongly decrease the maximum mass of the star~\cite{Li:2008zzt,Schulze:2006vw,Carroll:2008sv,Djapo:2008au,Schulze:2011zza}; therefore, in this situation, the presentation of a nonbaryonic phase like the quark matter could be a feasible way to stiffen the EOS and reach to a massive NS. Thus, a heavy NS can be a hybrid star~\cite{Chen:2012zx}. It would have been ideal if there had been a unified theory which could have treated both the hadronic and quark phases simultaneously in all ranges of temperatures and densities, but, unfortunately, there is no such reliable theory as of now. However, at finite temperature and zero baryon density, a numerical study on lattice formalism in QCD has provided some reliable results for physics of the deconfinement transition~\cite{Karsch:2000ps,Karsch:2008fe}. In this case, lattice calculations predict that the deconfinement happens via a smooth crossover transition~\cite{Aoki:2006we} at a temperature $T \approx180-200 $  MeV~\cite{Aoki:2006br,Bhattacharyya:2009fg}. However, studies at finite baryon densities on the lattice are very difficult. Some progress has been made in recent years in extending the calculations to finite quark chemical potential; however, they have not yet provided reliable results ~\cite{Gavai:2008zr,deForcrand:2003vyj}. Therefore, studying the EOS and phase transition of nuclear matter to the deconfinement quark phase at zero temperature and high densities, which is the case in NSs, from the first principles, is a difficult task due to the nonlinear and nonperturbative nature of the QCD governing on the behavior of such systems. 

Therefore, as a starting point, one can use some phenomenological models for describing the quark matter. Over the past few decades, many authors have intensively studied various aspects related to the formation of exotic degrees of freedom in neuron stars and proposed observational tests which confirm the existence of such constituents in the interior of compact stars (see, Ref.~\cite{Alaverdyan:2014tqa} and references therein). The structure and properties of hybrid stars (HSs) have been studied in several papers with different hadron and quark models and various types of phase transitions. In Refs. ~\cite{Baldo:2006bt,Maieron:2004af,Maruyama:2007ey,Lenzi:2010mz,Logoteta:2013ipa,Baldo:2008en,Alford:2015dpa,Yang:2008am}, a few of them are reported.

We employ the MIT bag  and Nambu-Jona-Lasinio (NJL) models for describing quark matter in this paper. The MIT bag model builds confinement and asymptotic freedom via a phenomenological model and is essentially an enhanced version of Bogoliobov's model for quarks that are considered three massless quarks in a vacuum cavity of radius R with a finite, spherical, square well potential. Bogoliobov's model has a few shortcomings such as the violation of energy-momentum conservation~\cite{Bogolubov:1968zk}. The MIT bag model solves this problem by the inclusion of phenomenological confining pressure which is named the bag constant, B~\cite{Chodos:1974je}. This prescription provides a mechanism for natural confinement and also causes the model to become a Lorentz-covariant model. To describe the quark phase, an improved version of the MIT bag model, in which interaction between $ u, d$ and $ s $, quarks inside the bag are taken in a one gluon-exchange approximation, was used~\cite{Farhi:1984qu,Alcock:1986hz,Glendenning:1990wj,Burgio:2002sn}. 

Besides, we adopt the three-flavor version of the NJL model to describe the deconfined quark phase. The NJL model contains some of the basic symmetries of QCD, namely, chiral symmetry. The most important feature of the NJL model is its nontrivial vacuum by breaking the chiral symmetry dynamically by spontaneous mass generation. In the NJL model, at the low-energy scale the gluon acquires a large effective mass that can be integrated out to a good approximation, leaving a local contact four-fermion interaction between the quarks. Upon this procedure, the confinement was lost, because the local color symmetry of QCD was reduced to a global symmetry. This drawback of the NJL model is not an issue when modeling  quark matter at high densities, since the quark matter is deconfined at high densities. The NJL model has been very successful in describing the vacuum properties of low-lying meosns and predicts at sufficiently high densities or temperatures a phase transition to a chiral symmetric state ~\cite{Hatsuda:1994pi,Ruivo:1999pr,Buballa:1998pr,Buballa:2003qv}. Strictly speaking, the NJL model is usable in vacuum and at high densities but not in the hadronic phase in between.

 Contrary to the quark matter case, microscopic theories of the nucleonic EOS have reached a high degree of sophistication. We employ the lowest-order constraint variational (LOCV) method for describing the nucleonic sector. The LOCV method is a well-known many-body technique that was originally used to study the properties of cold symmetric nuclear matter~\cite{Owen:1975xh,Modarres:1979jk} by using the Ried-type potential~\cite{Reid:1968sq,Green:1978fj} as the bare two-body forces (2BF). Later on, this approach was extended to finite temperature~\cite{Moshfegh:2005rom}, and also calculations of the EOS of asymmetric nuclear matter~\cite{Moshfegh:2007mxh}, pure neutron matter, and $ \beta $-stable matter~\cite{Modarres:2000nk,Modarres:2002ns} were carried out within this framework by using more sophisticated potentials. Moreover, relativistic corrections have been considered in calculating thermodynamic properties of nuclear matter within this  model at both zero and finite temperatures~\cite{Zaryouni:2010p,Zaryouni:2014fsa}. Recently, this technique has been extended by adding three-body forces (TBF) to this formalism~\cite{Goudarzi:2015dax,Goudarzi:2016uos}, and has been used to study the structure of the NS~\cite{Goudarzi:2015dax} as well as protoneutron star~\cite{Goudarzi:2016uos}. This model is successful in reproducing the correct saturation point parameters such as $ E_{sym}(\rho_{0},L$ and $ K_{sym}) $, by using a revised version of TBF which is based on an isospin-dependent parametrization of coefficient in the Urbana-type (UIX) forces. Within the LOCV formalism employing $ AV_{18} $ supplemented by TBF in Urbana type ~\cite{Goudarzi:2016uos} and chiral symmetry~\cite{Goudarzi:2019orb}, the maximum NS mass is obtained above $2 M_{\odot}$. Recently, the LOCV method is reformulated to extract the EOS of Hyper nuclear matter~\cite{Shahrbaf:2019bef,Shahrbaf:2019wex}.

 For the region of phase transition, a detailed study employing the Wigner-Seitz cell approach~\cite{Maruyama:2006jk} suggests that the mixed phase behaves more in accordance with the Maxwell construction than the Gibbs construction. It may happen that a hadron-quark mixed phase is unlikely to be stable for a reasonable value of surface tension~\cite{Maruyama:2006jk,Alford:2001zr,Neumann:2002jm}; then, the situation is closer to the Maxwell construction case, in which two pure phases are in direct contact with each other. Therefore, we restrict ourselves to analyzing the sharp  hadron to quark matter phase transition. Maxwell construction and Gibbs construction describe the first-order phase transition. The existence of the "mass twins" in the mass radius relationship for the compact star also seems to support the Maxwell construction viewpoint~\cite{Benic:2014jia}. Some authors argue that the phase transition could be crossover, which may lead to interpolation/percolation construction~\cite{Whittenbury:2015ziz,Li:2017xlb}. As we mentioned above, the lattice QCD calculation shows that the transition line for low baryon densities is a crossover~\cite{Ejiri:2012rr,deForcrand:2014tha,Endrodi:2015oba,Braguta:2015zta} but it is model dependent for high densities and low temperatures. Thus, it should completely be treated  phenomenologically in this case.

 Nowadays, an EOS should not only fulfill the maximum mass constraints,  $ 2.01^{+0.04}_{-0.04}\leq{M}_{TOV}/{M\odot}\lesssim 2.16^{+0.17}_{-0.15}$, but should additionally fulfill the new constraint on radii and tidal deformability of compact stars set by the binary NS system, GW170817~\cite{Most:2018hfd}. The brand new era of multimessenger astronomy started on August 17, 2017 with the first direct detection of both gravitational and electromagnetic radiation from the binary NS merger GW170817 which was recorded by the Advanced LIGO and Virgo network of gravitational-wave recorders~\cite{Abbott:2018wiz,Abbott:2018exr}.
 GW170817, established for the first time the association of short gamma ray bursts with NS mergers which can help solve the long-standing puzzle of the origin of these phenomena~\cite{Negele:1971vb,Drischler:2016djf,Annala:2017llu,Drischler:2017wtt}. GW170817 gives fundamental new insights into the nature of dense matter by adding the tidal deformability  (polarizability) constraints on EOS. The tidal deformability, $ \Lambda $, encodes to the response of NS to the external tidal field produced by its companion similar to response of a polar molecular to an external electrical field. We calculate the tidal deformability and radii of the NSs and HSs that exist in this paper with the mass of $ 1.4M_{\odot} $ and new constraints on these quantities are checked. 
 
 The paper is organized as follows. In Sec.~\ref{II-A}, we will address the nucleonic matter and briefly review the determination of baryonic EOS in beta equilibrium in the LOCV approach at zero temperature. Section~\ref{II-B} is concerned the quark matter EOS according to the MIT and NJL  models. In Sec.~\ref{III-A}, by using these models a hybrid equation of state is obtained, assumming a Maxwell construction. In Sec.~\ref{III-B}, the structure of the hybrid star is presented. Section~\ref{III-C},  is concerned with calculating the tidal deformability, and Sec.~\ref{IV} is devoted to the summary of the results and conclusions.

\section{Equation of state}\label{II}
The neutron star outer and inner crust exist at densities between $ 10^{4} \leq \epsilon_{crust} \leq   10^{14} $ gr cm$ ^{-3} $~\cite{Weber:2006ep}. Matter in the inner crust consists mostly of nuclei in a Coulomb lattice that is immersed in a gas of electrons and, above neutron drip(greater than or equal to $ 4*10^{11}  $ gr cm$^{-3} $), free neutrons. In densities above the saturation density(greater than or equal to $  10^{14}$gr cm $^{-3}  $), the relevant degrees of freedom are hadrons. At higher densities (several times the saturation density), baryons begin to overlap and lose their individuality, and to describe the medium, the quark degrees of freedom need to be included. In this work, we use the Harrison-Wiheeler equation of state for the neutron star crust. 

\subsection{Confined hadronic phase}\label{II-A}
Different theoretical approaches can be used to calculate empirical properties of infinite nuclear matter. In recent years, experimental observations, together with theoretical efforts for explaining and analyzing them, have provided reliable microscopic models for describing nuclear matter. In this research, we use the LOCV model, which is a microscopic model based on cluster expansion and is in a good agreement with empirical properties. In this section we briefly review the LOCV method, the details of which can be found in the  references in the Introduction. First, we restrict our attention to the baryonic matter and the procedure of adding TBF to the LOCV formalism; then, we employ this formalism to extract the EOS of the $ \beta $ stable matter.

\subsubsection{Asymmetric nuclear matter}

For the first step in the LOCV formalism a trial wave function of the $ N $-body interacting system at zero temperature is produced as
\begin{equation}
\Psi(1\ldots N)=F(1\ldots N)\Phi(1\ldots N),
\end{equation}
where $\Phi(1\ldots N) $ is a noninteracting ground-state wave function of $ N $ independent nucleons and $F(1\ldots N)$ is a $ N $-body correlation operator. The correlation operator is obtained in the Jastrow approximation, which is the symmetrized product of two-body correlation function operators, which is written as
\begin{equation}
F(1\ldots N)=\mathcal{S}\prod _{i>j}f(ij) ,
\end{equation}
where $\mathcal{S}$ is the symmetrizing operator. $f(ij)$ is read as
\begin{equation}
f(ij)=\sum _{\alpha ,p=1}^{3} f_{\alpha }^{p} (ij)O_{\alpha }^{p} (ij),
\end{equation}
where $\alpha=\lbrace J,L,S,T,T_{z}\rbrace$ [ Total (J) , orbital (L) , Spin (S) angular momentum, and Isospin (T) and third component of Isospin $ T_{z} $] and $p=$ 2, 3 is used for coupled channels with $J=L\pm1$. Otherwise, $p$ is set to unity. The operators $O_{\alpha }^{p} (ij) $ are written as 
\begin{equation}
O_{\alpha }^{p=1-3} =1,(\frac{2}{3} +\frac{1}{6} S_{12} ),(\frac{1}{3} -\frac{1}{6} S_{12} ) ,
\end{equation}
where $S_{12} =3(\sigma _{1} .\hat{r})(\sigma _{2} .\hat{r})-\sigma _{1} .\sigma _{2}$  is the usual tensor operator.
In general, the nuclear Hamiltonian is read as sum of the nonrelativistic single-particle kinetic energy and  potential
\begin{equation}
H=\sum _{i}\frac{p_{i} ^{2} }{2m_{i}}  +\sum _{i<j}V(ij)+... .
\end{equation}
The baryonic energy expectation value $ E_{B} $ is considered as the sum of one body energy, $ E_{1} $ and two-body energy, $ E2 $ and written as

\begin{equation}\label{E}
E_{B}[f]=\frac{1}{N} \frac{{\left\langle \Psi  \right|} H{\left| \Psi  \right\rangle} }{{\left\langle \Psi  \mathrel{\left| \vphantom{\Psi  \Psi }\right.\kern-\nulldelimiterspace} \Psi  \right\rangle} } =E_{1} +E_{MB} \cong E_{1} +E_{2}~,
\end{equation}
in which 
\begin{equation}
E_{1} =\sum _{i}\frac{3\hbar ^{2} (k_{i}^{F}) ^{2} }{10m_{i}}~,
\end{equation}
where $ i=n$, $p $ and $ k_{i}^{F} $ is the corresponding nucleon momentum divided by $\hbar $
and $ E_{2} $  is defined as
\begin{equation}
E_{2} =\frac{1}{2N} \sum _{ij}{\left\langle ij \right|}  W(12){\left| ij-ji \right\rangle}~,
\end{equation}
with
\begin{eqnarray}\label{W}
W(12)=-\frac{\hbar ^{2} }{2m} [f(12),[\nabla _{12}^{2} ,f(12)]]\nonumber\\+f(12)V(12)f_{T}(12)~.
\end{eqnarray}
Higher-order terms in the cluster expansion series are neglectable~\cite{Moshfegh:2005rom}. This expression is now minimized with respect to the channel correlation functions but subjected to the normalization constraint, which is considered as ~\cite{Moshfegh:2005rom,Bordbar:1998xv,Owen:1977uun}
\begin{equation}
\frac{1}{N} \sum _{ij}\left\langle ij|h_{Tz }^{2} (12)-f^{2} (12)|ij-ji\right\rangle  =0~.
\end{equation}
The condition of healing the correlation functions to the Pauli function $ h_{Tz } (12) $, which for the asymmetric matter takes the following form, is also imposed~\cite{Owen:1977uun},

\begin{eqnarray}
h_{Tz}(r)&&=[1-\frac{9}{2} (\frac{J_{1}(k_{i}^{F}r)}{k_{i}^{F}r} )^{2} ]^{-\frac{1}{2} },~~~~~~T_{z} =\pm 1
\nonumber\\
&&=1 ,~~~~~~~~~~~~~~~~~~~~~~~~~~~~~~~~T_{z} =0~,
\end{eqnarray}

with $J_{1} (x)$  being the spherical Bessel function of order $1$. The normalization constraint introduces the Lagrangian multiplier parameters in the LOCV formalism. The procedure of minimizing Eq.~(\ref{W}) provides a number of Euler-Lagrange differential  equations for functions  $ f_{\alpha }^{p} (ij) $ . Solving these equations leads to the determination of correlation functions and then the two-body cluster energy. In the nuclear matter calculations, the saturation properties of cold symmetric nuclear matter fail to be reproduced correctly, if just 2BF is used. This deficiency can be resolved by inclusion of a TBF in the nuclear Hamiltonian. To avoid the full three-body problem, the TBF (semiphenomenological UIX interaction ) is included via an effective two-body potential derived after averaging out the third particle, which is weighted by the LOCV two-body correlation functions$f_{\alpha }^{p} (ij)$ at a given density $\rho_{B}$. For more details, see Refs. \cite{Goudarzi:2015dax,Goudarzi:2016uos}.

\subsubsection{Beta-stable matter}
As the density of hadronic matter increases beyond the saturation density, nuclei dissolve to form an interacting system of nucleons and leptons. If this system survives longer than the timescale of weak interactions, $ t\approx 10^{-10}s $, it is able to reach equilibrium with respect to the $ \beta $ decay $ n=p+e+\nu\bar{}_{e} $ and its inverse, which is called beta stable matter. Therefor, we have to consider the NS as an object of which the matter contains neutrons, protons, electrons, and muons. The $ \tau $ lepton is ignored because of its large rest mass compared with two other leptons. For such matter, the $ \beta $ equilibrium conditions (without trapped neutrinos) are
\begin{equation}\label{mun}
\mu _{n}=\mu _{p} +\mu _{e}
\end{equation}

\begin{equation}\label{mue}
\mu _{e}=\mu _{\mu}
\end{equation}
where $ \mu _{i} $ stands for the chemical potential of each particle. Chemical potentials of leptons at zero temperature can be expressed as
\begin{equation}
\mu _{i} =\sqrt{(p_{Fi}c)^{2}+(m_{i}c^{2})^{2}}~.
\end{equation}
The charge neutrality condition in the NS matter requires the following equality
\begin{equation}\label{rhop}
\rho _{p}=\rho _{\mu}+\rho _{e}~.
\end{equation}
Solvig the coupled equations.~(\ref{mun}),~(\ref{mue}),~(\ref{rhop}) self- consistently at any given baryon density $  (\rho _{B}=\rho _{n}+\rho _{p})  $, the energy of $\beta $-stable matter, which is written as the sum of the baryonic part energy $ E_{B} $ and leptonic part energy $ E_{L} $, can be determined, $ (E=E_{B}+E_{L}) $. The energy of the baryonic part is calculated using Eq.(~\ref{E}). Leptons are supposed to be noninteracting highly relativistic particls, so at zero temperature the energy of leptonic part can be written as
\begin{equation}
E_{L}=\frac{2}{h^{3}\rho_{B}}\sum_{i=e,\mu}\int_{0}^{P_{Fi}}d^{3}p_{i}\sqrt{(p_{i}c^{2})+(m_{i}c^{2})^{2}}~.
\end{equation}
The pressure of the NS matter as a function of baryonic density is calculated by using the following thermodynamic relation

\begin{equation}
P=\rho_{B}^{2}(\dfrac{\partial (E/N)}{\partial \rho_{B}})~.
\end{equation}

\subsection{Deconfined quark phase }\label{II-B}
Because of the nonperturbative and nonlinear nature of QCD, that is, governs the strongly interacting particles in the deconfined quark phase, describing such a system has to be done via different models. In this study we  employ two well-known models, namely, the MIT bag model and the NJL model, which are briefly reviewed in the following subsections.

\subsubsection{MIT bag model}\label{MIT}
 An improved version of the MIT bag model in which interaction between $ u$, $ d$, and $ s $ quarks inside the bag are taken in a one-gluon-exchange approximation was employed~\cite{Farhi:1984qu,Alcock:1986hz,Glendenning:1990wj,Burgio:2002sn}. At this stage, the thermodynamic potential $ (\Omega) $ includes the quark kinetic energy, as well as, one-gluon-exchange energy in which the fine structure constant of QCD is entered,
\begin{equation}
\Omega=\sum_{f}\Omega_{f}+B~,
\end{equation}
where $ B $ is the energy density difference between the perturbative vacuum and true vacuum, i.e., the bag constant which is the free parameter of the model, and at zero temperature $ \Omega_{f} $ takes the form

\begin{eqnarray}
&&\Omega_{f}(\mu_{f})=\frac{-1}{4\pi^{2}}\times
\nonumber\\
&&\Bigg[\mu_{f}(\mu_{f}^{2}-\frac{5}{2}m_{f}^{2})\sqrt{\mu_{f}^{2}-m_{f}^{2}}+\frac{3}{2}m_{f}^{4}\ln\Big(\frac{\mu_{f}+\sqrt{\mu_{f}^{2}-m_{f}^{2}}}{m_{f}}\Big)\Bigg]
\nonumber\\
&&+\frac{\alpha_{c}}{2\pi^{3}}\Bigg[3\Big(\mu_{f}\sqrt{\mu_{f}^{2}-m_{f}^{2}}-m_{f}^{2}\ln\big(\frac{\mu_{f}+\sqrt{\mu_{f}^{2}-m_{f}^{2}}}{m_{f}}\big)\Big)^{2}\nonumber\\&&-2(\mu_{f}^{2}-m_{f}^{2})^{2}-3m_{f}^{4}\ln\left(\frac{m_{f}}{\mu_{f}}\right)
\nonumber\\
&&+6\ln\frac{\sigma}{\mu_{f}}\Big\{\mu_{f}m_{f}^{2}(\mu_{f}^{2}-m_{f}^{2})^{\frac{1}{2}}-\mu_{f}^{4}\ln\Big(\frac{\mu_{f}+\sqrt{\mu_{f}^{2}-m_{f}^{2}}}{m_{f}}\Big)\Big\}\Bigg]~.\nonumber\\
\end{eqnarray}

in which $ m_{f} $ and $ \mu_{f} $ are the current quark mass and chemical potential, respectively, with $ f=u $,  $d $, $s $,  and  $ \alpha_{c} $ denotes the QCD fine structure constant and $ \sigma=\frac{m_{N}}{3}=313 $MeV is the renormalization point, where $ m_{N} $ is the nucleonic mass. The masses of $ u$ and $ d $ quarks are neglected, and we take $ m_{s}=150$, $300 $  MeV . Thermodynamic quantities can be derived in the standard way:

\begin{equation} 
n_{f}=\dfrac{\partial \Omega}{\partial \mu_{f}}
\end{equation}
\begin{equation} 
P=-\Omega    
\end{equation}
\begin{equation} 
\epsilon=\Omega+\sum_{f}\mu_{f}n_{f}~.
\end{equation}

\subsubsection{NJL model}\label{NJL}

We adopt the three-flavor version of the NJL model. The Lagrangian is given by~\cite{Buballa:2003qv}
\begin{equation}
L=\bar{q}(i\slashed{\partial}-\hat{m})q+L_{sym}+L_{det}~,
\end{equation}

\begin{equation}\label{L1}
L_{sym}=G\sum_{a=0}^{8}[(\bar{q}\lambda_{a}q)^{2}+(\bar{q}+i\gamma_{5}\lambda_{a}q)^{2}]
\end{equation}

\begin{equation}\label{L2}
L_{det}=-K[det(\bar{q}(1+\gamma_{5})q)+det(\bar{q}(1-\gamma_{5})q)]
\end{equation}
in which $q=(u,d,s)^{T}$ is a quark field with three flavors ($ N_{f}=3 $) and three colors ($ N_{c}=3 $), and $\hat{m}$=diag $(m_{u},m_{d},m_{s})$ is the the corresponding quark mass matrix. As $ m_{s} \neq m_{u} = m_{d}$, the isospin symmetery has been applied in this paper while SU(3) flavor symmetry explicitly broken. $L_{sym}$ is a $ U(3)_{L}*U(3)_{R} $ symmetric four-point interaction,in which $\lambda_{a}$ , $a=1,...,8$ denotes the Gell-Man matrices, the generators of $SU(3)$. $ L_{det} $ concerns the 't hooft interaction that is a $ SU(3)_{L}*SU(3)_{R} $ symmetric $ 2N_{f}=6 $-point interaction but it breaks the $ U(1) $ symmetry, which was left unbroken by $L_{sym} $. $ G $ is the four-point coupling constant,  and $ K $ is the six-point coupling constant. In the Hartree-Fock approximation, the quark self-energy, which arises from the interaction terms, is local and only implies a constant shift in the quark mass, which leads to the gap equation in the NJL model that is the relation between the constituent quark mass $ M $ and the current quark mass $ m $~\cite{Buballa:2003qv}
\begin{equation}\label{M}
M_{i}=m_{i}-4G\varphi_{i}+2K\varphi_{j}\varphi_{k} 
\end{equation}
where $ (i,j,k)$ is equal to any  permutation of $(u,d,s) $ and $ \varphi_{i}=<\bar{q}_{i}{q}_{i}> $ is the quark condensate. NJL is a nonrenormalizable model, since there are divergent integrals in it, so there are different regularization schemes to regularize the divergencies, which are  part of the NJL model. In this paper, the regularization has been done by using a sharp 3-momentum cutoff, $ \Lambda_{c} $ . There are five parameters that should be fixed in the $SU(3)$ NJL model: cutoff $ \Lambda _{c}$; the bare  quark mass $m_{u}=m_{d}$,  $m_{s}$; and the coupling constants G and K. The parameters are fixed by fitting to the empirical values of five observables, namely, the pion decay constant$ f_{\pi} $, the pion mass $ m_{\pi} $, and the mass of three pseduscalar mesons $ k,\eta$  $a,{\acute{\eta}} $. In Table~\ref{t1}~\cite{Buballa:2003qv}, we list three different parameter sets, which correspond to the fits of (Rehberg,Klevansky and Hufner)(RKH)~\cite{Rehberg:1995kh}, of Hatsuda and Kunihiro (HK)~\cite{Hatsuda:1994pi}, and of Lutz, Klimt, and Weise (LKW)~\cite{Lutz:1992dv}, together with related quantities in the quark and meson sectors and their empirical values~\cite{Hagiwara:2002fs}. The empirical quark masses listed have been rescaled to a renormalization scale of 1 GeV by multiplying them by 1.35 . The values given for the light quarks correspond to the average $ (m_{u}+m_{d})/2 $ . In RKH and HK parameter sets, the parameters were determined by fitting $ f_{\pi},  m_{\pi},m_{k},$ and $ m_{\acute{\eta}} $ to their empirical values , while the mass of the $ \eta $ meson is understimated by $ 6\%  $ in RKH and $ 11\% $ in HK. In the LKW parametrization, a vector and
axial-vector interaction term is considered in addition to
Eqs.~(\ref{L1}) and (~\ref{L2}), which enables the authors of ref~\cite{Lutz:1992dv} to fit the vector-meson nonet  $ (\rho, \omega, K^{*} $ and $\phi) $  as well, that cause
LKW parameter set has  a relatively larg cut-off and small bare quark masses in comparison with RKH and HK parameter sets. In the pseudoscalar meson sector, all three parameter sets obtained similar results. In this paper,the numerical calculations have been done in all three RKH, HK, and LKW  parameter sets.

The mean field thermodynamics potential in presence of the quark condensates at zero temperature takes the form~\cite{Buballa:2003qv}
\begin{eqnarray}\label{omega}
&&\Omega(\mu_{f},\varphi_{f})=\sum_{f=u,d,s}\Omega_{M_{f}}(\mu_{f})+2G(\varphi_{u}^{2}+\varphi_{d}^{2}+\varphi_{s}^{2})
\nonumber\\
&&-4k\varphi_{u}\varphi_{d}\varphi_{s}+\Omega_{0}~,
\end{eqnarray} 
where $ \Omega_{M_{f}} $ corresponds to the contribution of a gas of a quasiparticle with  mass $ M_{f} $, which at zero temperature is written as 
\begin{equation}\label{omegam}
\Omega_{M_{f}}(\mu_{f})=\frac{-N_{c}}{\pi^{2}}\int_{p_{F,f}}^{\Lambda}E_{p,f}p^{2}dp-\mu_{f}n_{f}
\end{equation} 
in which $E_{p,f}=\sqrt{p^{2}+M_{f}^{2}}  $ , $ p_{F,f}=\sqrt{\mu_{f}^{2}-M_{f}^{2}} $  and $ n_{f}=\dfrac{(p_{F,f})^{3}}{\pi^{2}} $ are the on-shell energy, the Fermi momentum, and number density of a quark of flavor $ f $ with the constituent mass $ M_{f} $  and  3-momentum $ p $, respectively. The thermodynamically consistent solutions correspond to the stationary point of $ \Omega $, which is found by minimizing it with respect to the condensates $\varphi_{u}$,  $ \varphi_{d}$, and $\varphi_{s} $    ($ \dfrac{\delta\Omega}{\delta\varphi_{f}} =0$). By applying the chain rule $ (\dfrac{\delta\Omega}{\delta\varphi_{f}}=\dfrac{\partial \Omega}{\partial M_{f}}\dfrac{\delta M_{f}}{\delta\varphi_{f}}\dfrac{\partial \Omega}{\partial \varphi_{f}}=0 )$ and using Eqs.~(\ref{M}),~(\ref{omega}),~(\ref{omegam}) one finds  $\varphi_{f} $ given by
\begin{equation}
\varphi_{f}=\frac{-N_{c}}{\pi^{2}}\int_{p_{F,f}}^{\Lambda}\frac{M_{f}}{E_{p,f}}p^{2}dp~.
\end{equation}

 $ \varphi_{f} $ has to be evaluated self-consistently with Eq.~(\ref{M}) and form a set of three coupled gap equations for the constituent quark masses. $ \Omega_{0} $ in Eq.~(\ref{omega}) is chosen such that the thermodynamic potential $ \Omega $ vanishes at zero $ \mu $ and $ T $.
 
Once the solutions of the gap equations for the constituent masses are found, other thermodynamic quantities can be derived in the standard way 
\begin{equation} 
P=-\Omega~~~~~~,~~~~~~\epsilon=\Omega+\sum_{f}\mu_{f}n_{f}~.
\end{equation}
The weak decays ( $ d\leftrightarrow u+e+\bar{\nu}_{e}\leftrightarrow s $ ) should be taken into account in the quark matter, so we have to include electrons (neutrinos have enough time to leave the system). The electrons are described by a noninteracting gas of massless fermions as often:
\begin{equation} 
P_{e}=\frac{\mu_{e}^{4}}{12\pi^{2}} ~~~~\rightarrow~~~~ \epsilon_{e}=\frac{\mu_{e}^{4}}{4\pi^{2}}
\end{equation}
Therefore, we will have
\begin{equation} 
P_{tot}=P+P_{e}~~~~~~,~~~~~~\epsilon_{tot}=\epsilon+\epsilon_{e}~ .
\end{equation} 
 in the $ \beta $-stable quark matter. The relations between chemical potentials of the particles take the form
\begin{eqnarray}
\mu_{d}=\mu_{s}=\mu
\nonumber\\
\mu=\mu_{u}+\mu_{e}~.
\end{eqnarray}

The charge neutrality condition implies ( $\frac{2}{3}n_{u}-\frac{1}{3}n_{d}-\frac{1}{3}n_{s}-n_{e}=0  $ ); thus, the system can be characterized by one independent variable, that is the baryon number density $ \rho_{B}=\frac{1}{3}(n_{u}+n_{d}+n_{s}) $.

\begin{table}[h!]\label{t1}
	\begin{center}
		\begin{tabular}{|l| c c c |c|}
			\hline
			&&&&\\[-3mm]
			& RKH\cite{Rehberg:1995kh}  & HK\cite{Hatsuda:1994pi}  & LKW\cite{Lutz:1992dv}
			& Empirical\cite{Hagiwara:2002fs}
			\\[1mm]
			\hline
			&&&&\\[-3mm]
			$\Lambda_{c}$ (MeV) &  602.3 & 631.4 & 750 &
			\\
			$G\Lambda_{c}^2$    & 1.835 & 1.835 & 1.82 &
			\\
			$K\Lambda_{c}^5$    & 12.36 & 9.29 & 8.9 &
			\\
			$m_{u,d}$ (MeV) & 5.5  & 5.5 & 3.6 & 3.5 - 7.5
			\\
			$m_s$ (MeV)     & 140.7 & 135.7 & 87 & 110 - 210
			\\
			$G_V/G$ & --- & --- & 1.1 &
			\\
			\hline
			$f_\pi$ (MeV) & 92.4 & 93.0 & 93 &  92.4
			\\
			$m_\pi$ (MeV) & 135.0 & 138 & 139 & 135.0, 139.6
			\\
			$m_K$ (MeV)   & 497.7 & 496 & 498 & 493.7, 497.7
			\\
			$m_\eta$ (MeV) & 514.8 & 487 & 519 & 547.3
			\\
			$m_{\eta'}$ (MeV) & 957.8 & 958 & 963 & 957.8
			\\
			$m_{\rho,\omega} $ (MeV)  & --- & --- & 765 & 771.1, 782.6
			\\
			$m_{K^*}$ (MeV) & --- & --- & 864 & 891.7, 896.1
			\\
			$m_\phi$ (MeV)  & --- & --- & 997 & 1019.5
			\\
			\hline
		\end{tabular}
	\end{center}
	\caption{\small Three parameter sets and related quark and meson properties in the three-flavor NJL model.} \label{t1}
\end{table}

\section{Results}\label{III}
\begin{figure*}[htb]
	\vspace{-0.70cm}
	\resizebox{0.45\textwidth}{!}{\includegraphics{{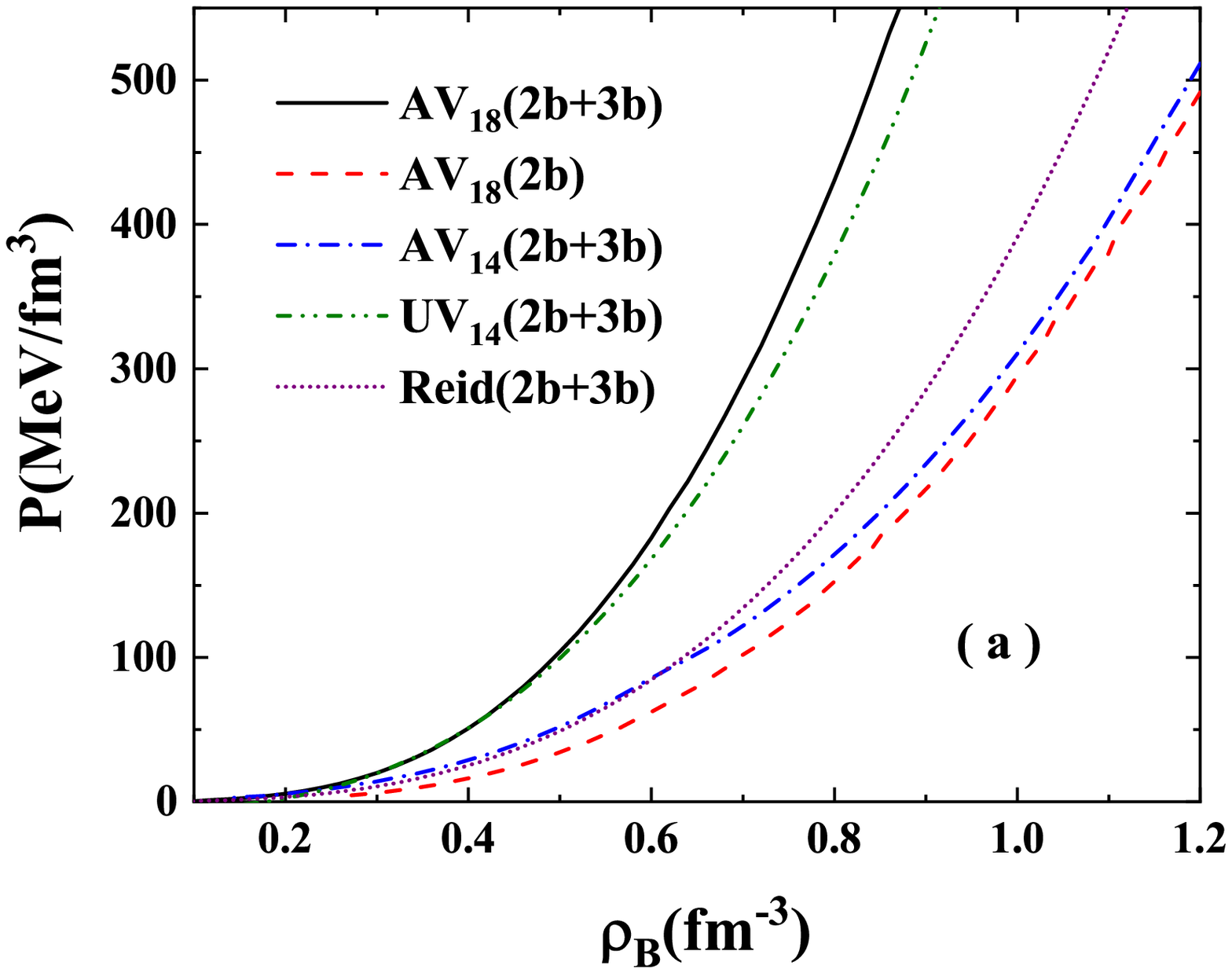}}}
	\resizebox{0.45\textwidth}{!}{\includegraphics{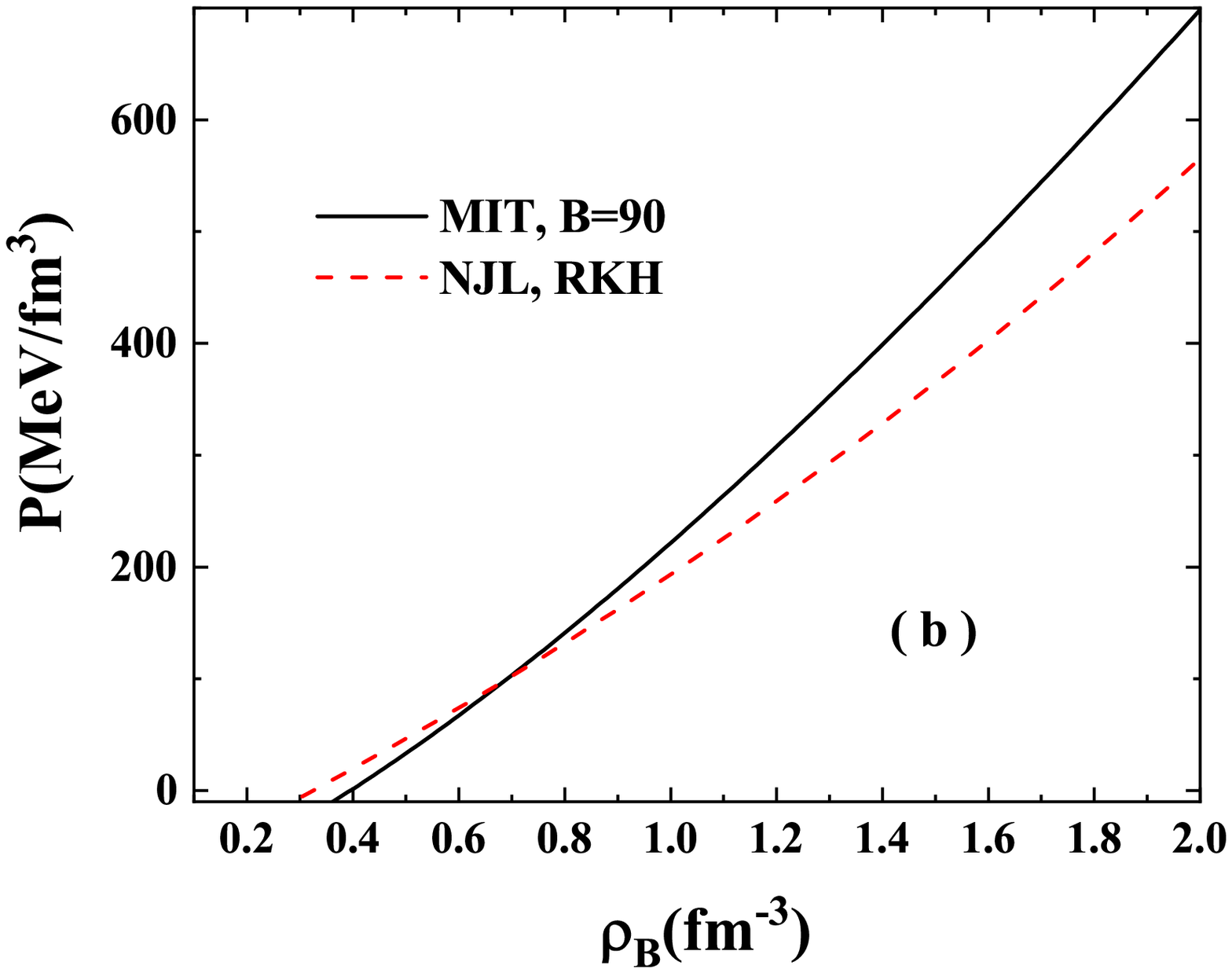}}
	\begin{center}
		\caption{{\small Panel a ( b ) : Pressure vs. baryon number density for nuclear ( quark) $ \beta $- stable matter within LOCV method for various interactions ( NJL model in parameter set RKH and MIT model with $ m_\text{{s}}=150 $ MeV and B= $90 $ {MeV}{fm$ ^{-3} $}).  } \label{fig1}} 
	\end{center}
\end{figure*}

Before starting to present the results for HSs,  we show the EOS of pure hadronic and quark mattes in Fig.~\ref{fig1}(a) and ~\ref{fig1}(b). In Fig.~\ref{fig1}(a) the pressures of nuclear $ \beta $-stable matter vs baryon number densities, $ \rho_{B} $, for various interactions are plotted. As seen in Fig~\ref{fig1}(a), the EOS of the $ AV_{18} $ supplemented by TBF is stiffer with respect to the others, so the higher maximum mass is expected in comparison to the other EOSs in the figure. In Fig.~\ref{fig1}(b) the pure quark matter pressure vs baryon number density is plotted for a sample parameter of the MIT bag model with $B= 90$ MeVfm$ ^{-3} $  and the NJL model with  RKH parameter set. Our calculation is in line with the original calculations in Refs~\cite{Alcock:1986hz,Buballa:2003qv}.

\subsection{Hadron-quark hybrid EOS}\label{III-A}
To stablish an EOS governs to whole HS, one needs to investigate the hadron-quark phase transition. As the quark-hadron mixed phase is unlikely to be stable for reasonable values of the surface tension ~\cite{Maruyama:2006jk,Alford:2001zr,Neumann:2002jm}, we restrict ourselves to analyzing the phase transition on Maxwell construction that is a sharp phase transition from neutral hadronic matter to homogeneous neutral quark matter. Each phase is considered to be in $ \beta $equilibrium and also charge neutrality. The requirement of the charge neutrality effectively reduces each phase to a one-component system controlled by the baryonic density or equivalently a baryonic chemical potential. The transition point in the Maxwell construction is identified by the conditions of thermal, mechanical and one-component chemical equilibrium, which at zero temperature takes the form
\begin{equation}\label{p}
P_{1}(\mu_{B})=P_{2}(\mu_{B})
\end{equation}
in which subscript 1 (2) stands for the hadronic (quark) phase. equation~(\ref{p}) means that Maxwell construction corresponds to constant pressure in the density interval between two phases.  $\mu_{B}  $ is the baryon chemical potential ($\mu_{B1}=\mu_{p} +\mu_{n}  $ and $ \mu_{B2}=\mu_{u} +\mu_{d} +\mu_{s} $). This equation also means that in this construction the baryon chemical potential $ \mu_{B} $ is continuous whereas the electron chemical potential $ \mu_{e} $ jumps at the interface between the two phases(  in Gibbs construction, the electron chemical potential is taken as continuous, too~\cite{Burgio:2002sn} ).
Maxwell construction can be considered as a limiting scenario in which the surface tension is large.
\subsubsection{MIT bag model }\label{hybridmit}

\begin{figure*}[htb]
	\vspace{-0.70cm}
	\resizebox{0.45\textwidth}{!}{\includegraphics{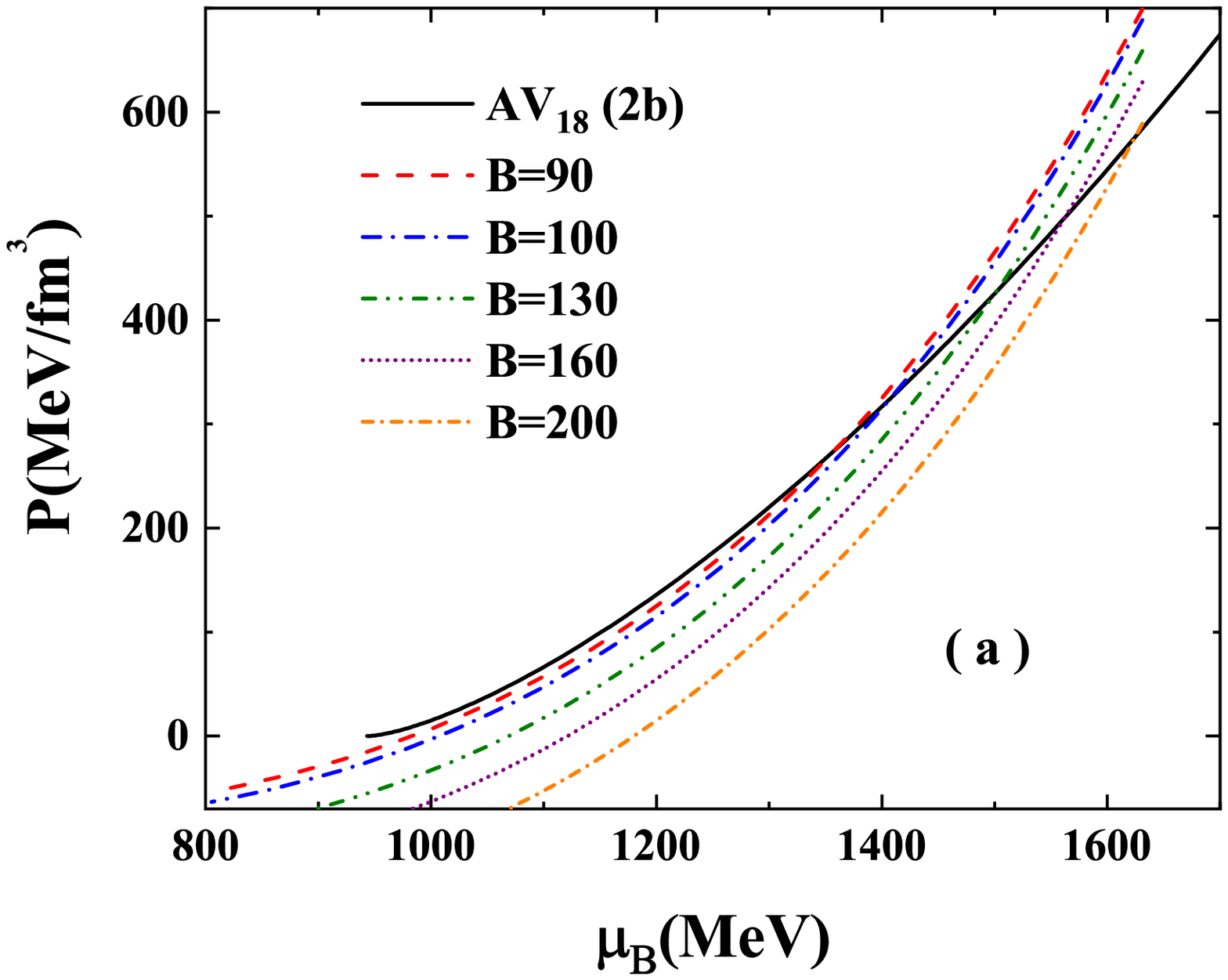}}
	\resizebox{0.44\textwidth}{!}{\includegraphics{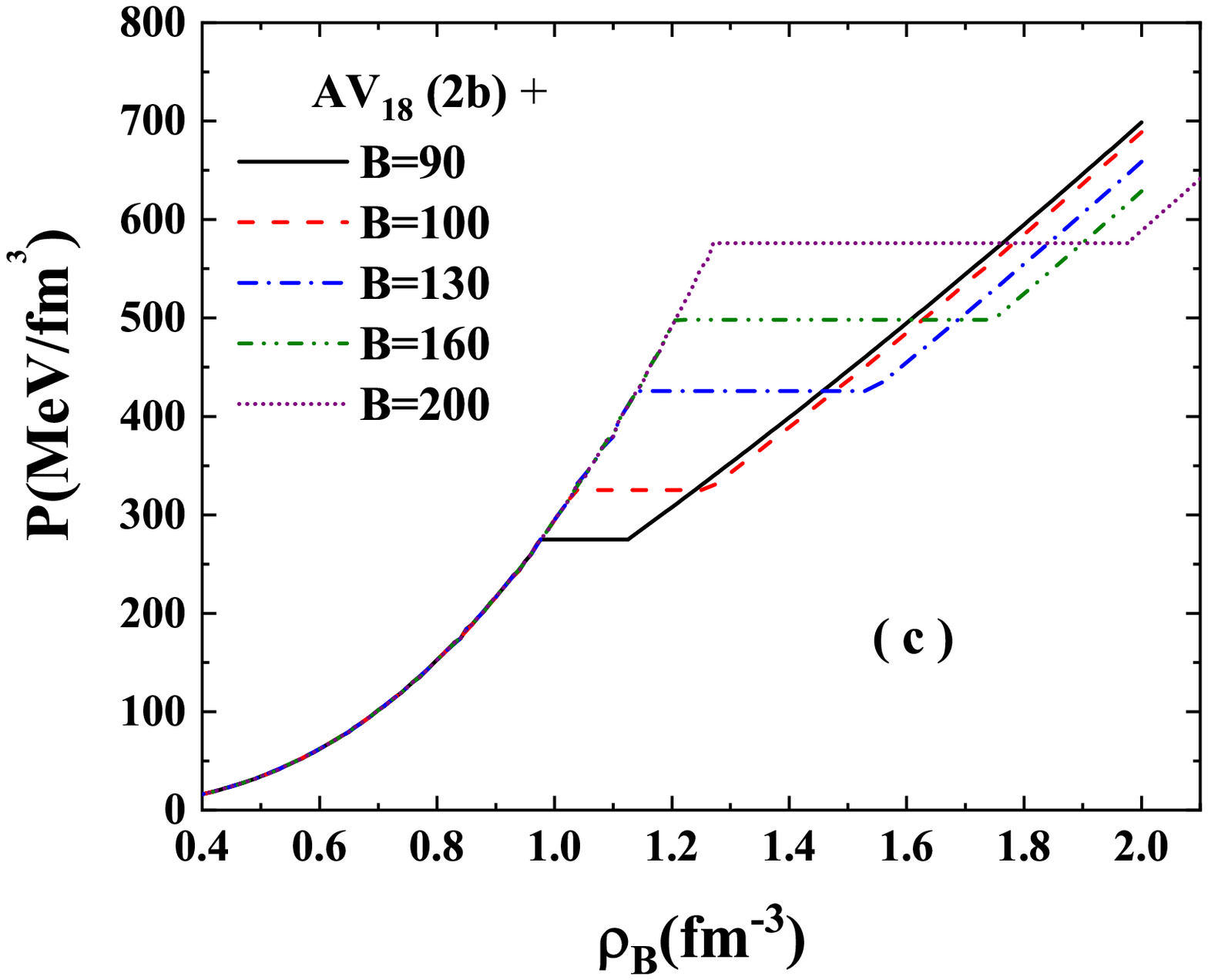}}
	\resizebox{0.45\textwidth}{!}{\includegraphics{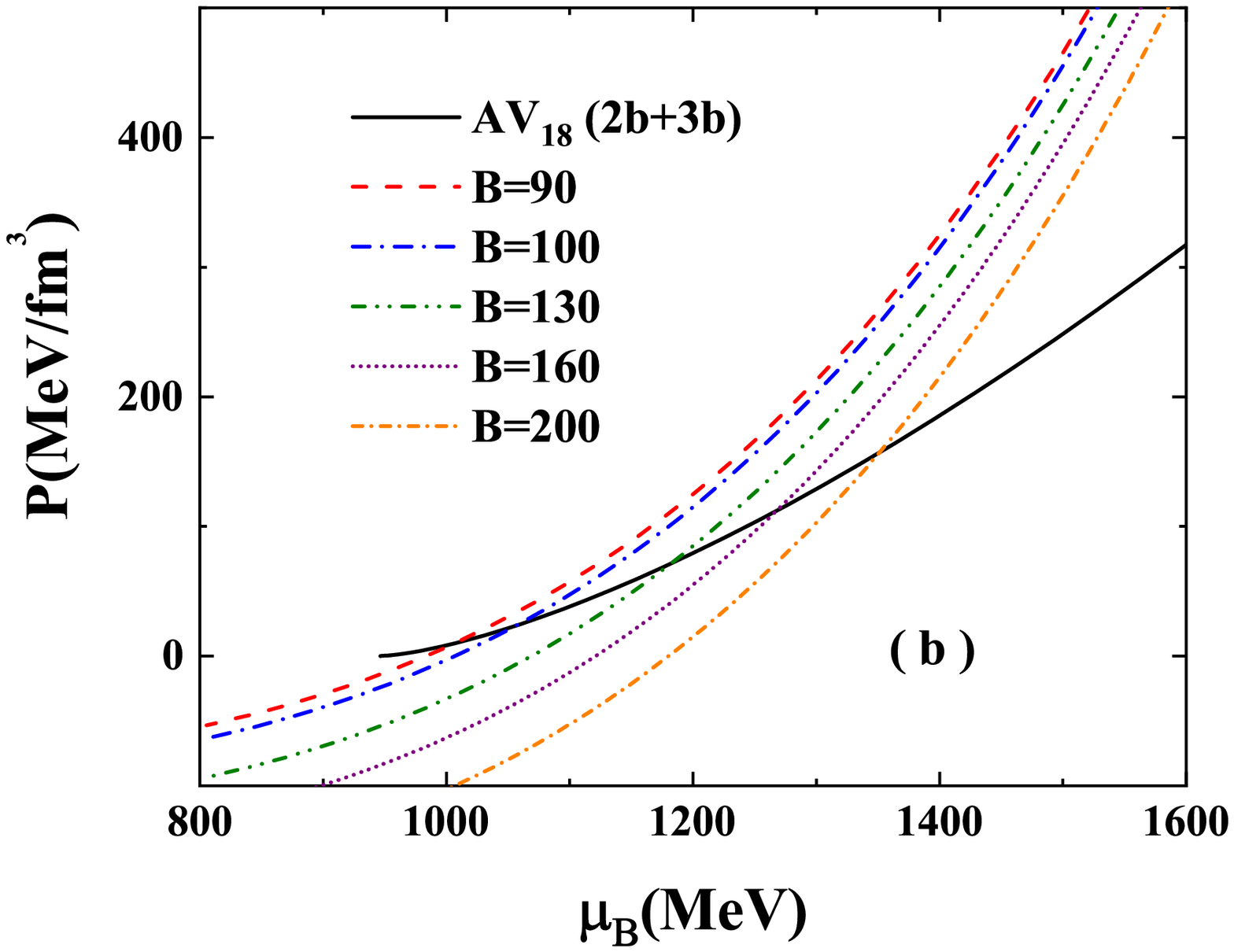}}
	\resizebox{0.44\textwidth}{!}{\includegraphics{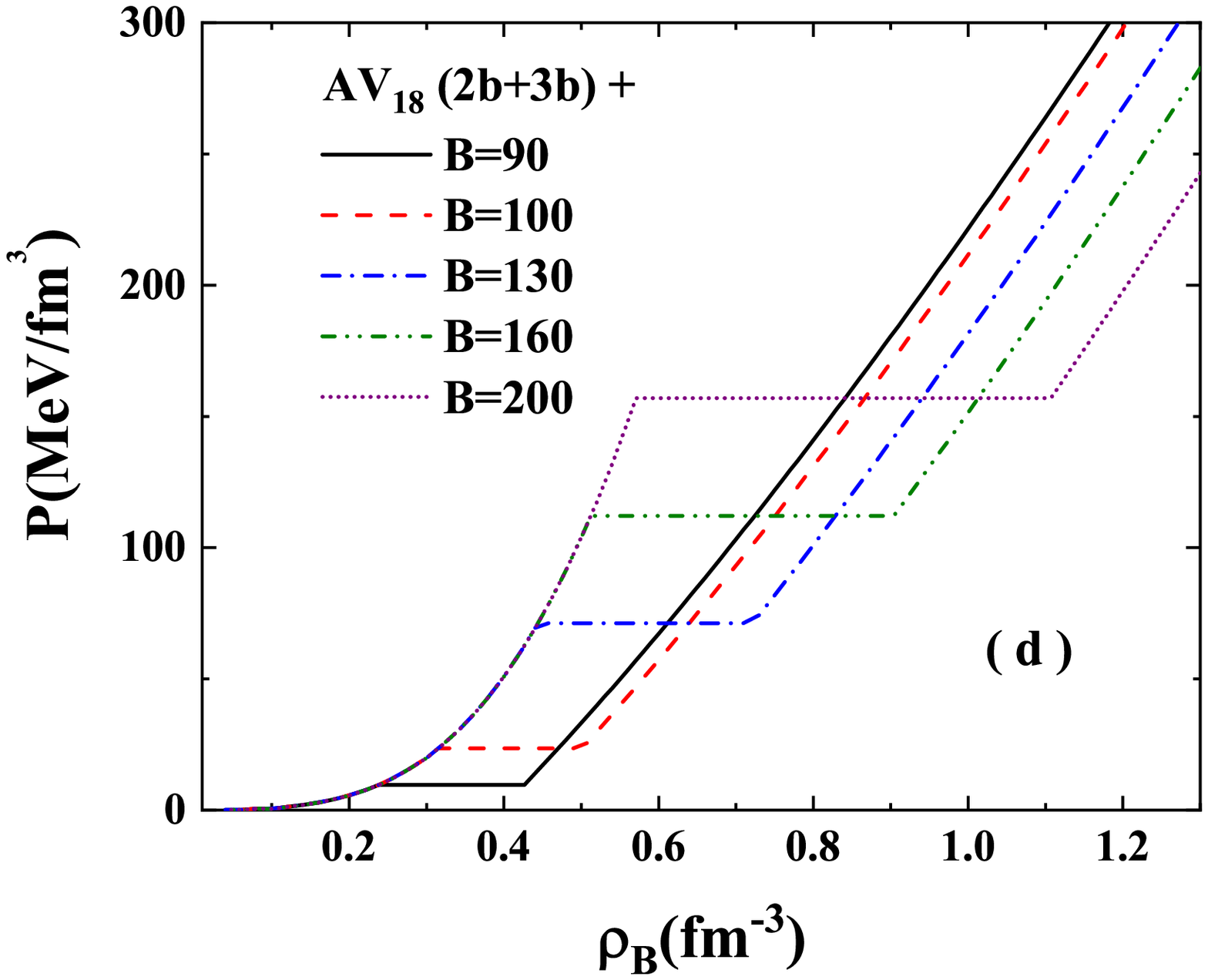}}
	
	\begin{center}
		\caption{{\small Panel a ( b ) : Pressure vs. baryon chemical potential for the MIT model with $ m_\text{{s}}=150 $ MeV and various bag constants and $ AV_{18} $ interaction supplemented without (with )TBF. Panel c ( d ) : The corresponding hadron-quark hybrid EoS's in Maxwell construction.}\label{fig2} } 
	\end{center}
\end{figure*}

We will first discuss the results obtained with the MIT bag model with a strange rest mass of $ m_{s}=150 $ MeV for the quark sector. In Fig.~\ref{fig2}(a) [\ref{fig2}(b)], we plot the pressure as a function of baryonic chemical potential in the MIT bag model with various bag constants and LOCV method with $ AV_{18} $ interaction supplemented without [with] TBF, respectively. In ~\ref{fig2}(c) [\ref{fig2}(d)], the corresponding hybrid star EOSs are displayed in Maxwell construction. For the bag constant less than  $78$ MeVfm$ ^{-3} $  there is no intersection between the hadron and quark pressure curves which means the nucleonic phase will remain stable with respect to the formation of quark phase droplets for noted bag constants. For  $ 78  $ {MeV}{fm$ ^{-3} $} $\leq$ B $\leq 84 $ {MeV}{fm$ ^{-3} $}  the hadron transition densities are less than nuclear saturation density, $ \rho_{0} $. Thus, we focus on the phase transition for B $  \geq 90  $ {MeV}{fm$ ^{-3} $}  in which the transition densities are more than $ 1.5\rho_{0} $. As the bag constant increases, the phase transitions from nuclear to quark matter take place at rather high baryon densities and also the transition region extends. When we consider just 2BF, the EOS becomes much softer and the phase transitions take place in much higher densities. In this case, the phase transition occurs in about $ 6\rho_{0} $ with $B= 90 $ {MeV}{fm$ ^{-3} $} . 

\begin{figure*}[htb]
	\vspace{-0.70cm}
	
	\resizebox{0.325\textwidth}{!}{\includegraphics{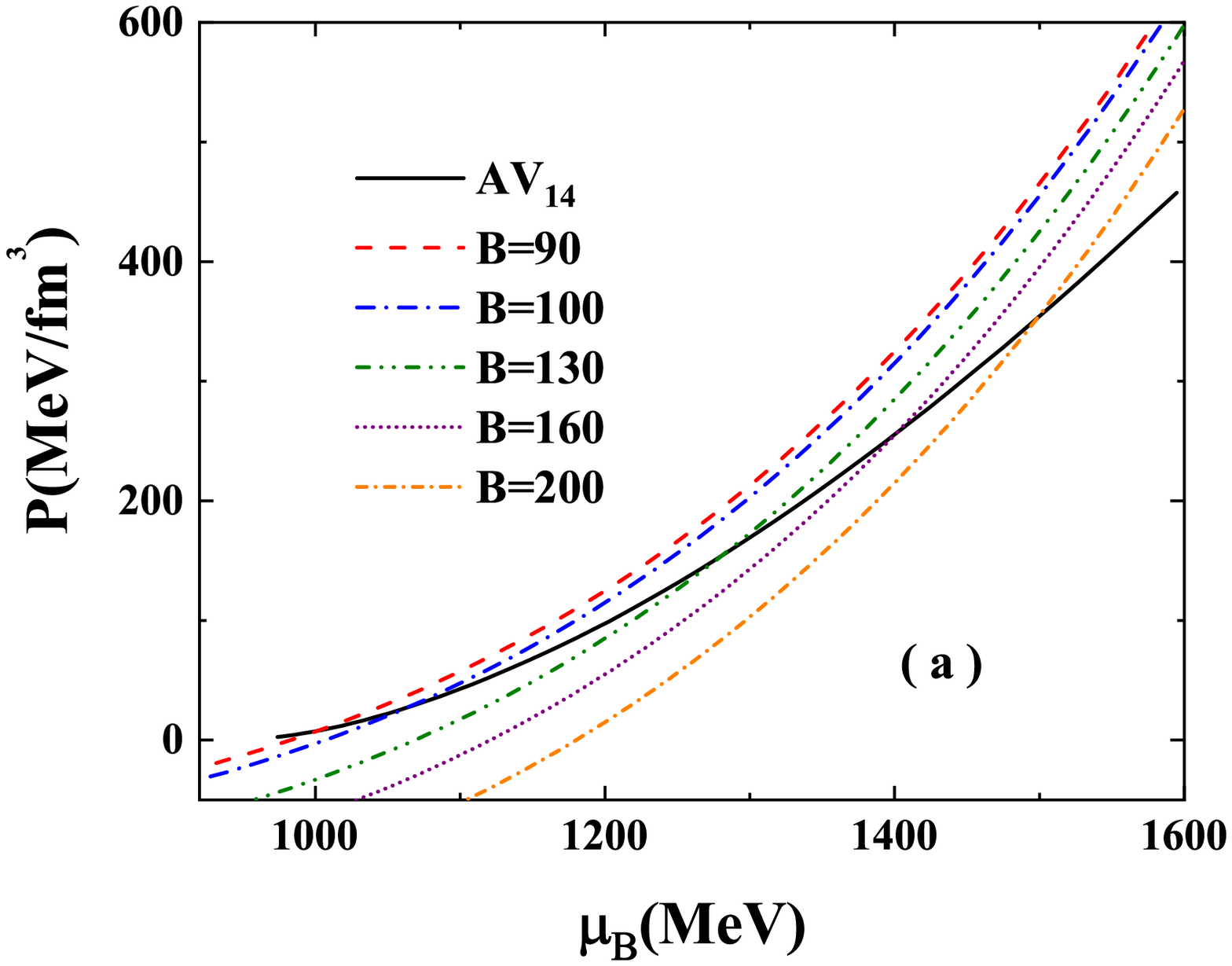}}
	\resizebox{0.325\textwidth}{!}{\includegraphics{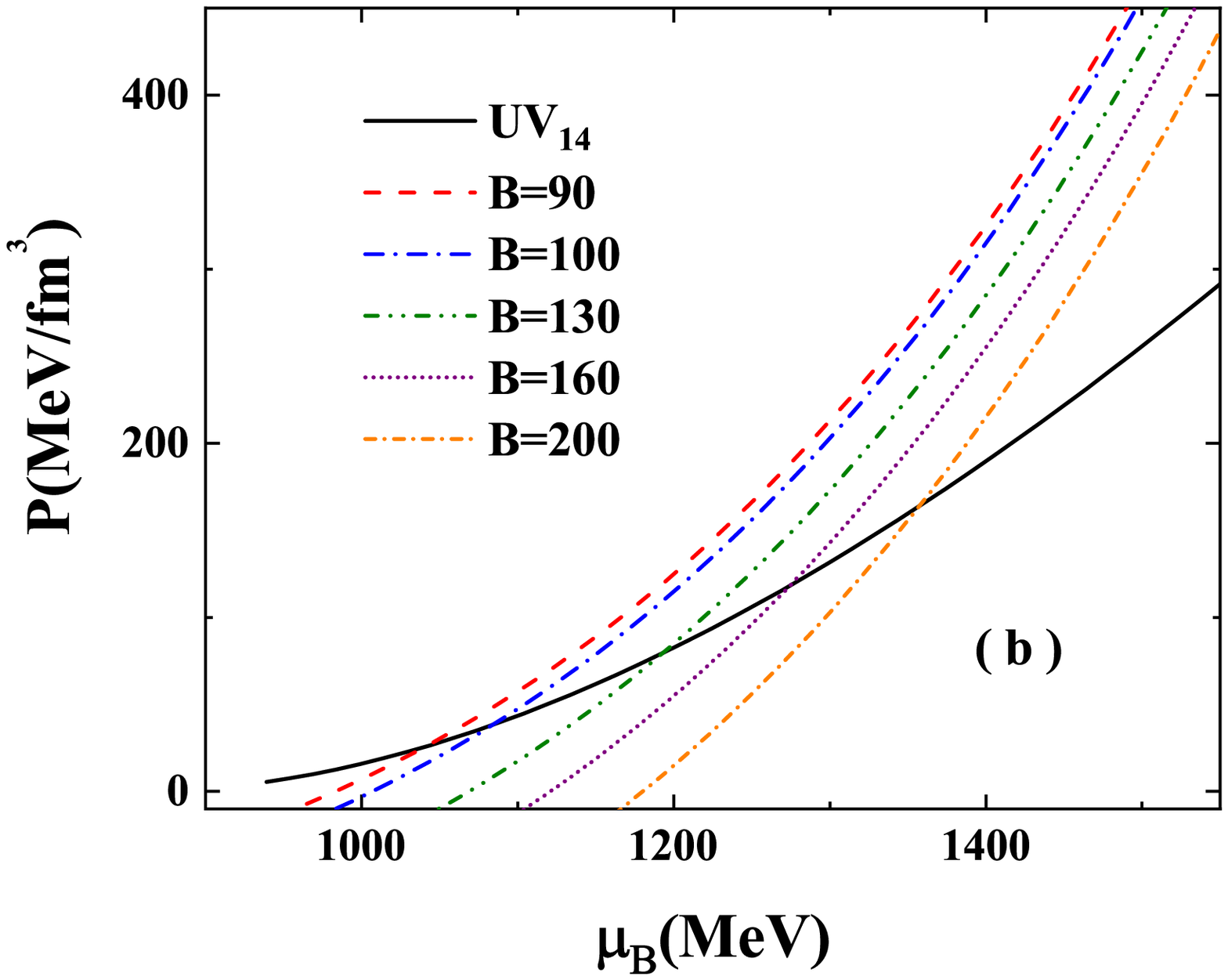}}
	\resizebox{0.325\textwidth}{!}{\includegraphics{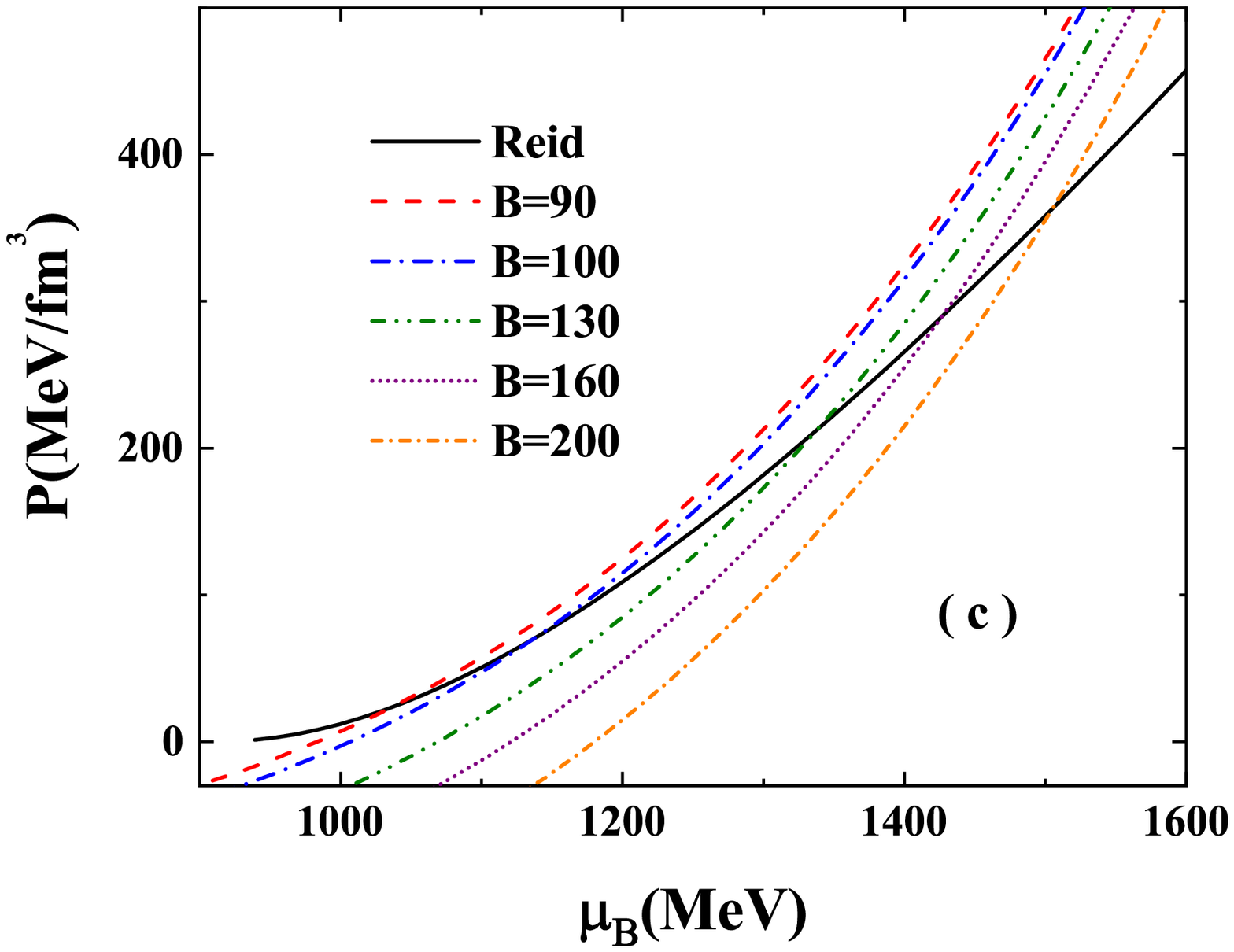}}
	\resizebox{0.325\textwidth}{!}{\includegraphics{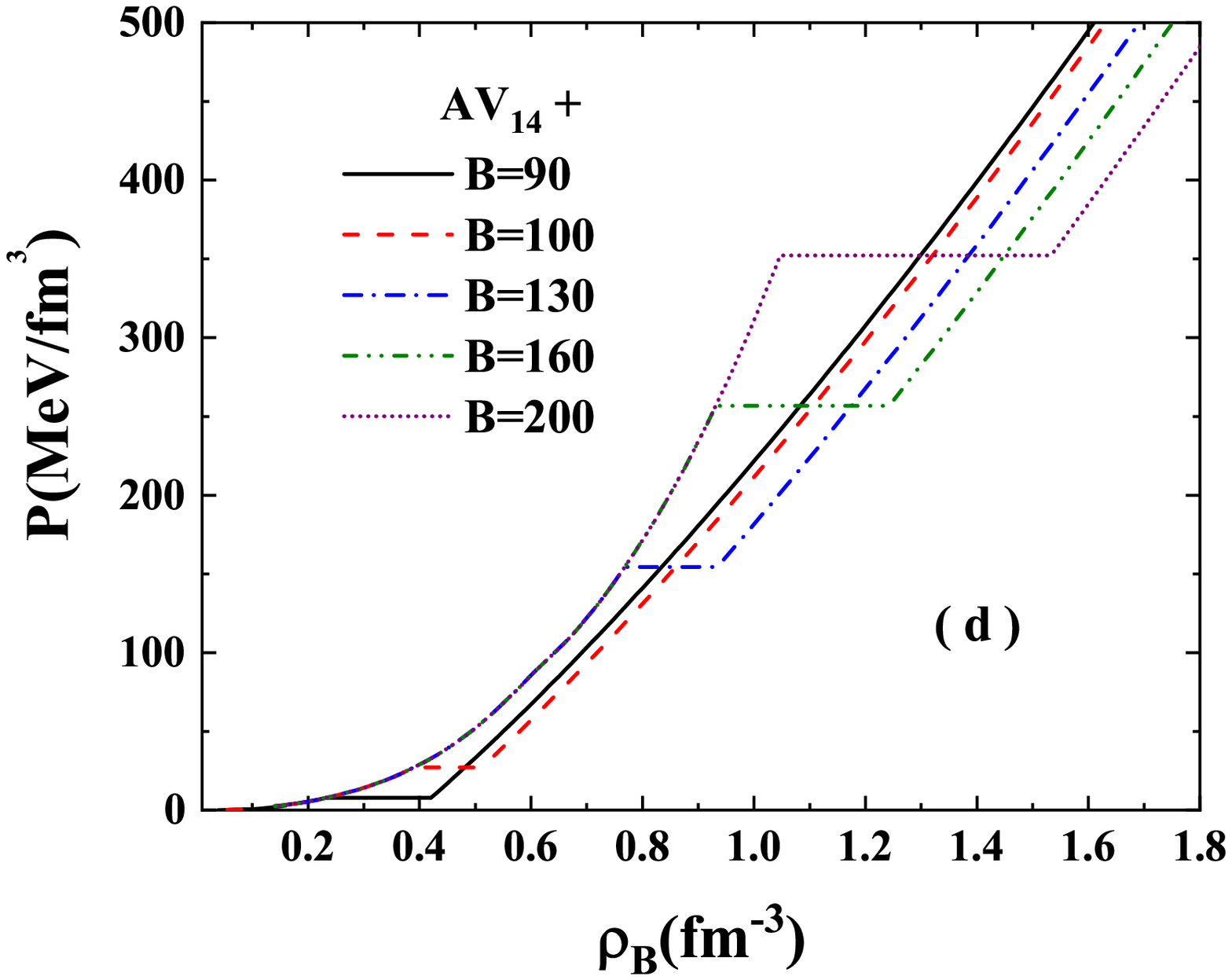}}
	\resizebox{0.325\textwidth}{!}{\includegraphics{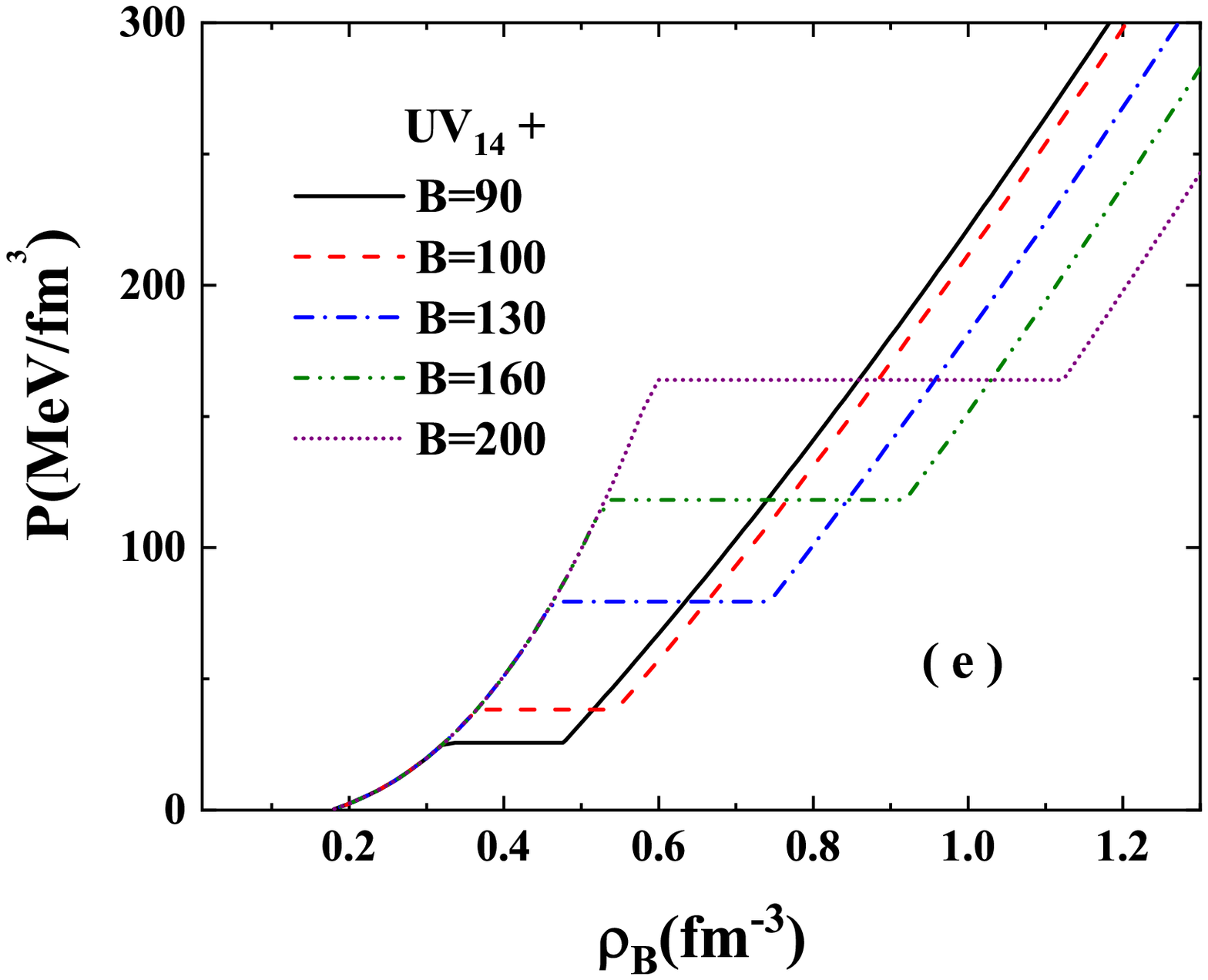}}
	\resizebox{0.325\textwidth}{!}{\includegraphics{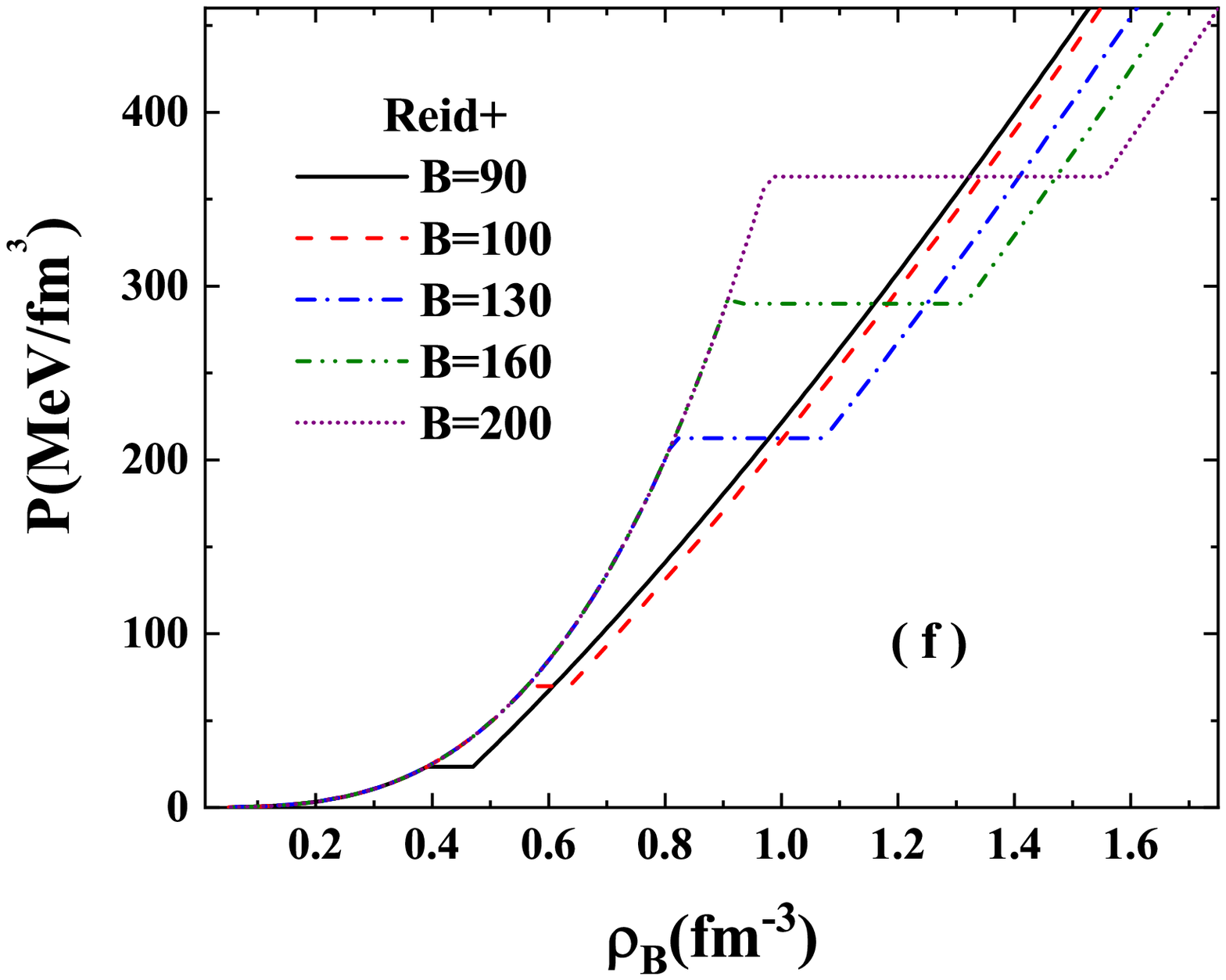}}
	\begin{center}
		\caption{{\small (a), (b), and (c)  : Pressure vs baryon chemical potential for MIT bag model with various bag constants and $ m_{s}=150 $ MeV combined with $ AV_{14} $ , $ UV_{14} $  and Reid 68 interactions supplemented by TBF respectively. (d), (e), and (f) : The corresponding hadron-quark hybrid EoS's in Maxwell construction with $ AV_{14} $ , $ UV_{14} $  and Reid 68 interactions supplemented by TBF. }\label{fig3} } 
	\end{center}
\end{figure*}

To examine the effect of nucleon-nucleon (N-N) interactions on the hadron-quark phase transition region, we have also carried out the calculations with some other bare two-body N-N interactions supplemented with TBF, namely, $ AV_{14}$,  $UV_{14} $, and Reid 68 the results of which are displayed  in Fig.~\ref{fig3}. In Figs.~\ref{fig3}(a), ~\ref{fig3}(b), and ~\ref{fig3}(c), we display quark matter EoSs in the MIT bag model with various bag constants and hadronic matter EOS with $ AV_{14} $,  $UV_{14}  $ and Reid 68 potential, respectively. In Figs.~\ref{fig3}(d), ~\ref{fig3}(e),and ~\ref{fig3}(f),  the corresponding hybrid EOSs are displayed. The transition densities in these cases are almost in similar ranges. Again, the phase transition moves to high baryon densities, and the transition region extends as the bag constant increases. 

 The hadron-quark phase transition properties for various N-N interactions and bag constants are summarized in Table~\ref{t2}, in which it is seen that the range of critical baryon chemical potential is almost similar in all hadron interactions supplemented by the TBF. We further observe that they are lower than the situation in which the TBF is not considered. In the latter, increasing the critical chemical potential increases the critical baryon density and also extends the transition region. As seen in Table~\ref{t2}, in this situation, we observe high-energy density discontinuity in comparison with considering the TBF in any bag constants. In all hadron interaction supplemented by TBF with $B= 90, 100$ MeVfm$^ {-3} $, the energy discontinuity is around $ 200  $ MeVfm$ ^{-3} $ while in the cases  $B= 130, 160, 200 $ MeVfm$ ^{-3} $,  the energy density discontinuities takes larger values. If the energy density discontinuity becomes too high, then the star becomes unstable as soon as the quark matter core appears, which itself is due to the fact that the pressure of the quark matter is unable to cancel out the additional downward force from the gravitational attraction that the additional energy in the core applies on the rest of the star (we will elaborate on this point in Sec.~\ref{structmit} ).

\subsubsection{NJL model }\label{hybridnjl}
\begin{figure*}[htb]
	\vspace{-0.70cm}
	\resizebox{0.325\textwidth}{!}{\includegraphics{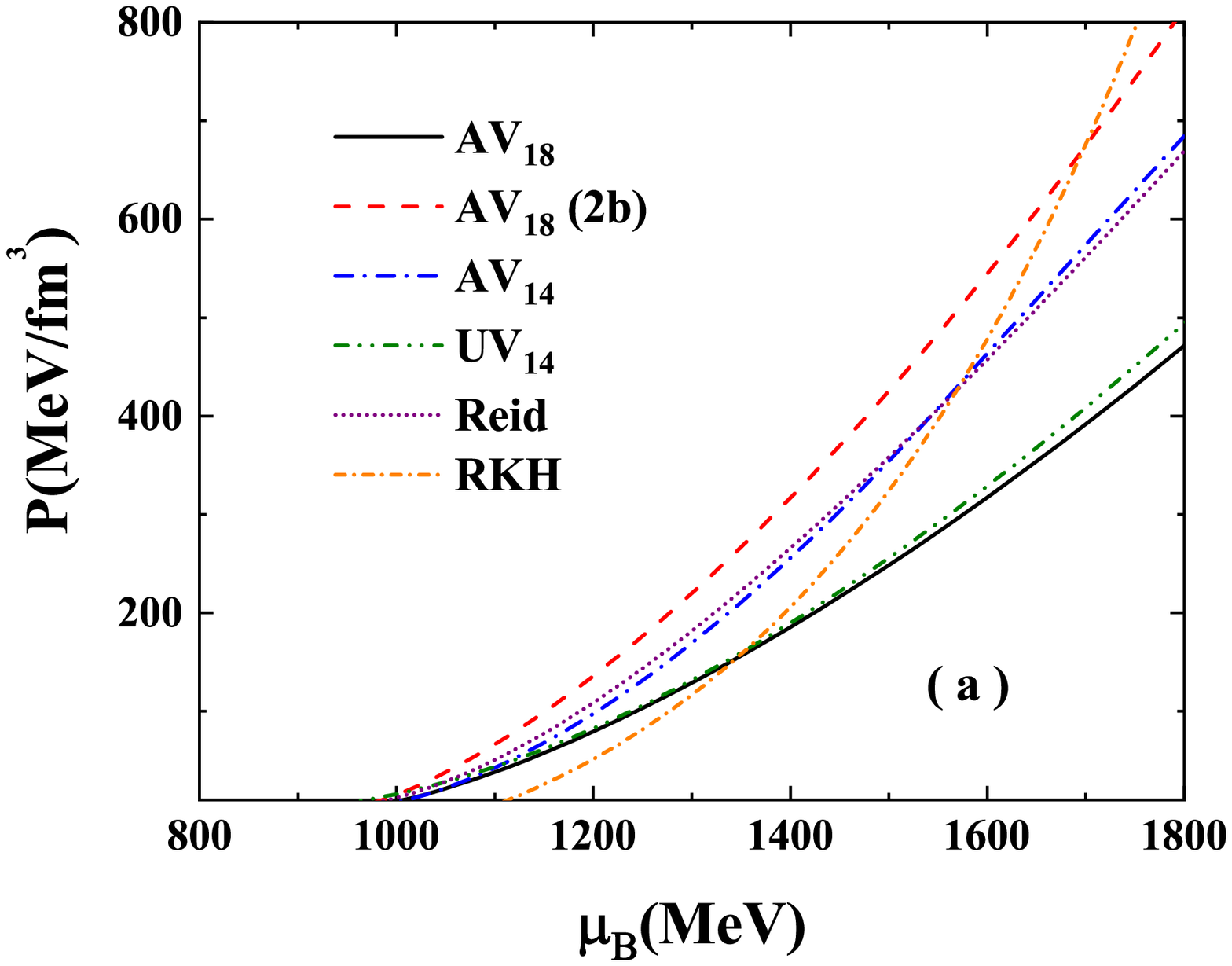}}
	\resizebox{0.329\textwidth}{!}{\includegraphics{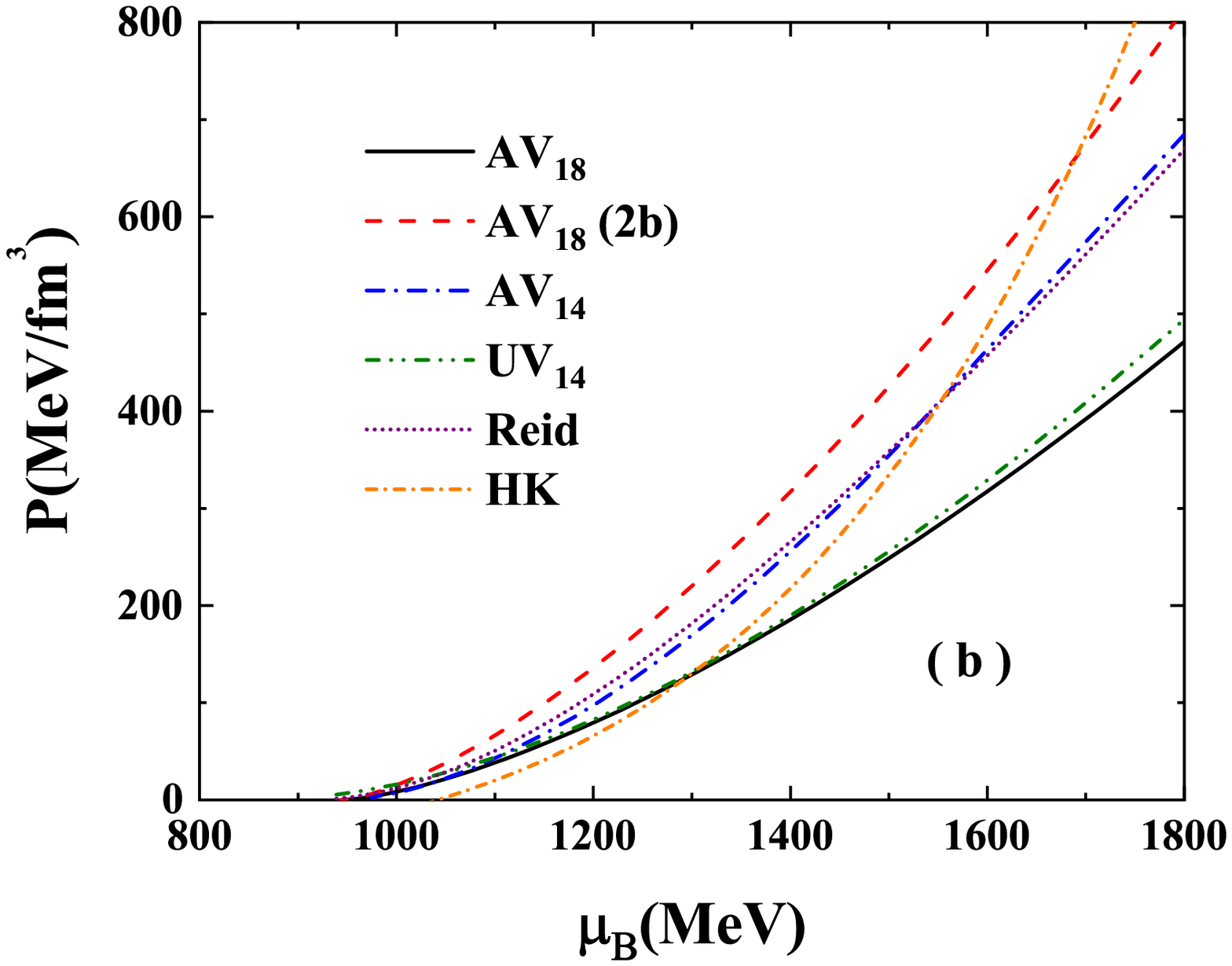}}
	\resizebox{0.329\textwidth}{!}{\includegraphics{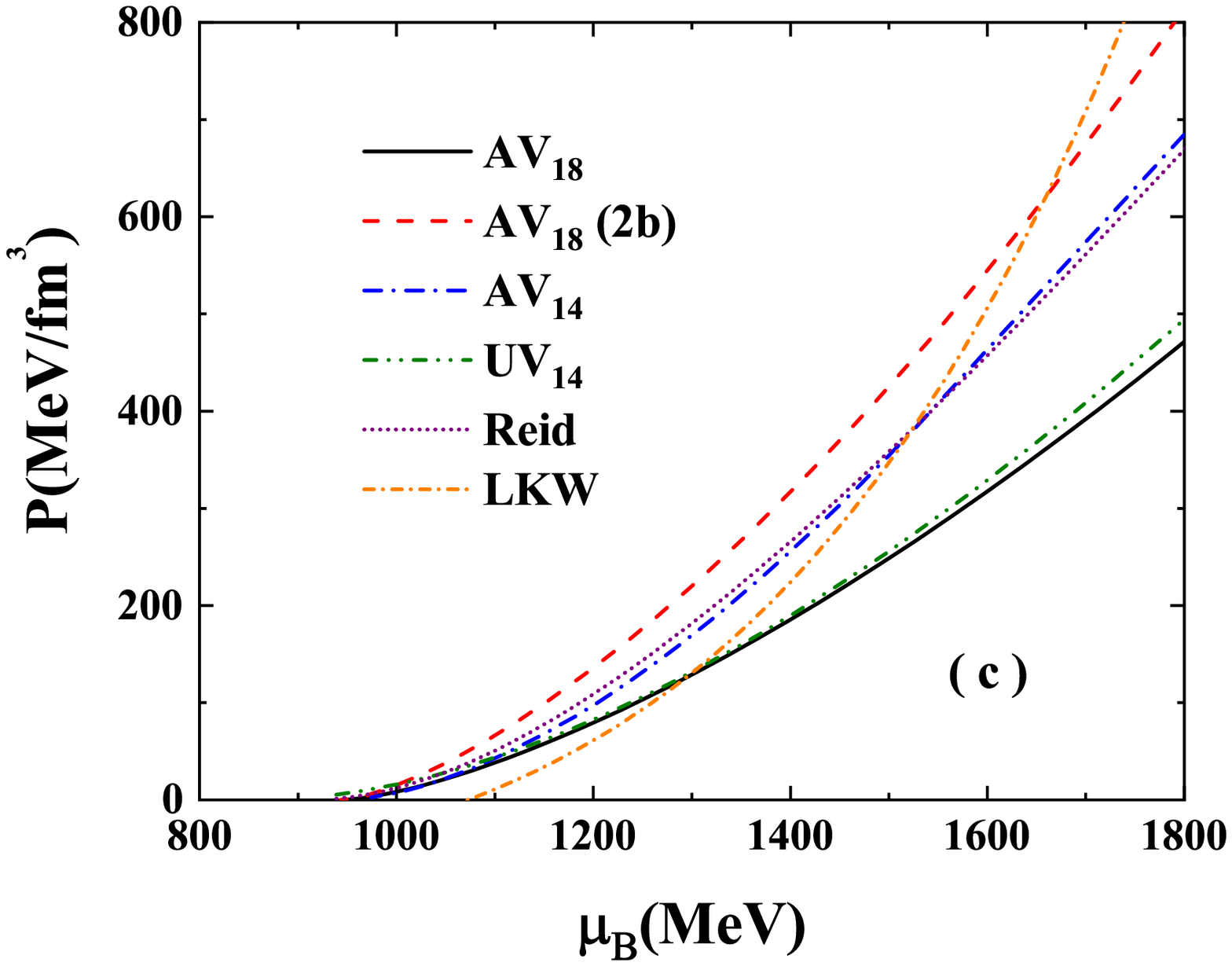}}
	\resizebox{0.329\textwidth}{!}{\includegraphics{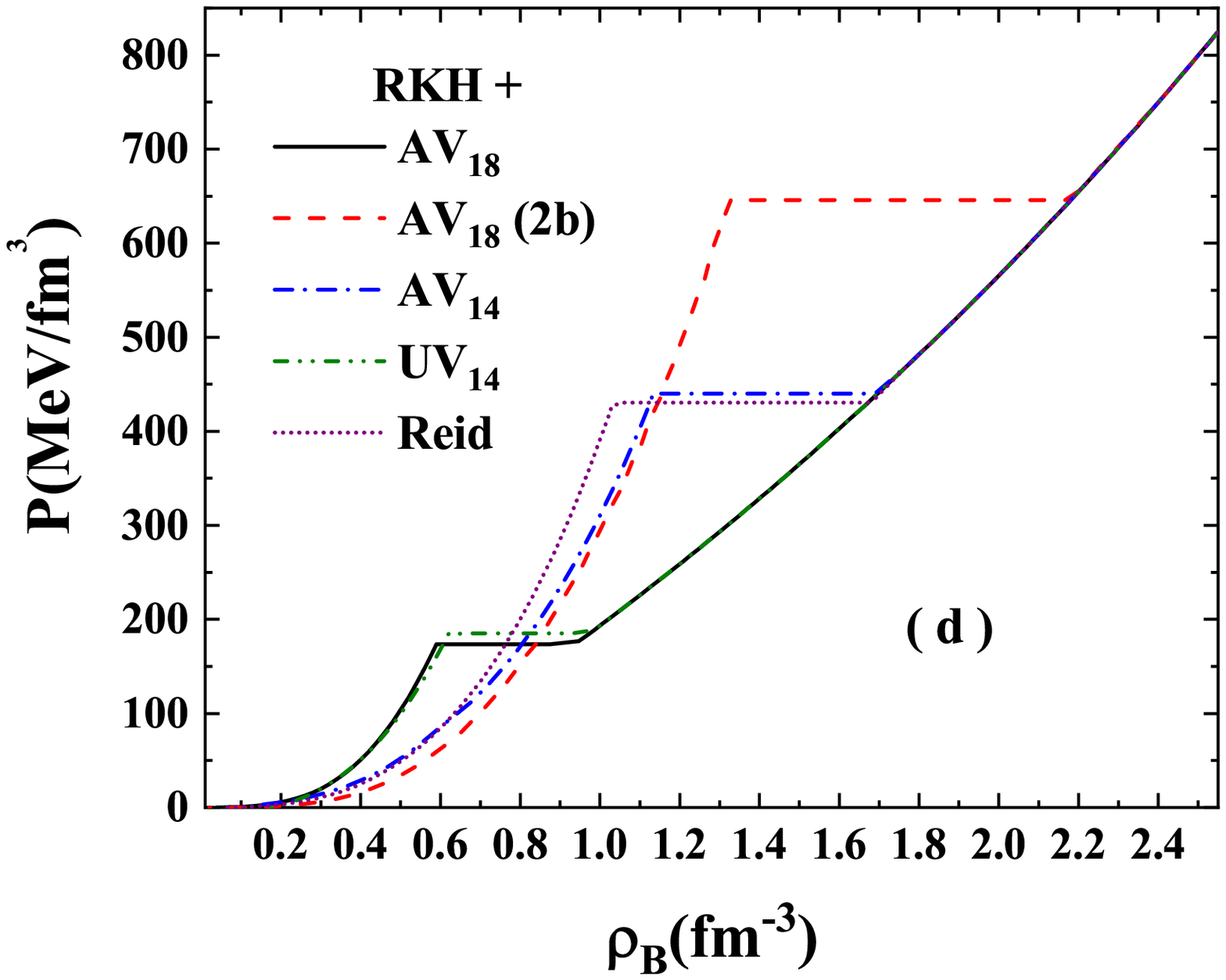}}
	\resizebox{0.329\textwidth}{!}{\includegraphics{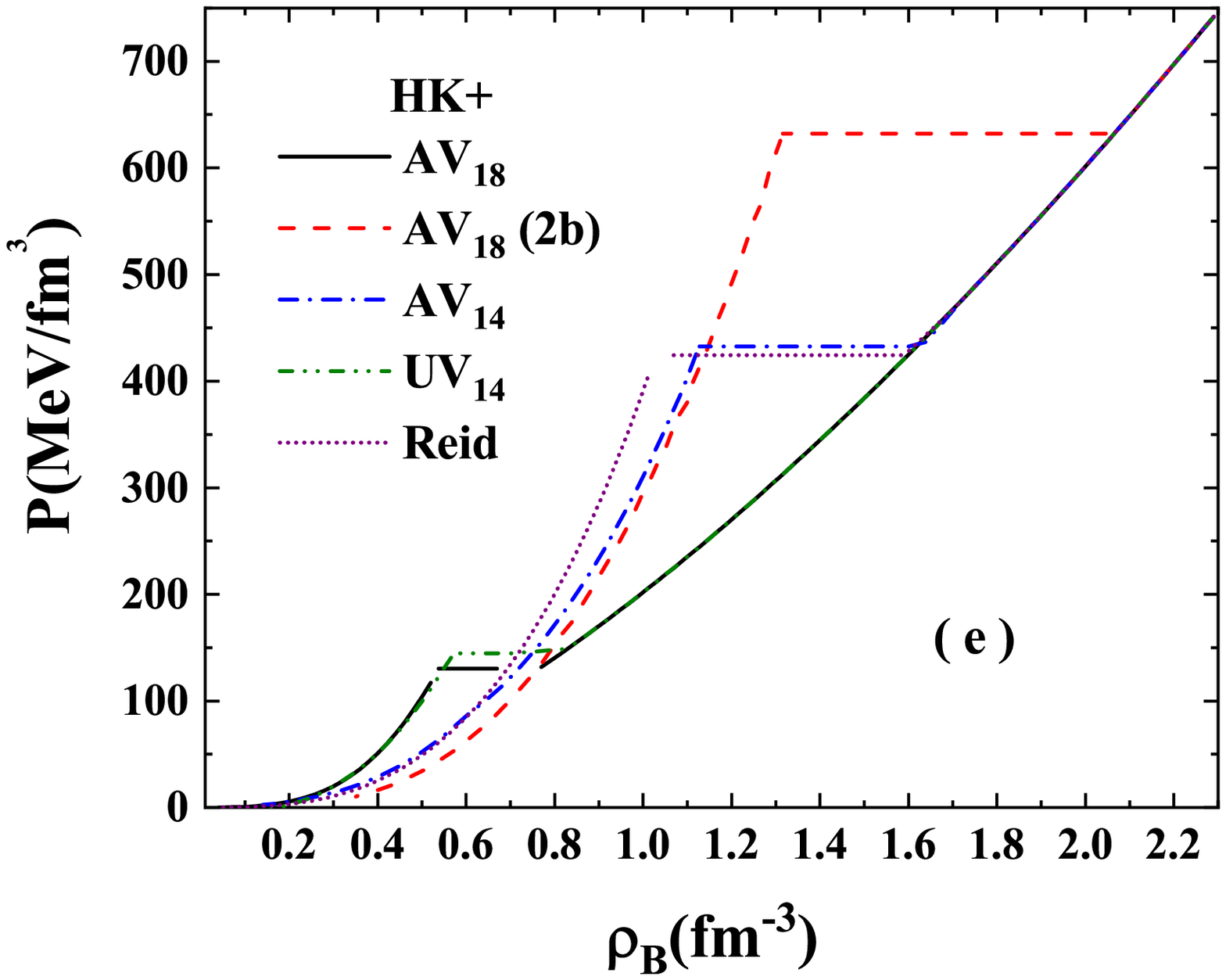}}
	\resizebox{0.329\textwidth}{!}{\includegraphics{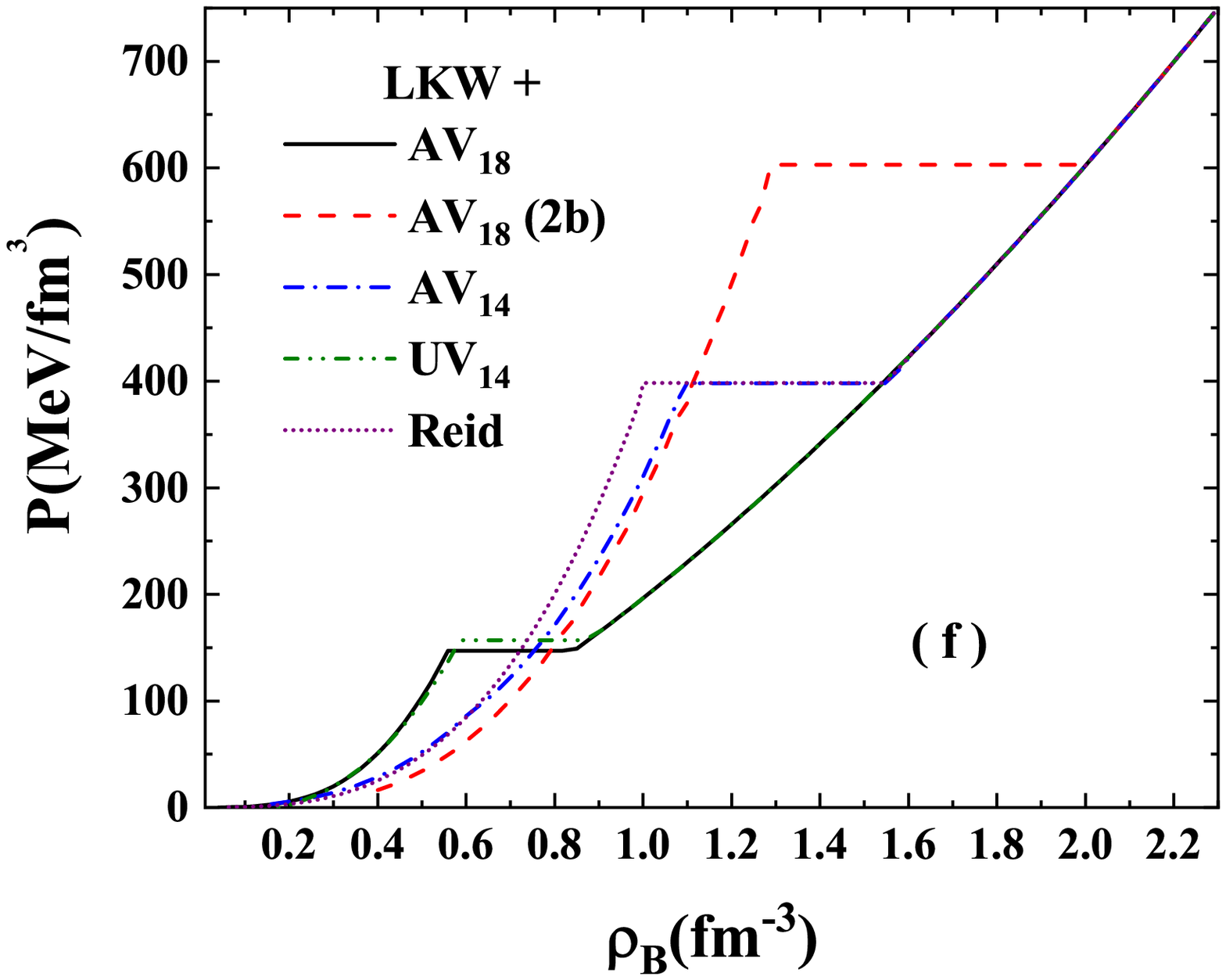}}
	\begin{center}
		\caption{{\small (a), (b), and (c): Pressure vs baryon chemical potential for various hadron interactions combined with NJL in RKH, HK, and LKW parameter sets, respectively. (d), (e), and (f): The corresponding hadron-quark hybrid EOS in Maxwell construction with NJL in RKH, HK, and LKW parameter sets.} \label{fig4}}
	\end{center}
\end{figure*}

Now, we present the numerical results for the NJL model for the quark matter sector. In Figs.~\ref{fig4} (a), ~\ref{fig4} (b), and ~\ref{fig4} (c), we display the pressure vs baryon chemical potential of the hadronic matter EOS with various hadron interactions and quark matter EOS within the NJL model for  RKH, HK and LKW, parameter sets respectively where as,Figs.~\ref{fig4} (d), ~\ref{fig4} (e), and ~\ref{fig4} (f), are the corresponding hybrid EOS. It is worth noting that in all above cases the phase transition  occurs. In RKH parameter set, the transition density occurs in a little higher  chemical potential than the cases in the  HK and LKW parameter sets for all hadronic EOS. By using only two-body N-N potential, the phase transition density occurs in considerably large densities compared with considering the three-body forces in all potentials, for any three parameter sets. Here, the starting point of phase transition occurs in about $ 7\rho_{0} $ but with TBF, it occurs in slightly more than $ 3\rho_{0} $  with  $ AV_{18} $ combined with TBF.  

Table~\ref{t3} summarizes the hadron-quark phase transition properties for various N-N interactions and various parameter sets of the NJL model. As seen in this table, and in comparison with Table~\ref{t2}, when we use the NJL as the quark model, the values of critical chemical potentials increase relative to the MIT bag model with the same hadron interactions and correspond to large values of the bag constant. With the increasing the chemical potential, the critical baryon density rises and the phase transition region extends  which enlarges the energy discontinuity for any parameter sets. Except for the HK parameter set combined with the $ AV_{18} $ and $ UV_{14} $  supplemented by TBF for which the energy discontinuity is lower than around $ 200$ MeVfm$ ^{-3} $, in the other hadron interactions with and without TBF, the energy discontinuities are around $ 400-1300 $ MeVfm$ ^{-3} $, and these values are too large to retain the stability of the star. As we mentioned earlier, if the energy density discontinuity becomes considerably too high, the star becomes unstable. Since the pressure of the quark matter is unable to counteract the additional downward force from the gravitational attraction (we will refer to this point in Sec.~\ref{structnjl}).

\subsection{Hybrid star structure }\label{III-B}

The structure of a hybrid star is calculated by numerical integration of the Tolman-Oppenheimer-Volkoff  (TOV) equations:
\begin{eqnarray}
\frac{dP(r)}{dr} =&&-\frac{GM(r)\epsilon (r)}{c^{2} r^{2} } (1+\frac{P(r)}{\epsilon (r)} )(1+\frac{4\pi r^{3} P(r)}{M(r)c^{2} } )
\nonumber\\
&&\times(1-\frac{2GM(r)}{rc^{2} } )^{-1},
\end{eqnarray}
\begin{equation}
\frac{dM(r)}{dr} =\frac{4\pi \epsilon (r)r^{2} }{c^{2} },
\end{equation}
in which $\epsilon (r)$ is the total energy density, $ M(r) $ is the star mass within radius $ r $, and $G$ denotes the gravitational constant.

\subsubsection{MIT bag model }\label{structmit}

\begin{figure*}[htb]
	\vspace{-0.70cm}
	\resizebox{0.45\textwidth}{!}{\includegraphics{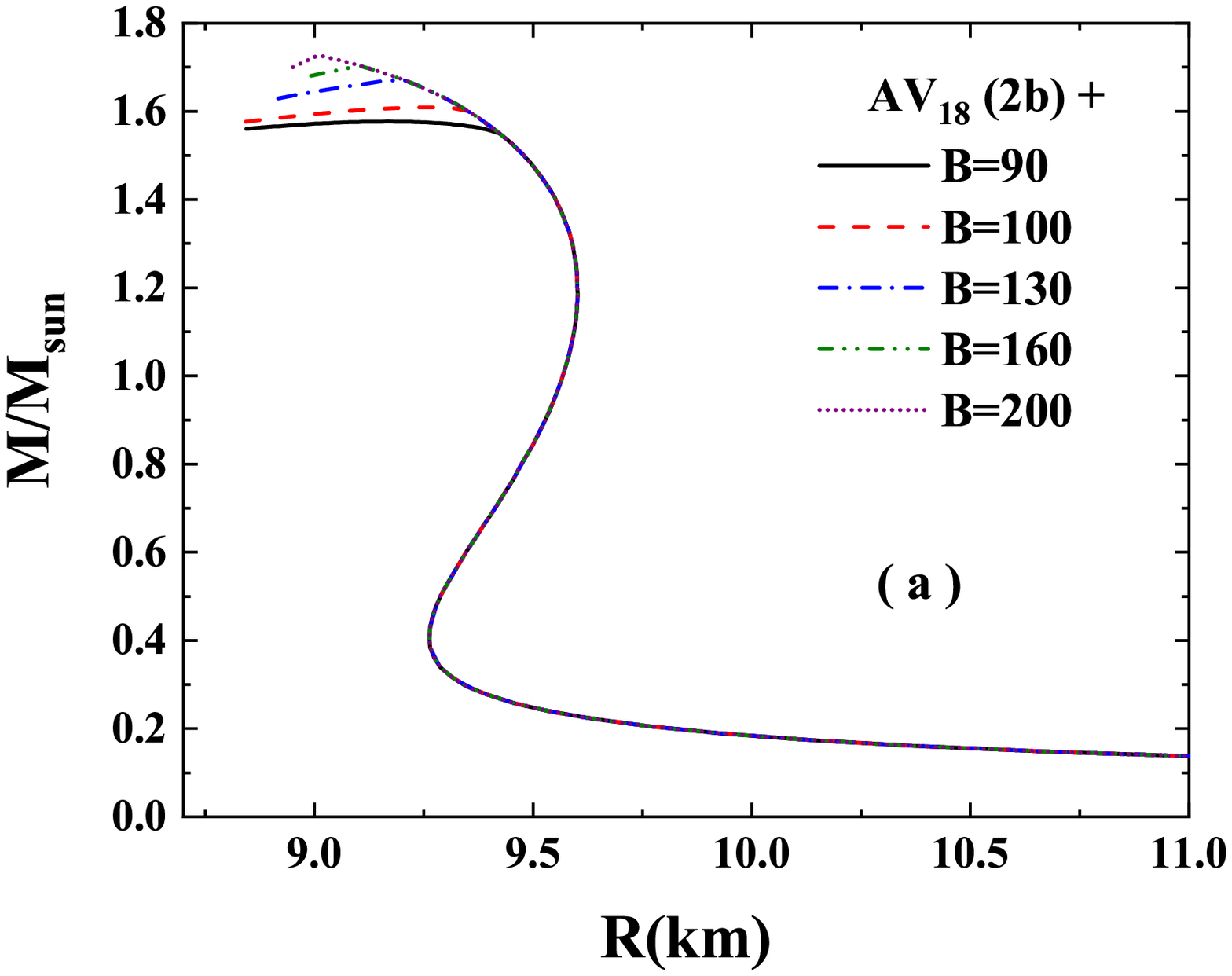}}
	\resizebox{0.44\textwidth}{!}{\includegraphics{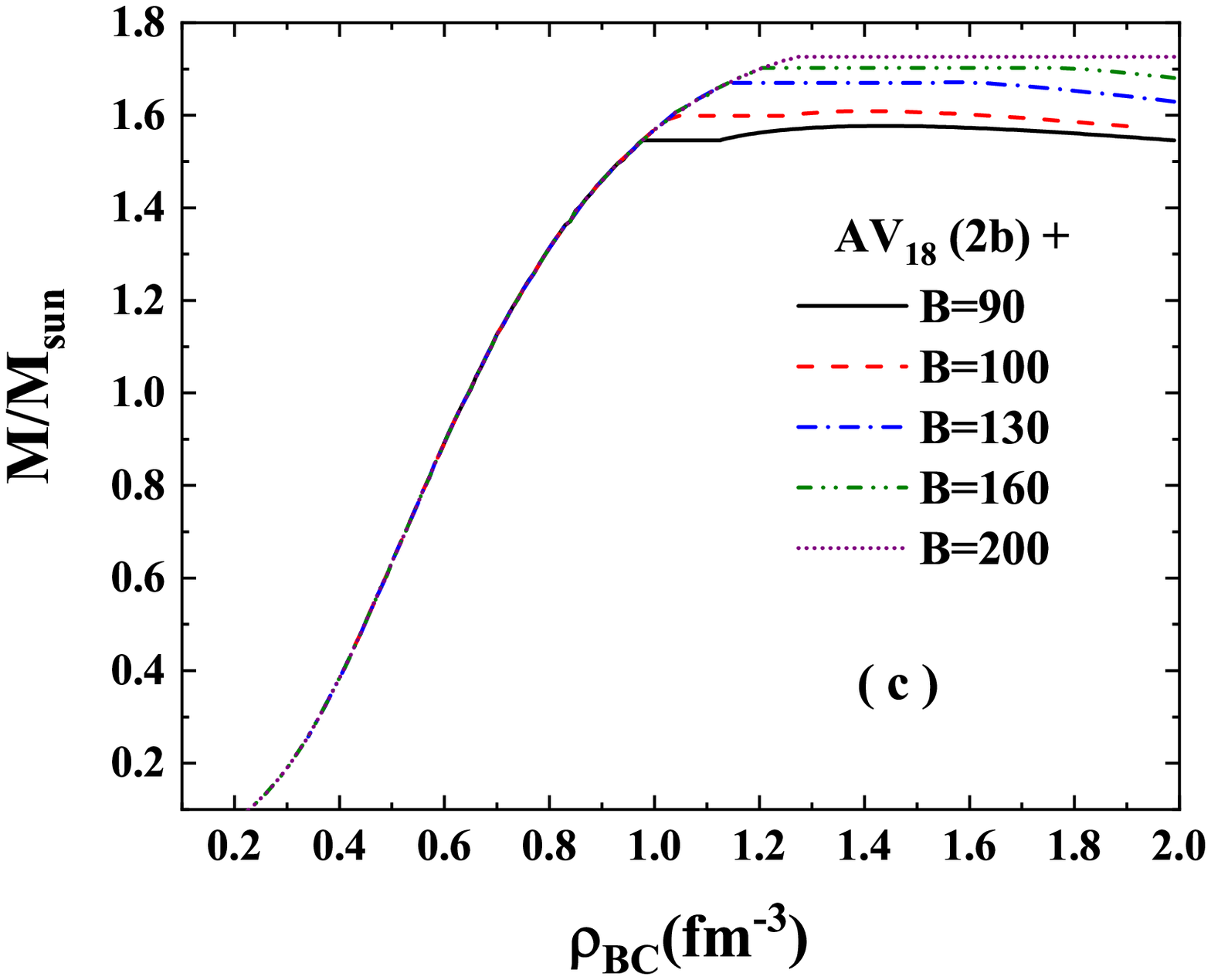}}
	\resizebox{0.45\textwidth}{!}{\includegraphics{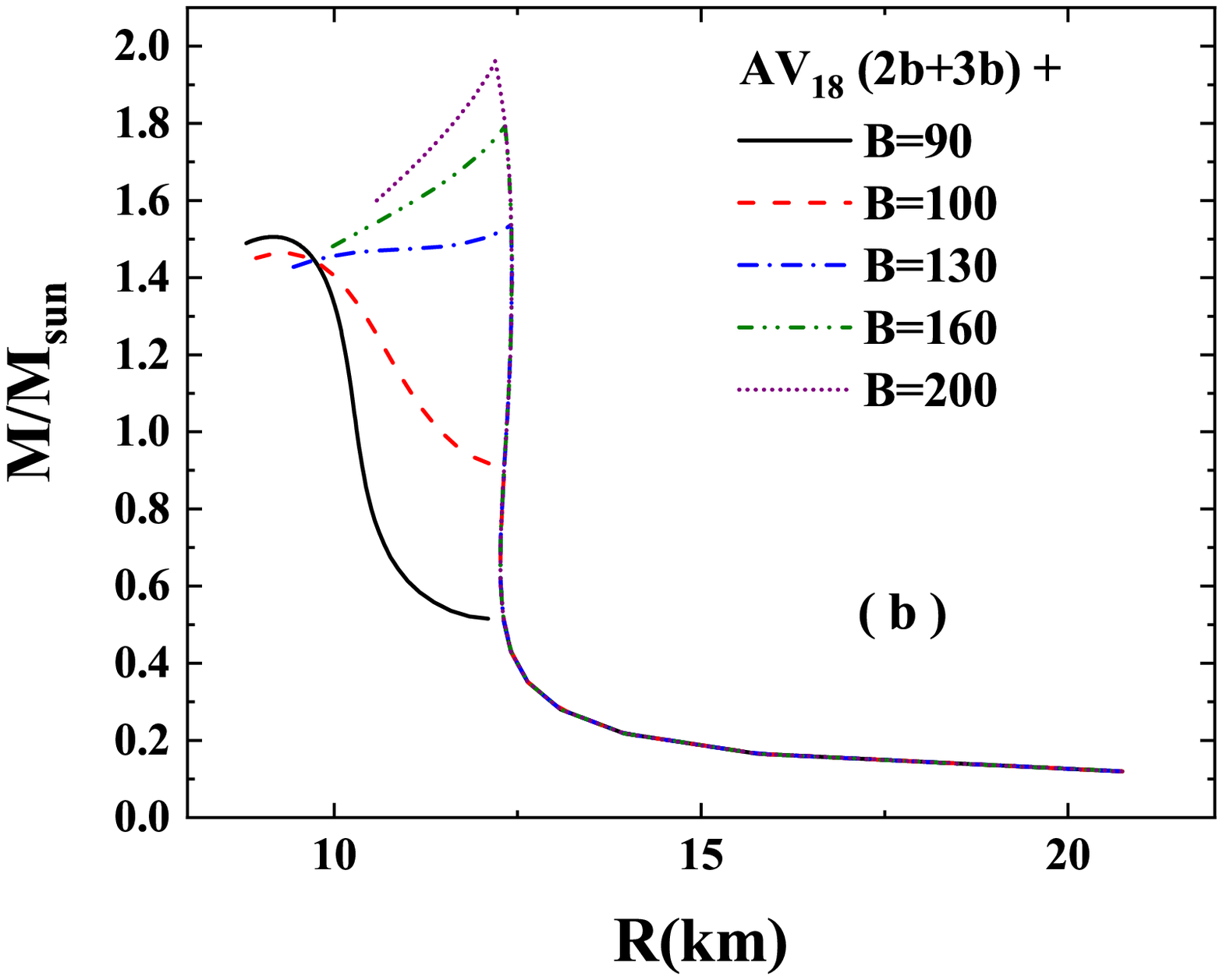}}
	\resizebox{0.44\textwidth}{!}{\includegraphics{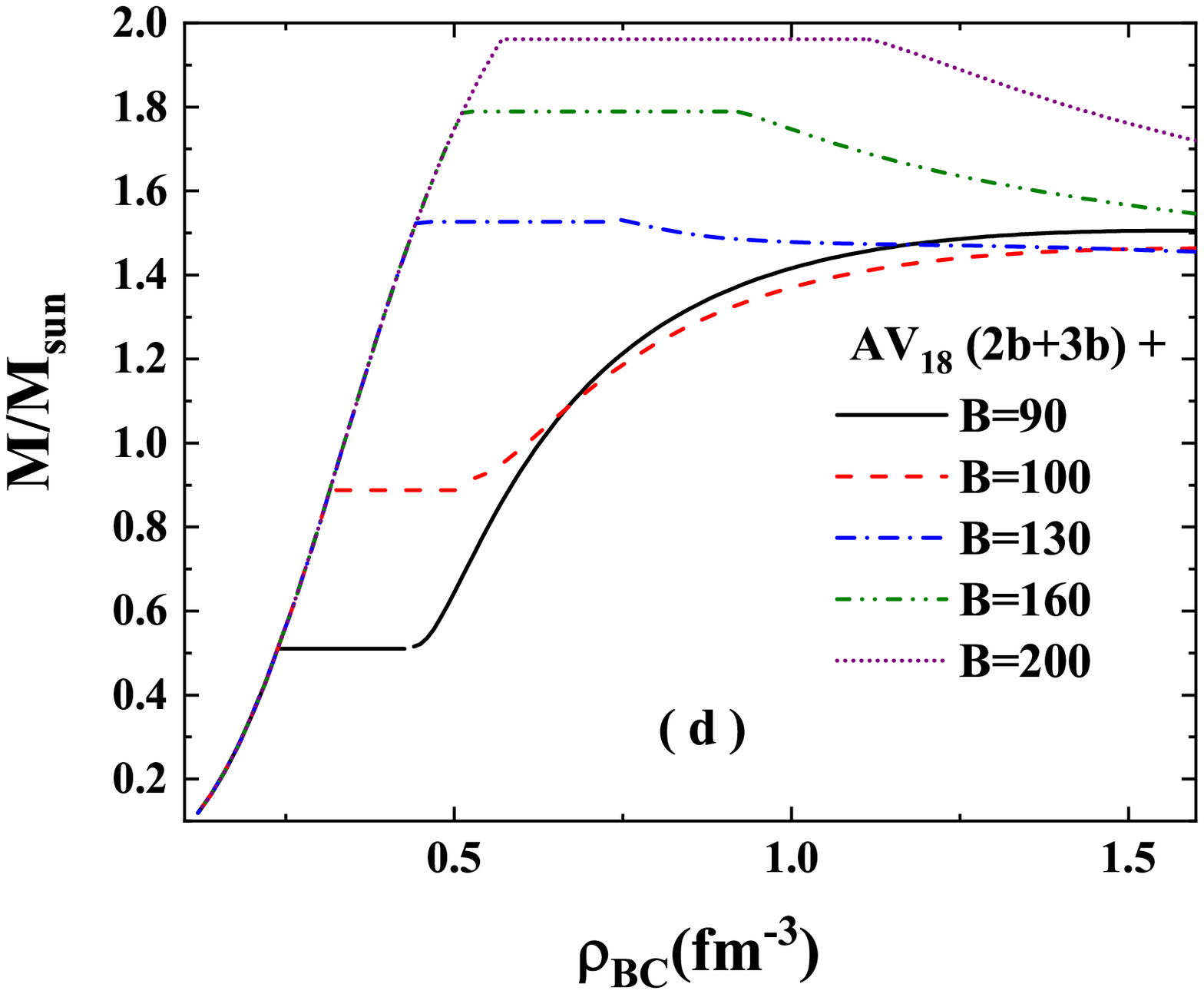}}
	\begin{center}
		\caption{{\small (a) [(b)]: The gravitational HS masses vs radius of the star for the MIT bag model with various bag constants and $ m_{s}=150 $ MeV combined with AV18 interaction supplemented without [with] TBF. (c) [(d)]: The corresponding gravitational HS mass vs central baryon density of the star without [with] TBF. } \label{fig5}}
	\end{center}
\end{figure*}

\begin{figure*}[htb]
	\vspace{-0.70cm}
	
	\resizebox{0.325\textwidth}{!}{\includegraphics{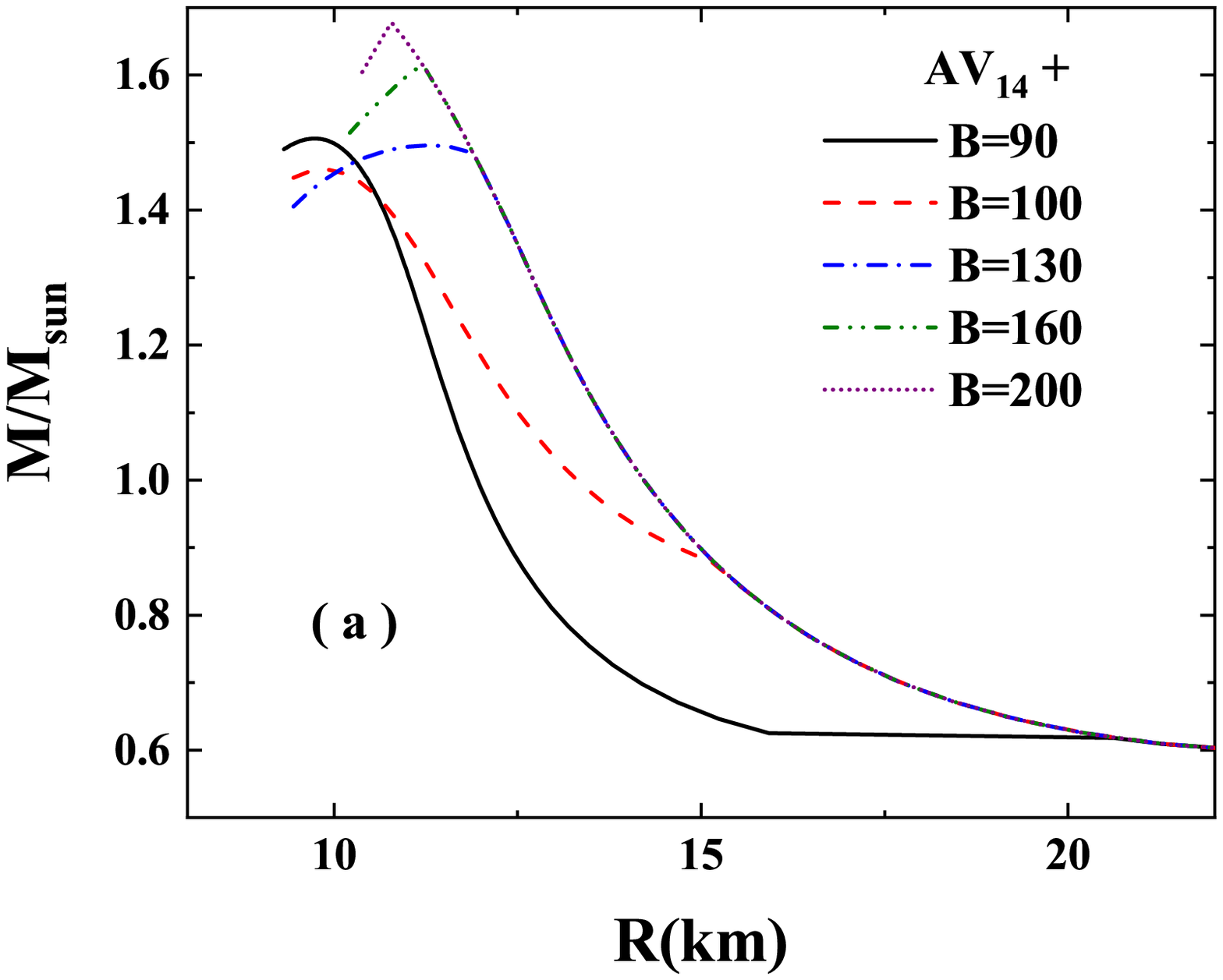}}
	\resizebox{0.325\textwidth}{!}{\includegraphics{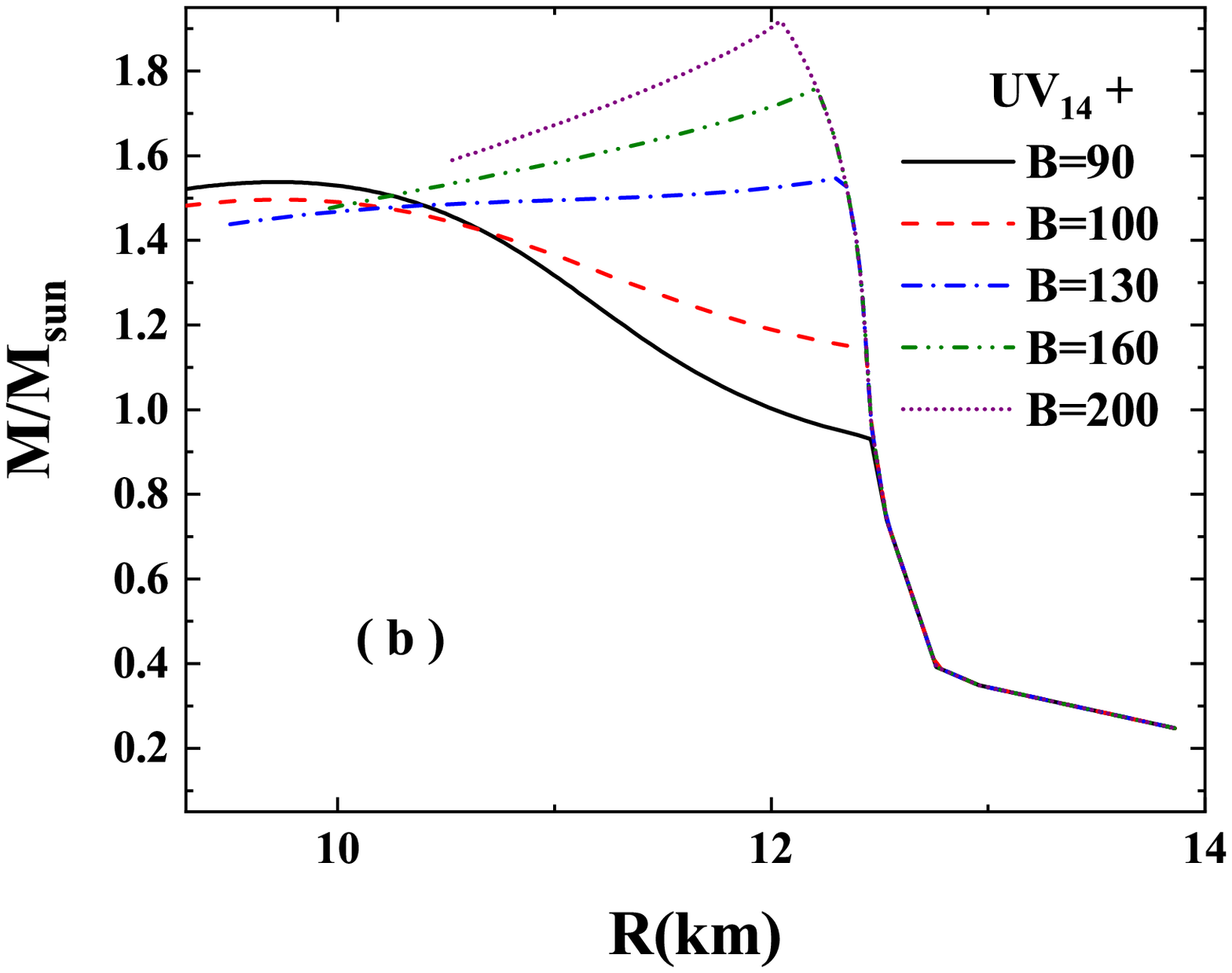}}
	\resizebox{0.325\textwidth}{!}{\includegraphics{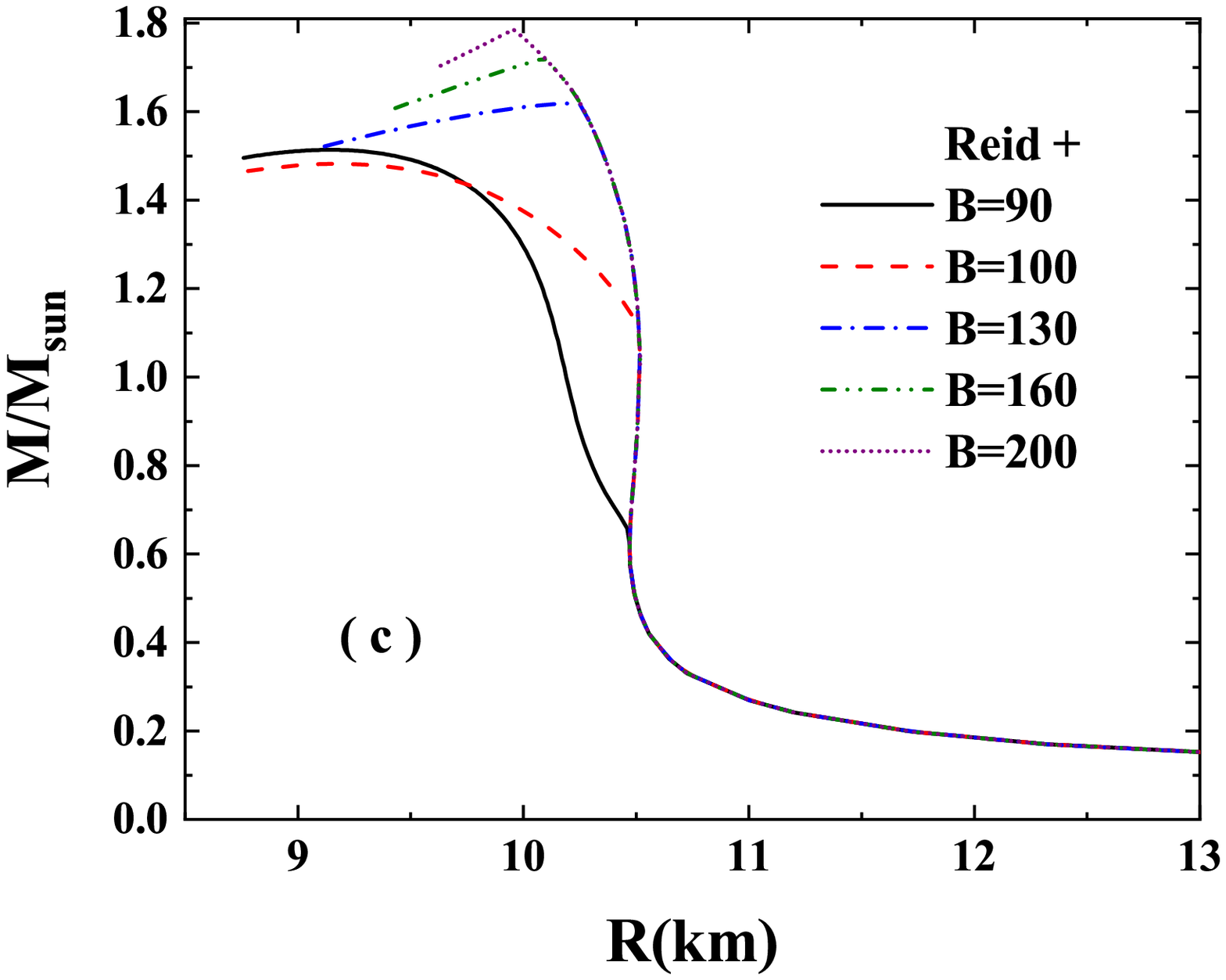}}
	\resizebox{0.325\textwidth}{!}{\includegraphics{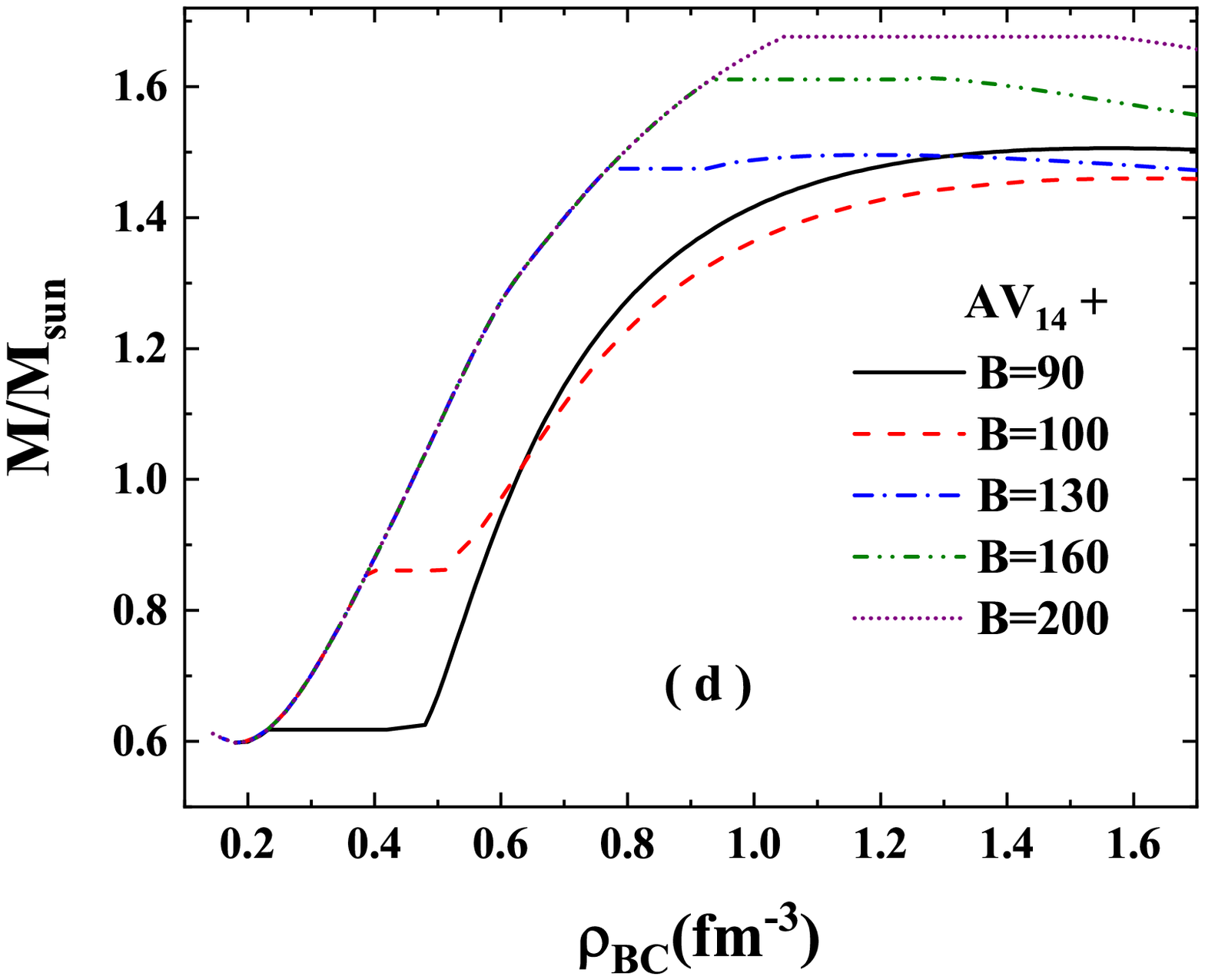}}
	\resizebox{0.325\textwidth}{!}{\includegraphics{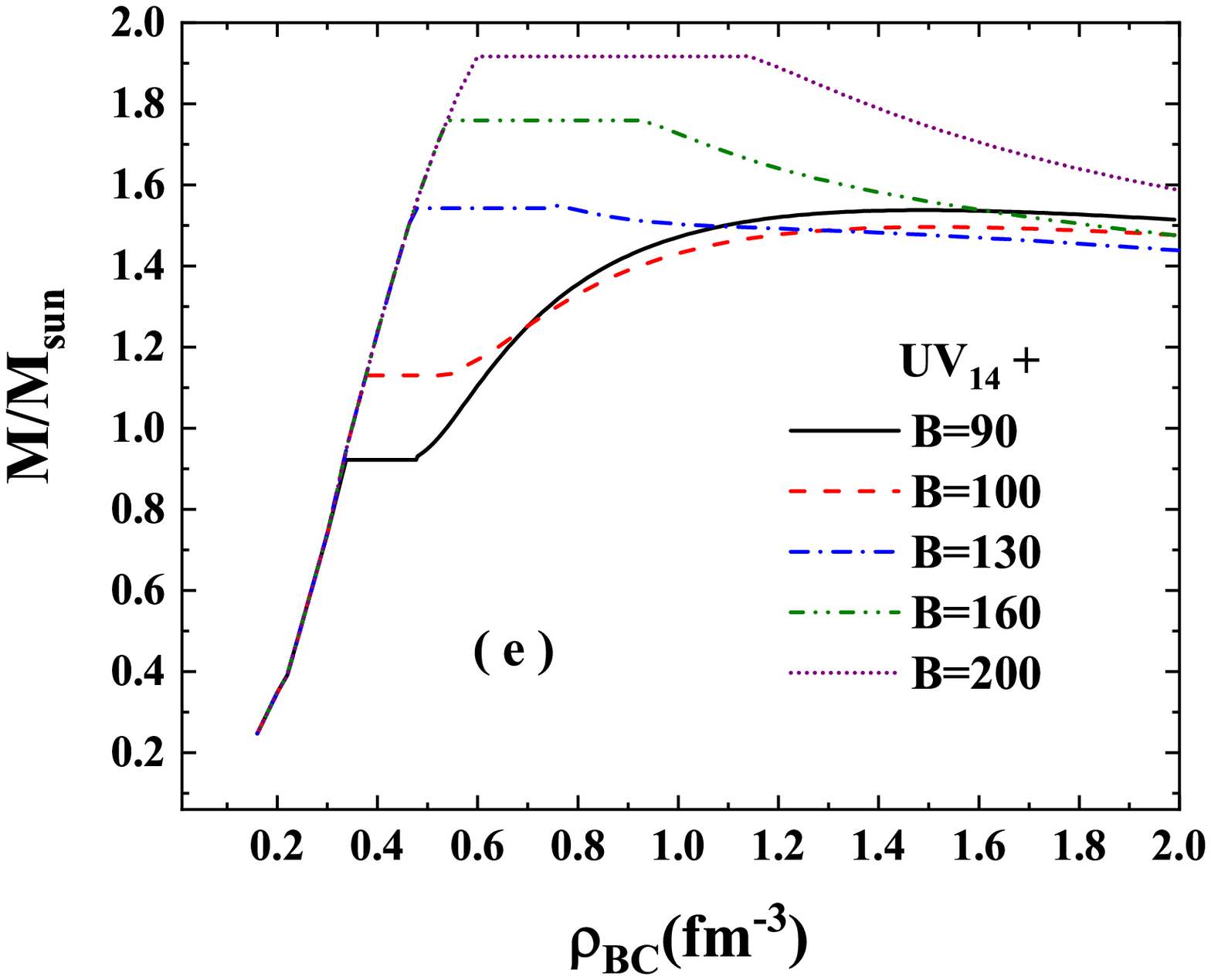}}
	\resizebox{0.325\textwidth}{!}{\includegraphics{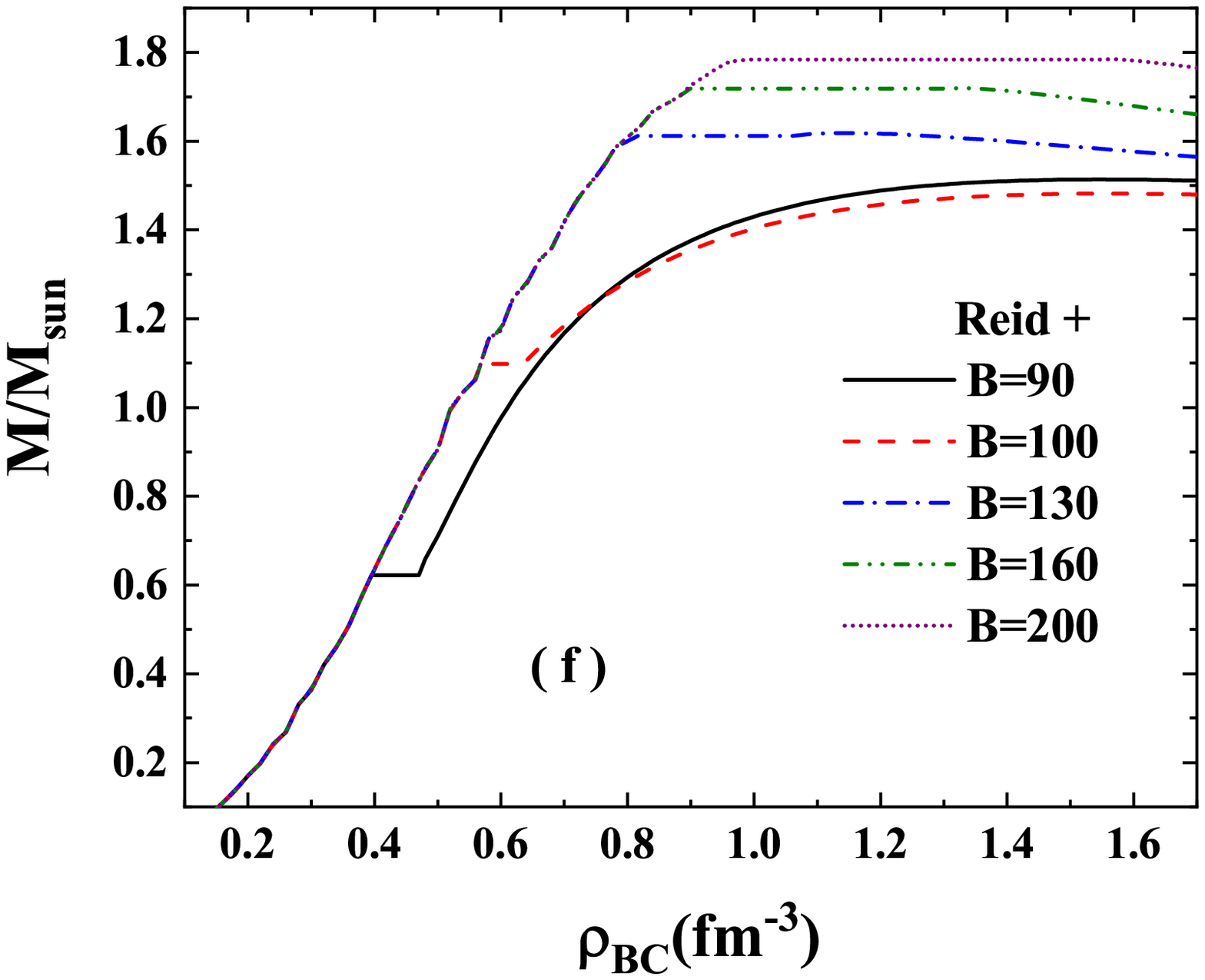}}
	\begin{center}
		\caption{{\small (a), (b), and (c): The gravitational HS masses vs the radius of the star within the MIT bag model with various bag constants and			$ m_{s}=150 $ MeV combined with $ AV_{14} $, $ UV_{14} $, and Reid 68 hadron interactions supplemented by TBF, respectively. (d), (e), and (f): The corresponding gravitational HS mass vs central baryon density of the star with $ AV_{14} $, $ UV_{14} $, and Reid 68 hadron interactions supplemented by TBF. } \label{fig6}}
	\end{center}
\end{figure*}

First, the result of a HSs structure concerning the MIT bag as quark model is presented. In Figs.~\ref{fig5}(a) [~\ref{fig5}(b)], we plot the gravitational HS mass vs radius within the MIT bag model with $ m_{s}=150 $ MeV and various bag constants, combined with $ AV_{18} $ interaction supplemented without [with] TBF and in Figs.~\ref{fig5}(c) [~\ref{fig5}(d)],  the corresponding gravitational HS mass vs central baryon density of the star, $ \rho_{BC} $, are displayed. The results of other N-N interactions are displayed in Fig.~\ref{fig6}. In Figs.~\ref{fig6}(a), ~\ref{fig6}(b), and ~\ref{fig6}(c), we plot the mass radius relation for HSs within  MIT bag model with $ m_{s}=150 $ and various bag constants combined with $ AV_{14} $ , $UV_{14} $ and Reid 68 interactions supplemented by TBF. In Figs.~\ref{fig6}(d), ~\ref{fig6}(e), and ~\ref{fig6}(f), the corresponding gravitational HS mass vs central baryon density of the star is displayed. 

 Stable hybrid star with pure quark core is predicted with $ B=90 ,100 $ MeVfm$ ^{-3} $ combined with all hadron interactions with or without TBF and also $ B=130 $ MeVfm$ ^{-3} $ combined with $ AV_{14} $ interaction supplemented by TBF. The strange mass is considered $ m_{s}=150 $ Mev in above cases. As seen in Figs.~\ref{fig5} and~\ref{fig6}, in these cases, the  mass vs central baryon density curves are obviously increasing after the onset of the pure quarks in the core of the star, and also mass-radius curves are smooth at the maximum point. In other cases, although the mass vs central baryon density curves show a slight increasing behavior after the onset of pure quarks in the core through a small density range, there is a cusp at the maximum point in mass-radius curve in all of them. However, maybe it is not clearly visible in some cases, but since the central baryon density in which the maximum mass occurs is a little higher than transition density of quark to hadron, the curves continue to increase through a very small density range. The values of densities are comparable in Tables ~\ref{t2} and~\ref{t4}. As seen in Table~\ref{t2},, in the cases of stable hybrid star with pure quark core, the energy density discontinuity is low,  while in other cases in which the energy density discontinuity extend more, a stable HS with a pure quark core is not accessible. In the latter, the large energy discontinuity at the transition point is reflected as a cusps on the mass radius relation. These cusps are clearly visible in mass-radius curves of Figs.~\ref{fig5} and~\ref{fig6}. The effects are strong enough to render the star unstable.
 The maximum mass of the stable HS with pure quark core is about $ 1.5M_{\odot} $ in all hadron interaction cases.For the HS with $ B=200 $MeVfm$ ^{-3} $, combined with $ AV_{18} $ interaction supplemented by TBF, however, the maximum mass reaches 1.962 which is compatible with the observations, the star becomes unstable as soon as the quark phase onset in the core of the star.
	
In Table~\ref{t4}, we summarize the structure properties of pure neutron and hybrid stars for various N-N interactions and bag constants of the MIT bag model with $ m_{s}=150 $ MeV. 
 
\begin{figure*}[htb]
	\vspace{-0.70cm}
	
	\resizebox{0.325\textwidth}{!}{\includegraphics{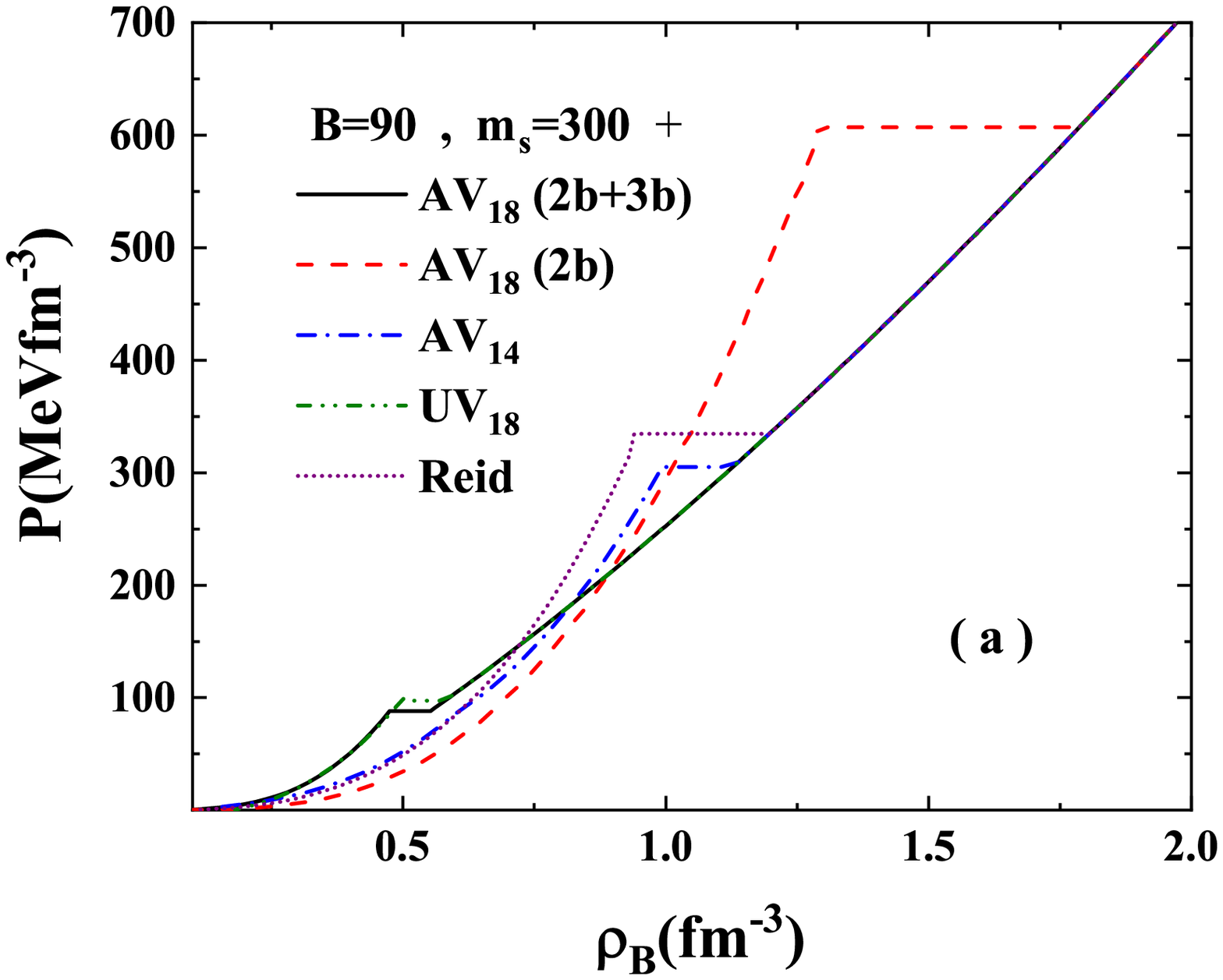}}
	\resizebox{0.329\textwidth}{!}{\includegraphics{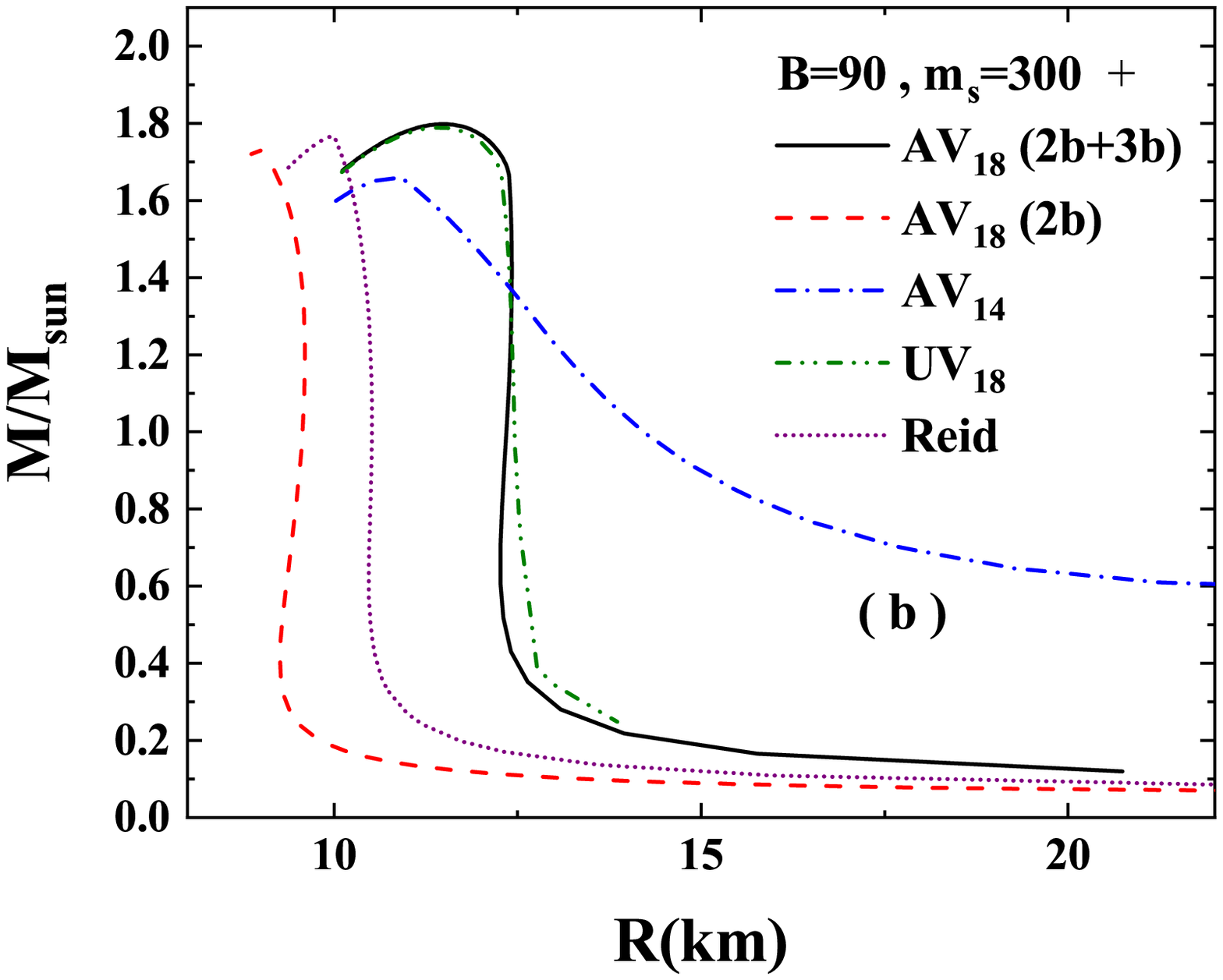}}
	\resizebox{0.329\textwidth}{!}{\includegraphics{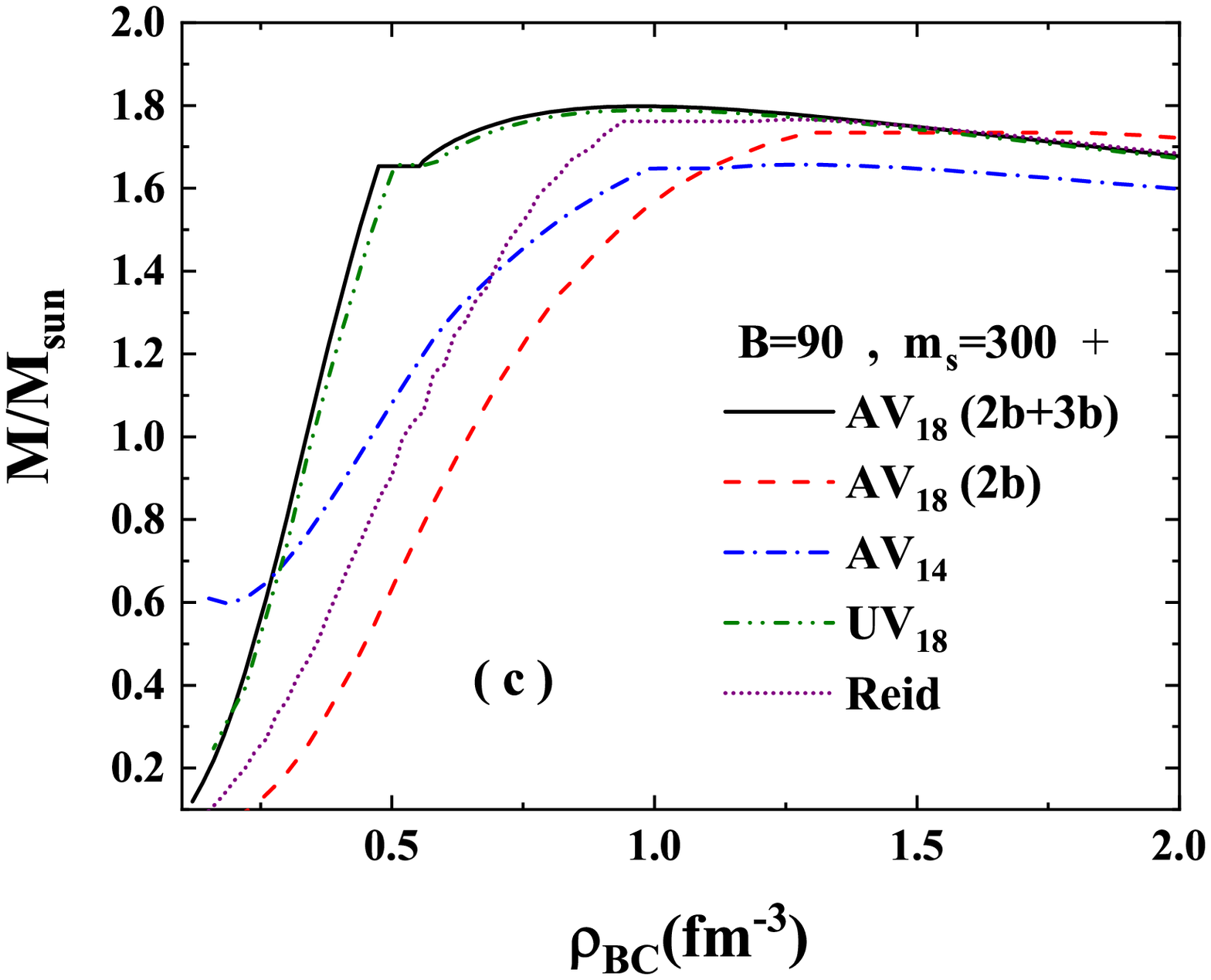}}
	\resizebox{0.329\textwidth}{!}{\includegraphics{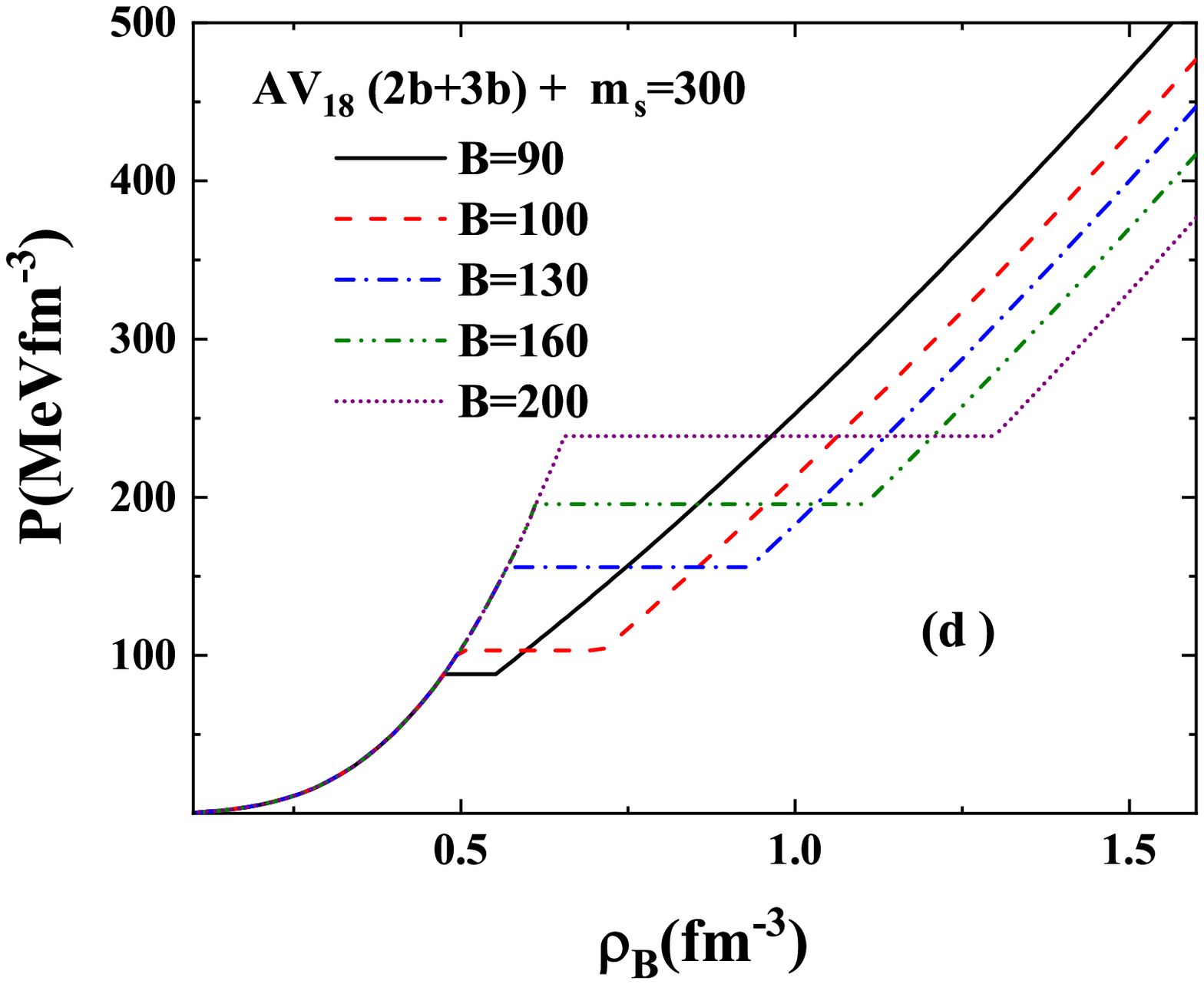}}
	\resizebox{0.329\textwidth}{!}{\includegraphics{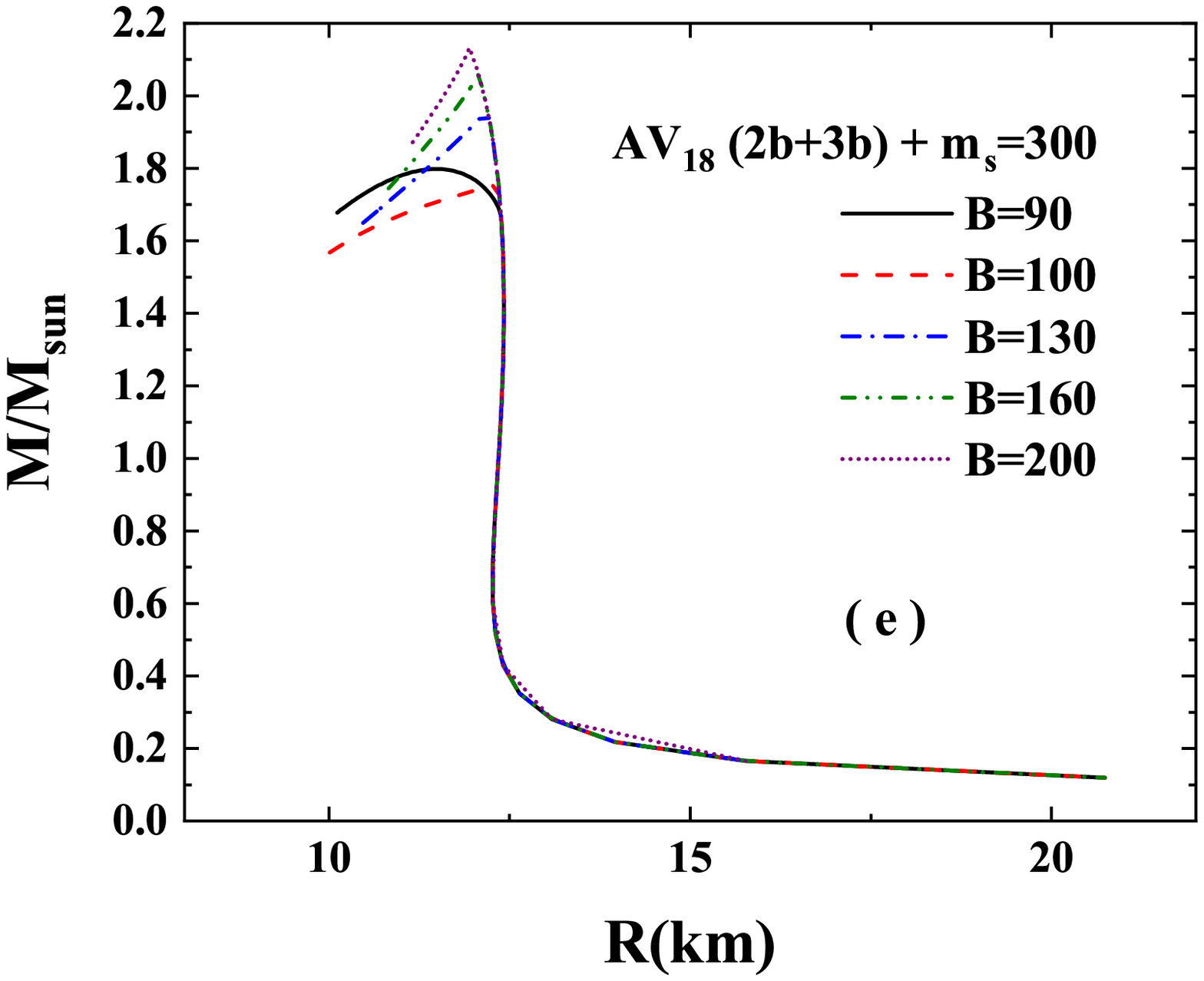}}
	\resizebox{0.329\textwidth}{!}{\includegraphics{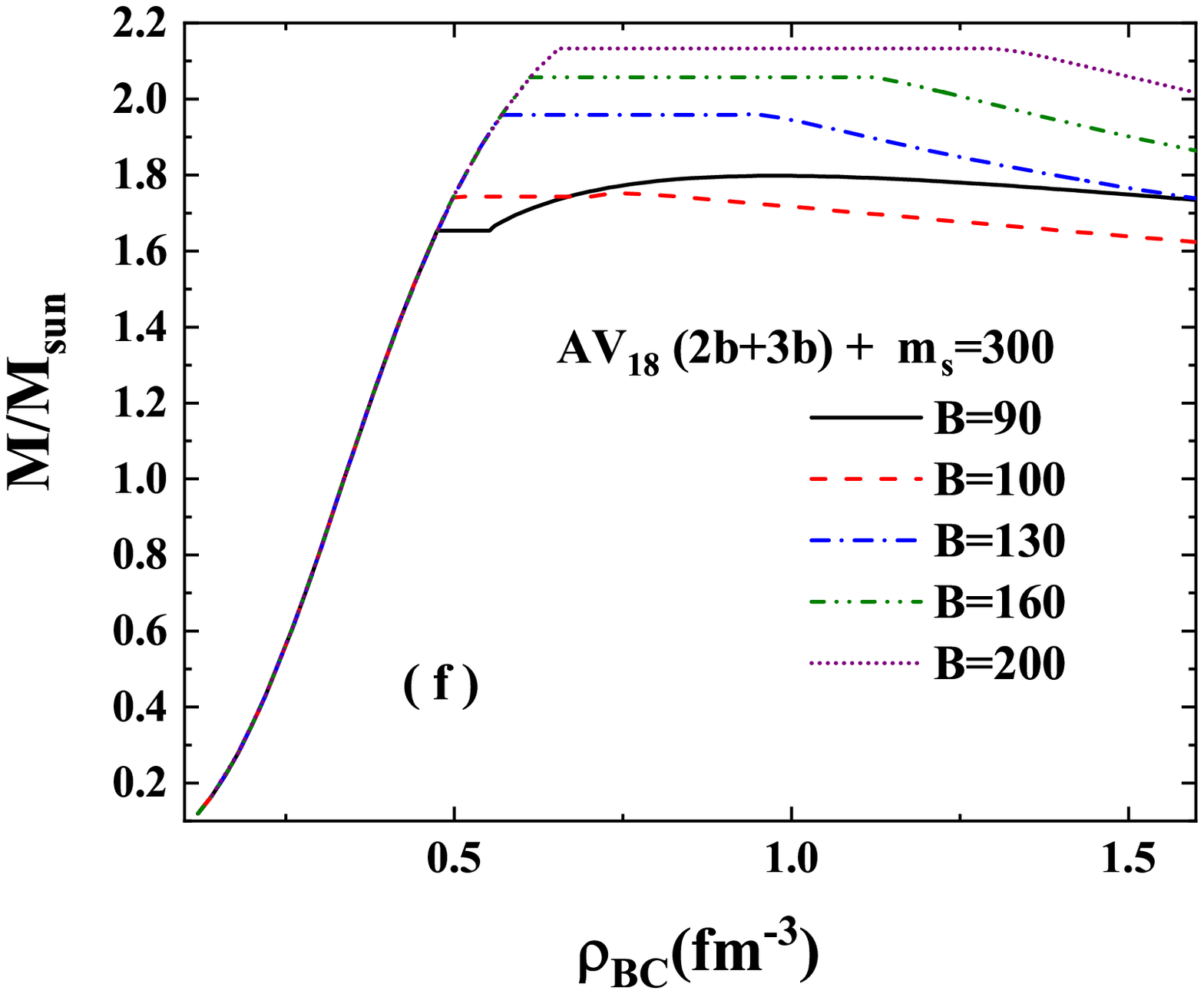}}
	\begin{center}
		\caption{{\small Panels a , b  and c : Pressure vs. baryon density, the gravitational HS masses vs. radius and central baryon density of the stars for variuos hadronic interactions and bag constant of B=$ 90 $ {MeV}{fm$ ^{-3} $} and $ m_{s}=300 $ MeV respectively. Panels d , e and f : The same as previous panels but for $ AV_{18} $ interaction supplemented with TBF and various bag constants of MIT bag model with $ m_{s}=300 $ MeV . } \label{fig7}}
	\end{center}
\end{figure*}

 We examine the effect of increasing the strange mass on the MIT model results which are displayed in Fig.~\ref{fig7}.
 In Fig.~\ref{fig7}(a), \ref{fig7}(b), and \ref{fig7}(c), we display pressure vs baryon density in Maxwell construction , mass vs radius, and mass vs central density of HSs for the bag constant of $B= 90$ MeVfm$ ^{-3} $ with $ m_{s}=300 $ MeV combined with several hadron interactions with or without TBF. As seen in Fig.~\ref{fig7}(a),  $ AV_{18} $ and $ UV_{14} $ potentials supplemented by TBF have almost the same phase transition densities while in $ AV_{14} $ and Reid potentials supplemented by TBF, these values are higher than them. In the case with just considering 2BF, the phase transitions take place in much higher densities. As seen in Fig.~\ref{fig7}(b), \ref{fig7}(c), the stable HS is predicted with  $ AV_{18} $ and $ UV_{14} $ potentials supplemented by TBF with the maximum masses of $ 1.796 M_{\odot} $  and $  1.778M_{\odot} $ respectively, since the gravitational mass continues to increase after the onset of the quark and also the maximum point in the mass-radius curve is stiff. The situation is better with respect to the results gained in the case with $ m_{s}=150 $ MeV in which the predicted stable HSs reach to the maximum mass of about $ 1.5 M_{\odot} $. The other HSs are predicted to be unstable because of the appearance of cusps in mass-radius curves, which render the star unstable. 
  In Fig.~\ref{fig7}(d), \ref{fig7}(e), and \ref{fig7}(f), we examine the effect of increasing the bag constant on the results. As seen in Fig.~\ref{fig7}(d), as the bag constant increases, the phase transitions take place at rather high baryon densities, and also the transition region extends which corresponds to the unstable HS since as it is seen in Fig.~\ref{fig7}(f), \ref{fig7}(e), however the gravitational mass slightly increases after the onset of quark through a small density range, there are cusps in all mass-radius curves which render the stars unstable. Nevertheless, the maximum masses of  $ 2.05 M_{\odot} $  and $  2.13M_{\odot} $ are calculated for$ B= 160$ MeVfm$ ^{-3} $ and $ B= 200$ MeVfm$ ^{-3} $ combined with $ AV_{18} $ potential supplemented by TBF . 
  
 The results of phase transition and the HS structure of the mentioned model are summarized in Tables ~\ref{t5} and ~\ref{t6}.
 By looking at Table~\ref{t5}, one can realize that the energy density discontinuity in $ AV_{18} $ and $ UV_{14} $ potentials supplemented by TBF is very low, around $ 100 $ MeVfm$ ^{-3} $, while in other cases this value is higher, and again we conclude that high-energy density discontinuity is responsible for the instability of HSs.
	   
 Our results are in line with other works in HS with other hadronic EOS such as the microscopic Brueckner-Hartree-Fock (BHF)many-body theory, relativistic mean field (RMF) model~\cite{Burgio:2002sn,Burgio:2001mk,Burgio:2004hz}, APR98 variational chain summation method
  and also Valecka models~\cite{Alford:2002rj}; all these EOS are combined with the MIT bag model in which HS's with bag constants of around $ B=90 $ MeVfm$ ^{-3} $ and $ m_{s}=150 $ MeVfm$ ^{-3} $ are stable with a pure quark core and the maximum mass is about $ 1.5M_{_{\odot}} $. It is to be noted that for a large bag constant, bigger than around B=$ 140 $ MeVfm$ ^{-3} $, the stability of the star will be lost.

\subsubsection{NJL model }\label{structnjl}

\begin{figure*}[htb]
	\vspace{-0.70cm}
	
	\resizebox{0.325\textwidth}{!}{\includegraphics{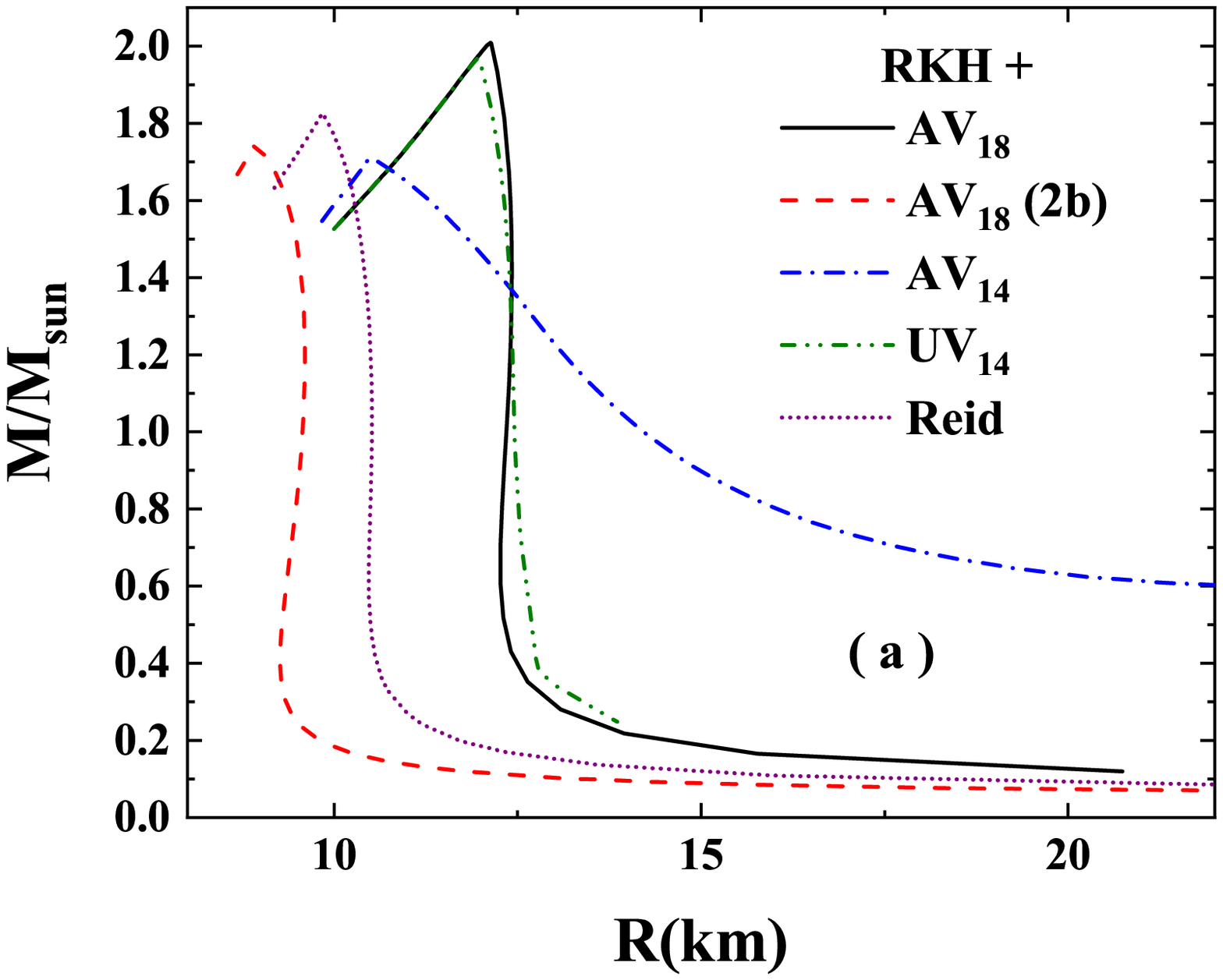}}
	\resizebox{0.329\textwidth}{!}{\includegraphics{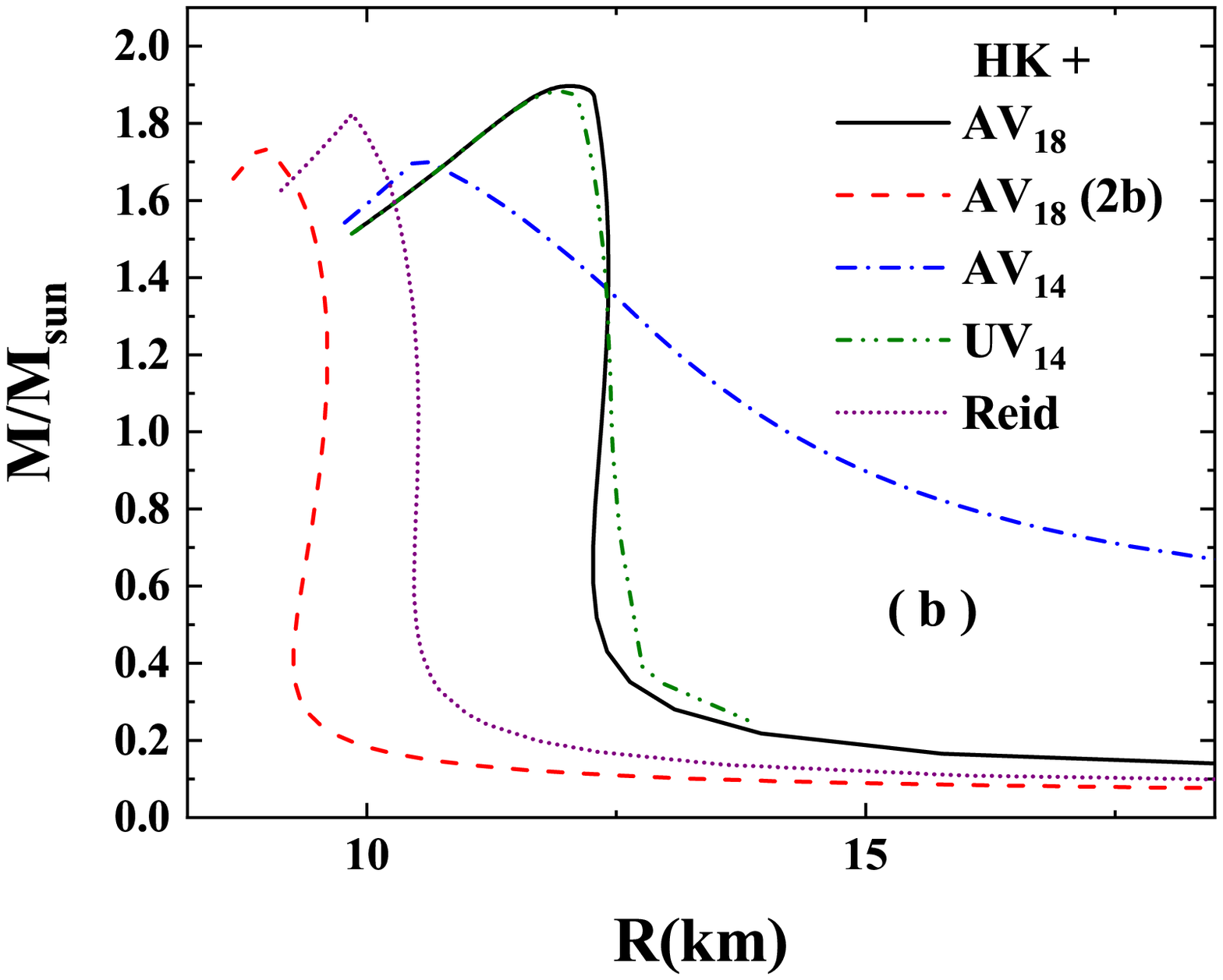}}
	\resizebox{0.329\textwidth}{!}{\includegraphics{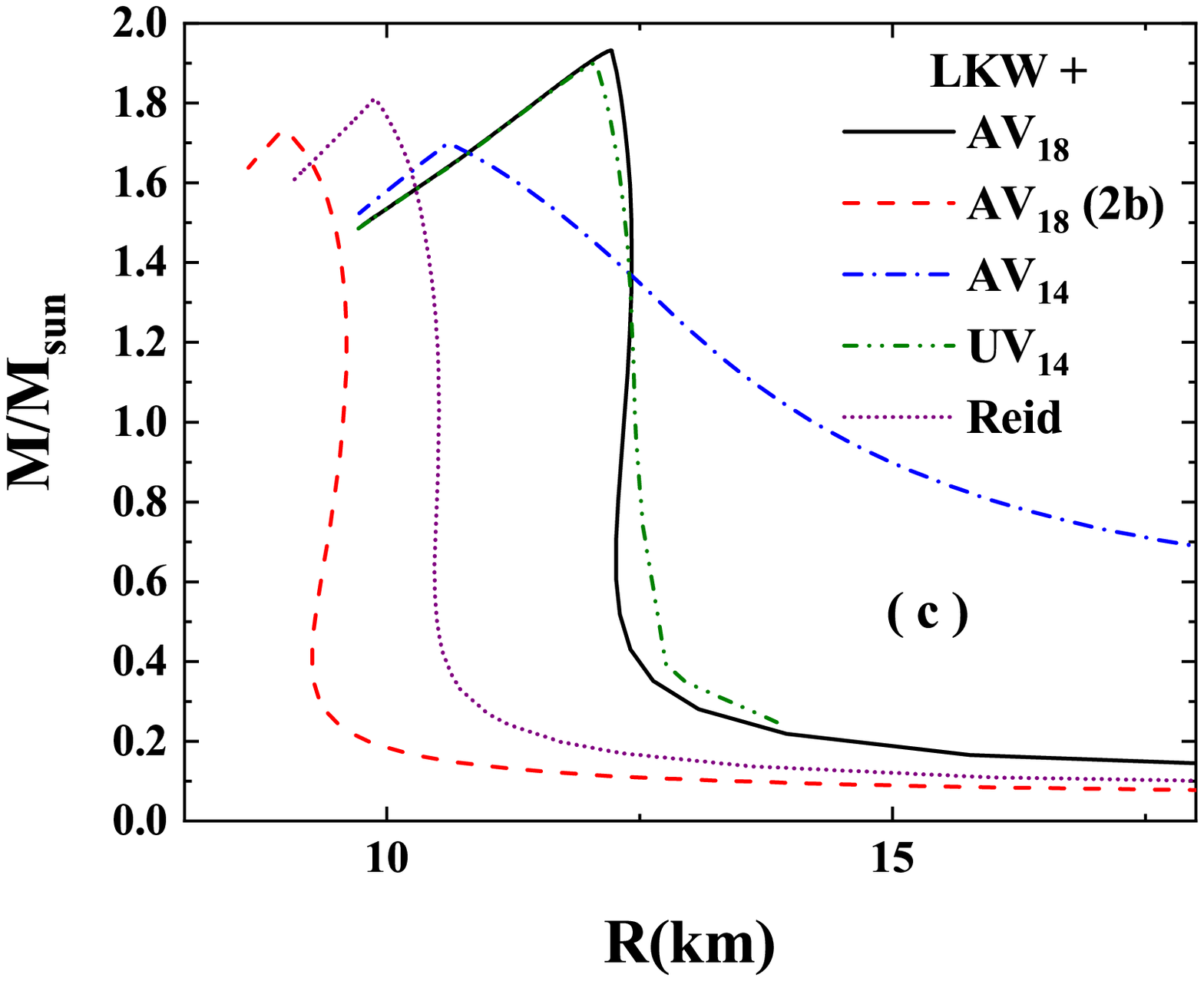}}
	\resizebox{0.329\textwidth}{!}{\includegraphics{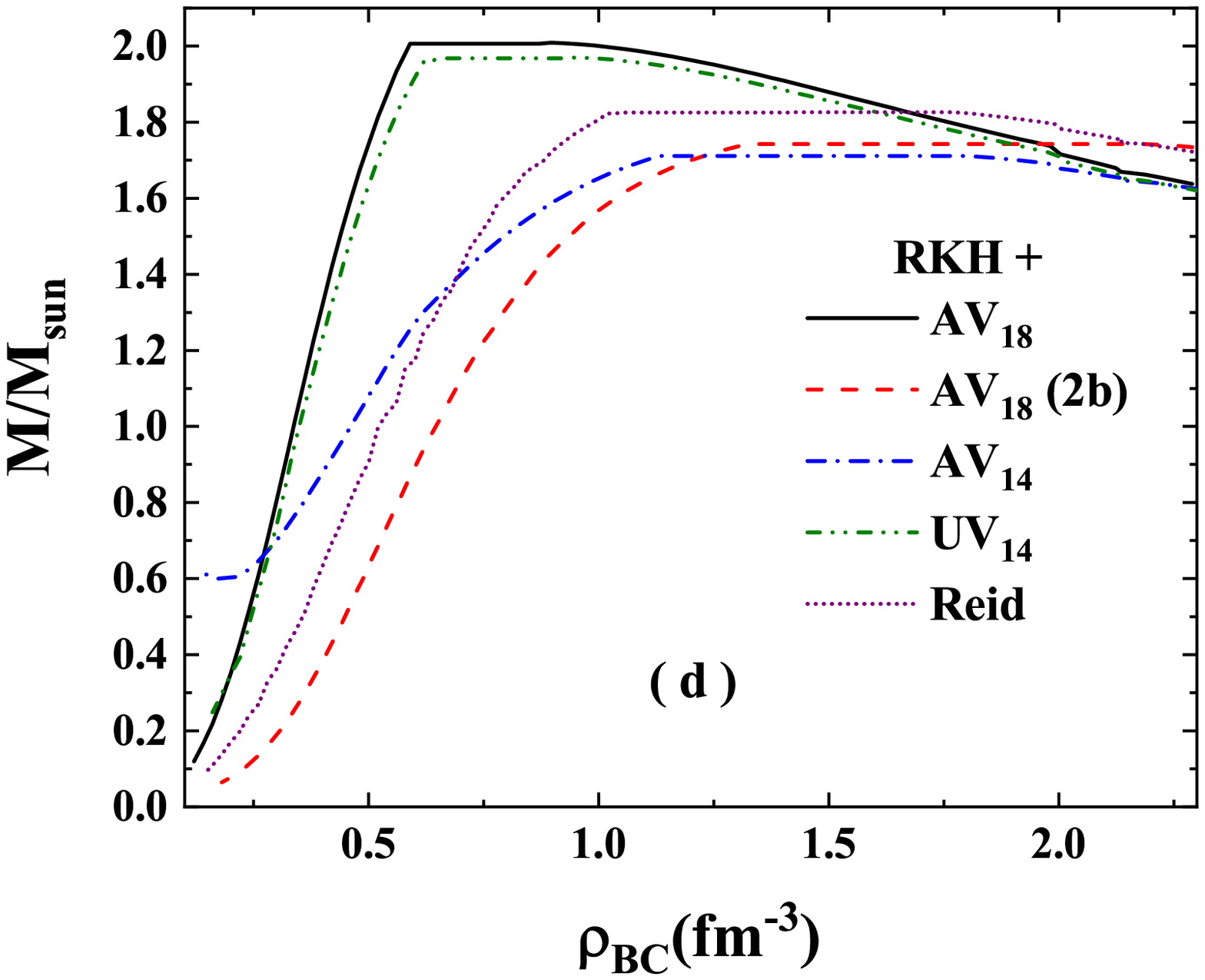}}
	\resizebox{0.329\textwidth}{!}{\includegraphics{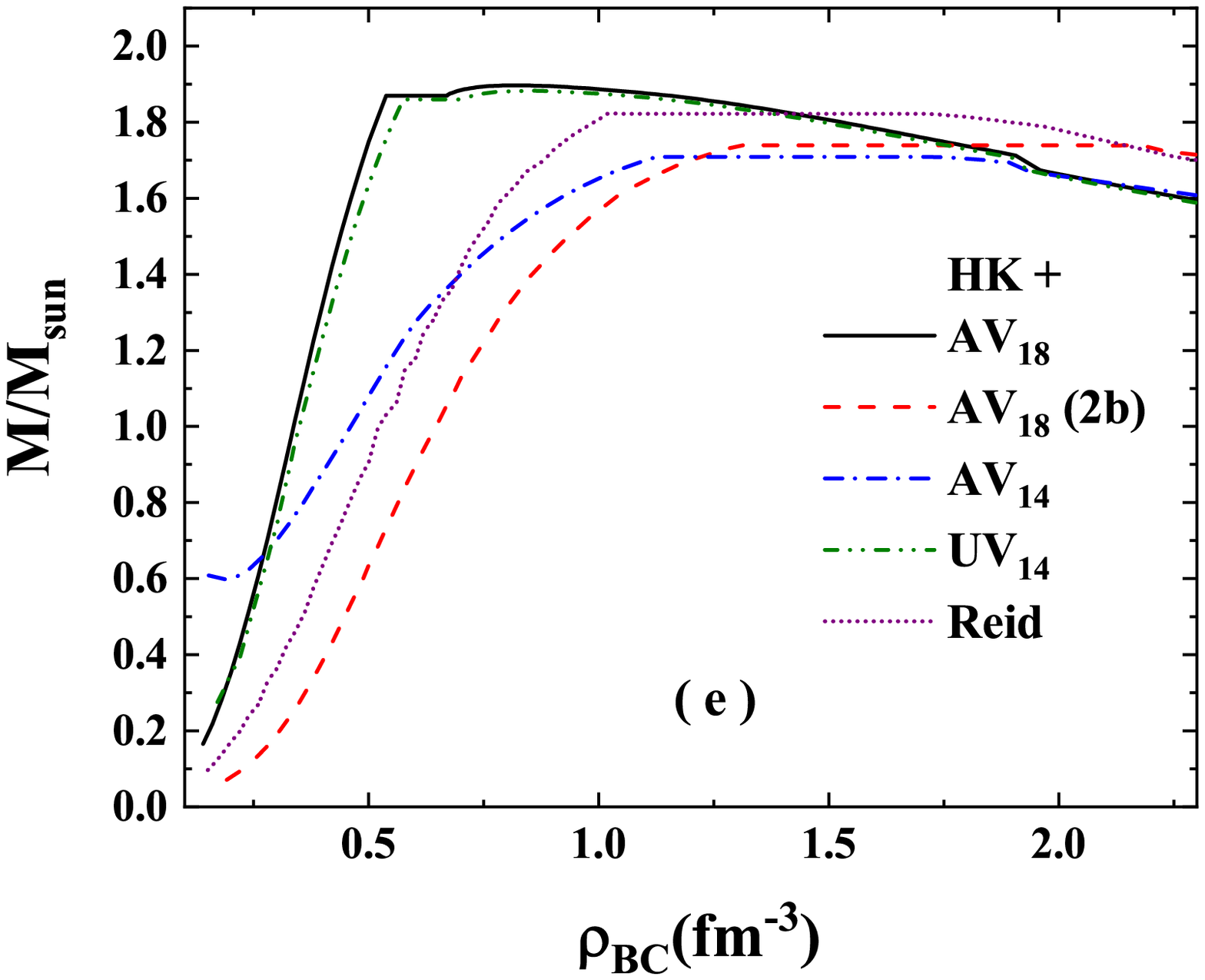}}
	\resizebox{0.329\textwidth}{!}{\includegraphics{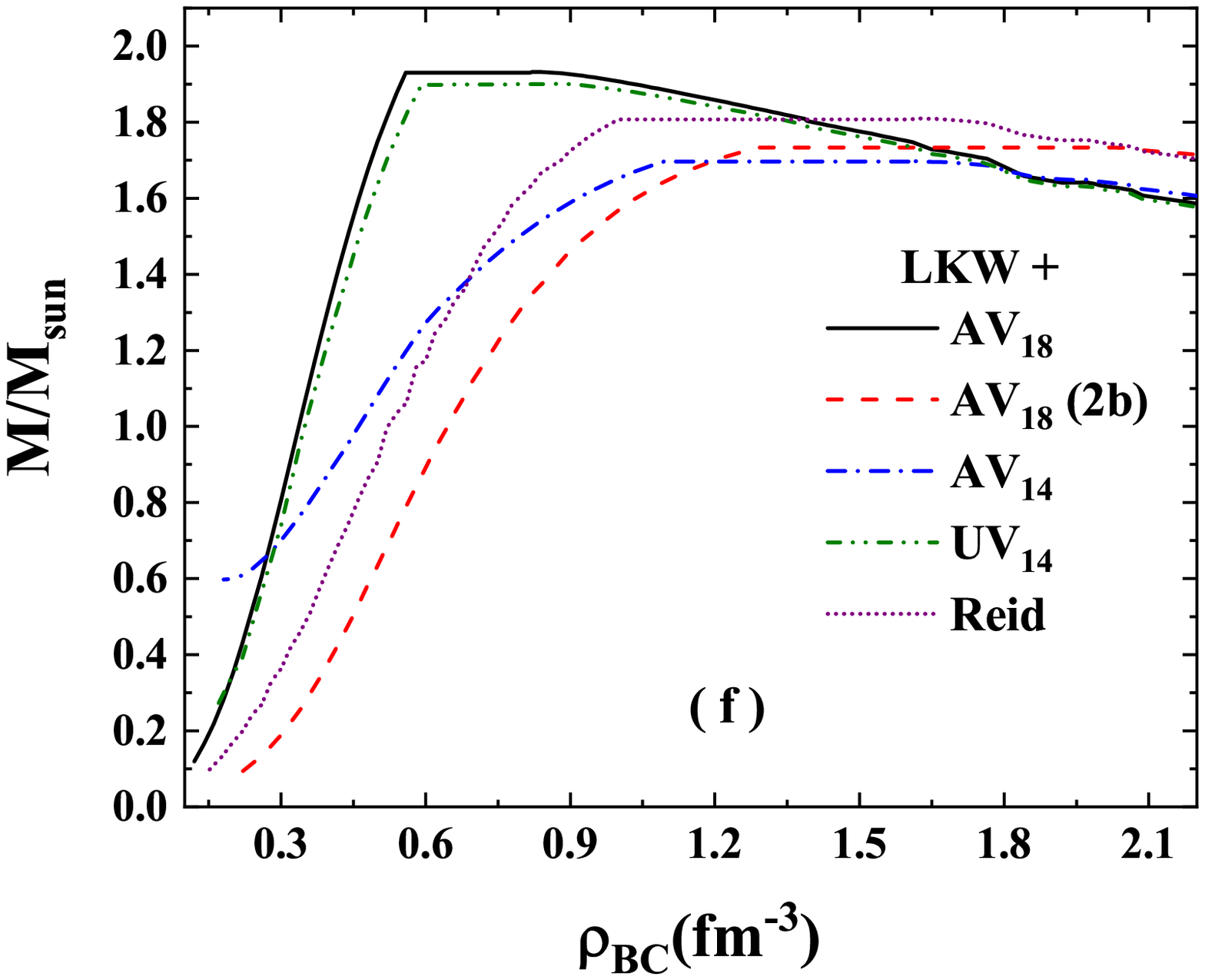}}
	\begin{center}
		\caption{{\small Panel a , b and c : The gravitational HS masses vs. radius of the star for variuos hadronic interactions and NJL with RKH , HK and LKW  sets of parameters respectively. Panel d , e and f : The corresponding gravitational HS masses vs. central baryon density of the star for  RKH , HK and LKW  sets of parameters. } \label{fig8}}
	\end{center}
\end{figure*}

In this subsection, we will present the results of the HS structure within the NJL model. In Fig.~\ref{fig8}(a), \ref{fig8}(b), and \ref{fig8}(c), we  plot the HS gravitational mass vs radius of the star for various hadronic interactions and RKH , HK, and LKW parameter sets respectively and Fig.~\ref{fig8}(d), \ref{fig8}(e), and \ref{fig8}(f), show the corresponding HS mass vs central baryon density of the star.
In all cases, the maximum masses of HSs are more than $ 1.693 M_{\odot} $. In the  RKH parameter set with $ AV_{18} $ combined with TBF, it reaches to $ 2.01 M_{\odot}$. As seen in Fig 8(b) in the HSs within the HK parameter set combined with $ AV_{18} $ and $ UV_{14} $ interactions, the cusps in mass-radius relations are slightly smoothed out and therefore the corresponding HSs are stable. As seen in Fig 8(e) the mass-central density curves obviously increases after the onset of quark phase. 

In other cases, although the curves increase slightly through a very small density range by the onset of the pure quark phase, in mass-radius curves, there exist cusps in maximum point which cause the instability of the HSs. However, maybe it is not clearly visible in some cases but since the central baryon density in which the maximum mass occurs is a little higher than the transition density of quark to hadron, the curves continue to increasing through a very small density range. The  values of densities are comparable in Tables ~\ref{t3} and ~\ref{t7} which will be presented in following.  In Table~\ref{t5}, we summarize the structure properties of pure neutron and hybrid stars for various hadron interactions and parameter sets of the NJL model.

In all cases, the energy density discontinuity in transition region is large enough to make a cusp in mass-radius curves. The only exception is HK parameter set combined with $ AV_{18} $ and $ UV_{14} $ interactions combined with TBF. It turns out, however, that in these cases the cusps are so strong that the stars are rendered unstable. In the HK parameter set with $ AV_{18} $ and $ UV_{14} $ supplemented by TBF, the cusps are slightly smoothed out, and the corresponding HSs are stable. The results in both quark models lead to the point that considerably larger energy densities in the quark core probably are the driving factor for the instability.

Our results are in acceptable concurrence with those of some works conducted in HSs within the original NJL model and other hadron models~\cite{Baldo:2002ju,Baldo:2006bt,Buballa:2003qv,Yang:2008am,Schertler:1999xn,Steiner:2000bi}. According to the conclusions of forgoing papers, other hadronic equations of state, namely,  BHF many-body theory and also the RMF model combined with NJL, the stable HS with a pure quark core is not predictable. This result seems to be rather insensitive to the choice of the hadronic EoS and must mainly be attributed to the quark EOS derived within the NJL model~\cite{Baldo:2002ju,Baldo:2006bt,Buballa:2003qv}. In Ref.~\cite{Baldo:2006bt}, the authors conclude that the instability is closely linked to the lack of confinement in the original NJL model. Besides, as it is pointed out in Ref.~\cite{Buballa:2003qv}, these results rely  heavily on the assumption that the NJL model parameters which have been fitted in vacuum can be applied to dense matter. However, according to all the findings in HSs, is also possible that applying the vacuum fitted parameters to dense matter is not the correct approach and maybe a considerable modification of the effective NJL-type quark interactions in dense matter is needed. 
	 
\begin{figure*}
	\vspace{-0.70cm}
	\resizebox{0.305\textwidth}{!}{\includegraphics{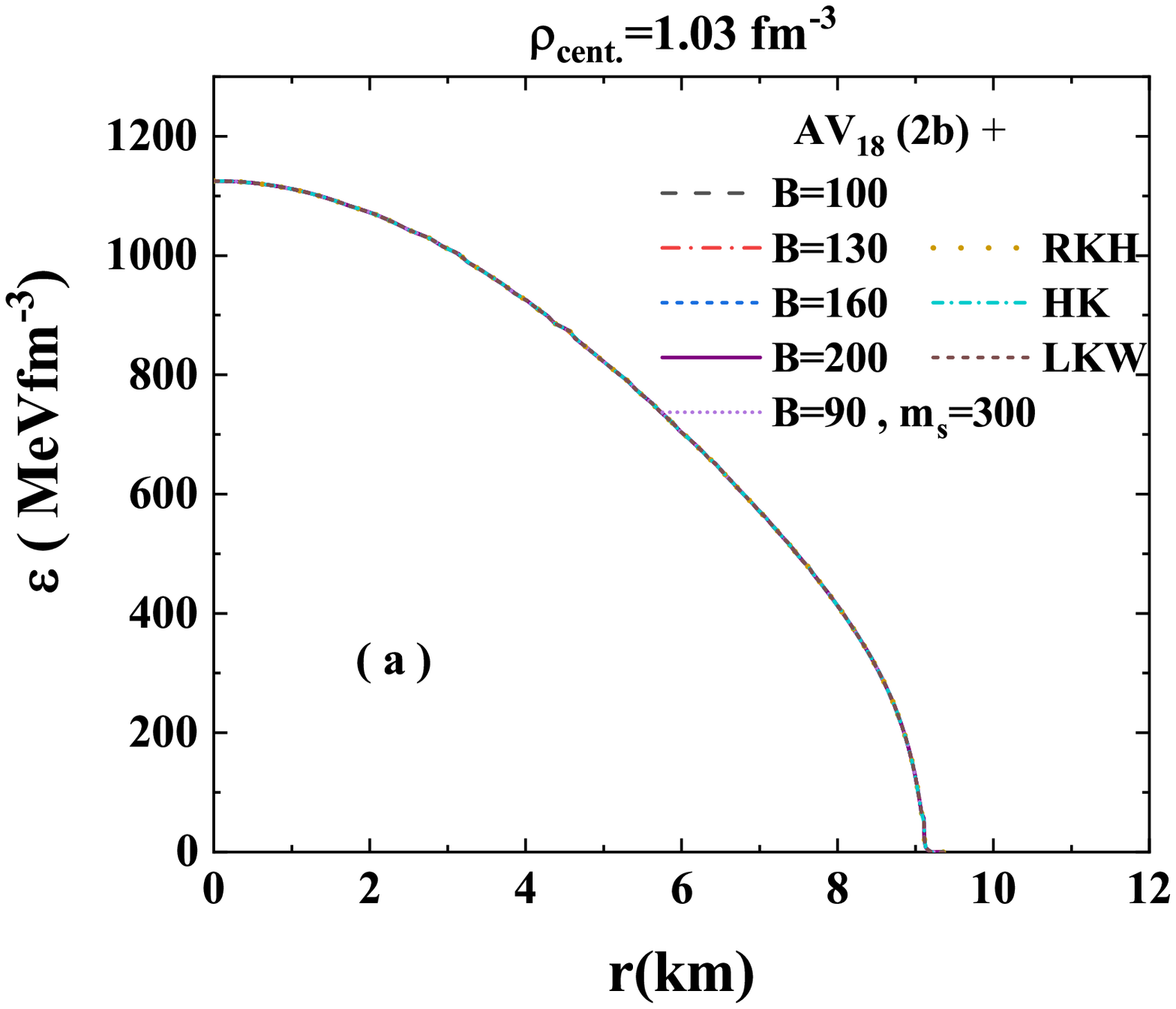}}
	\resizebox{0.305\textwidth}{!}{\includegraphics{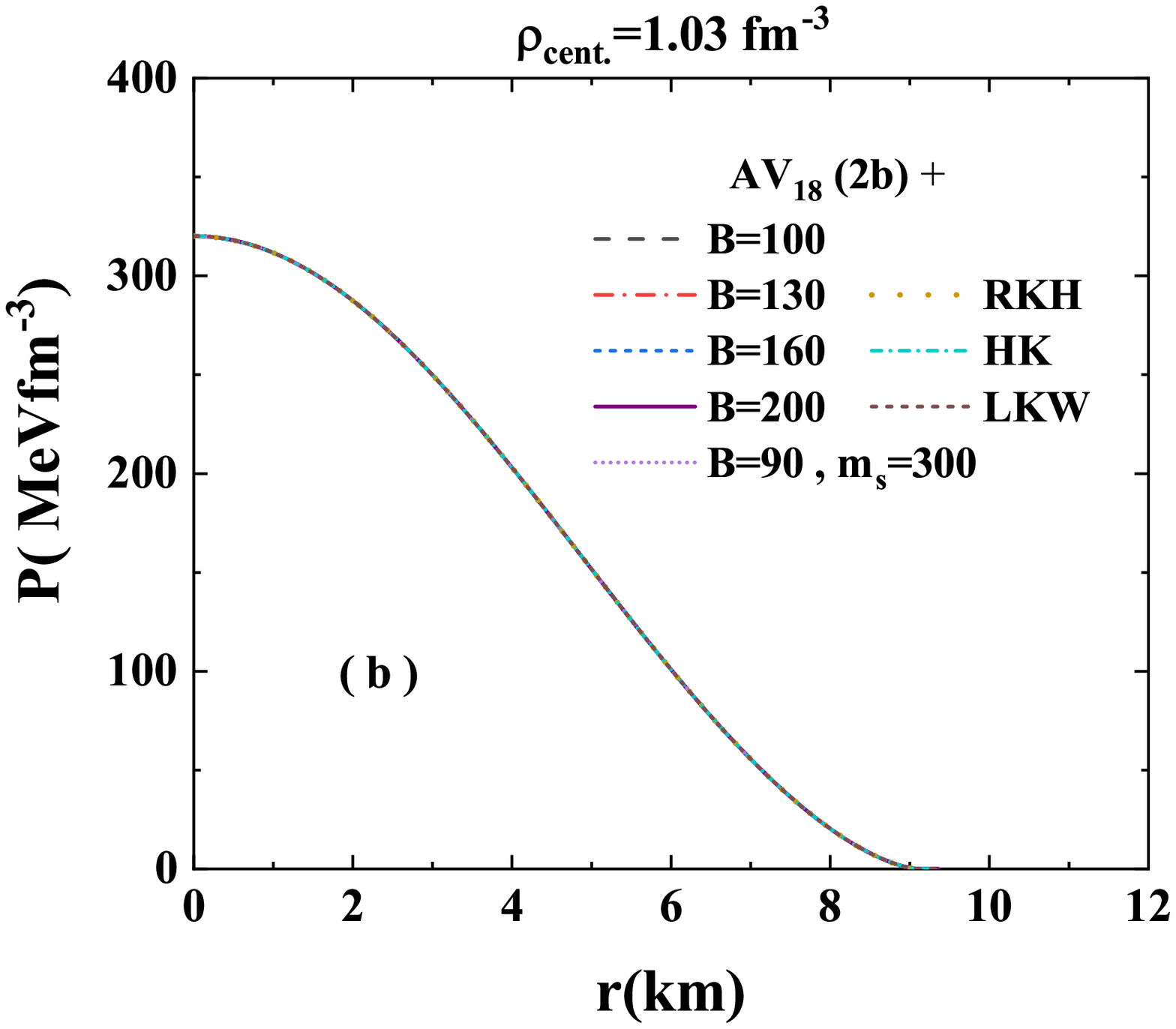}}
	\resizebox{0.305\textwidth}{!}{\includegraphics{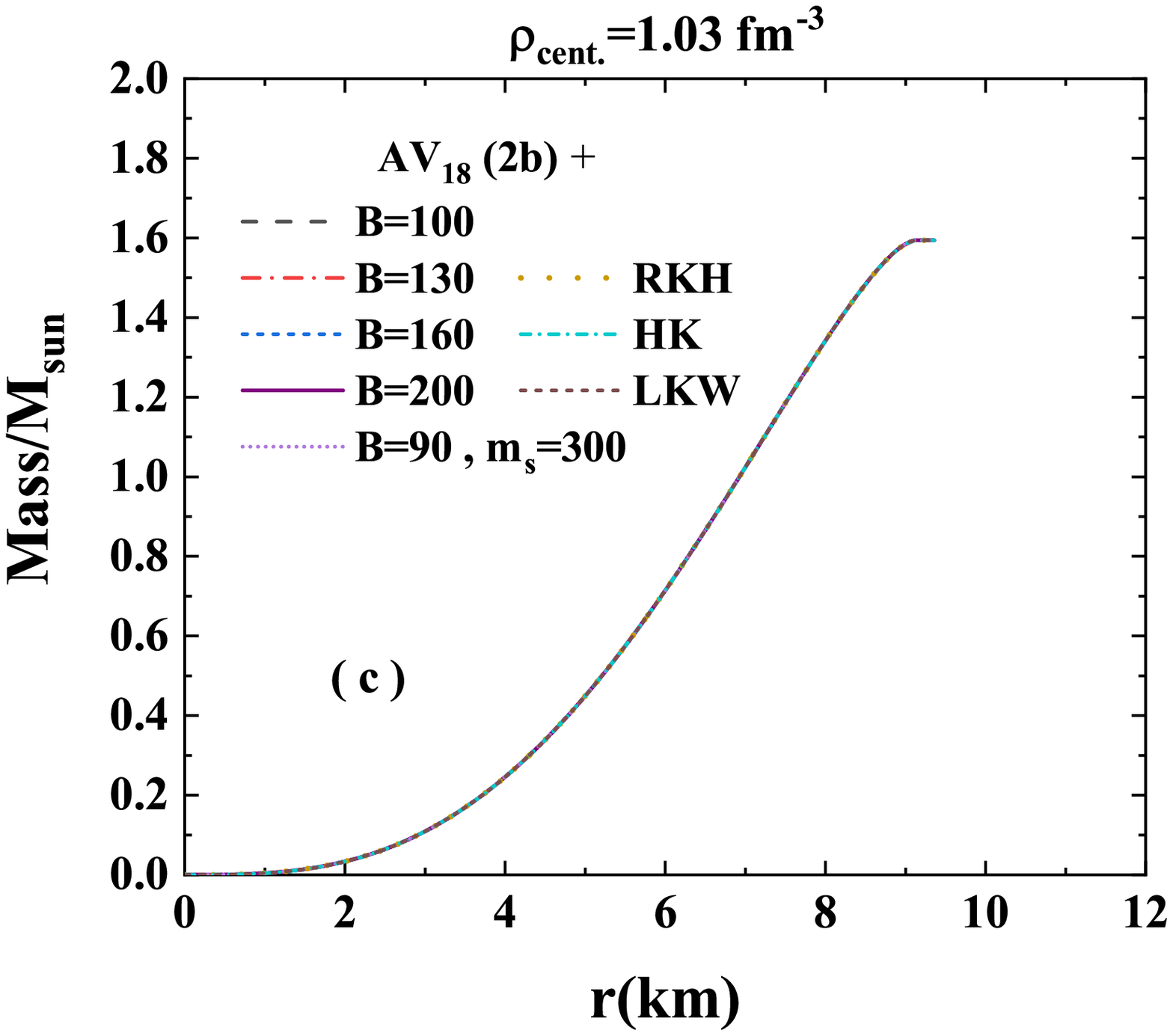}}
	\resizebox{0.305\textwidth}{!}{\includegraphics{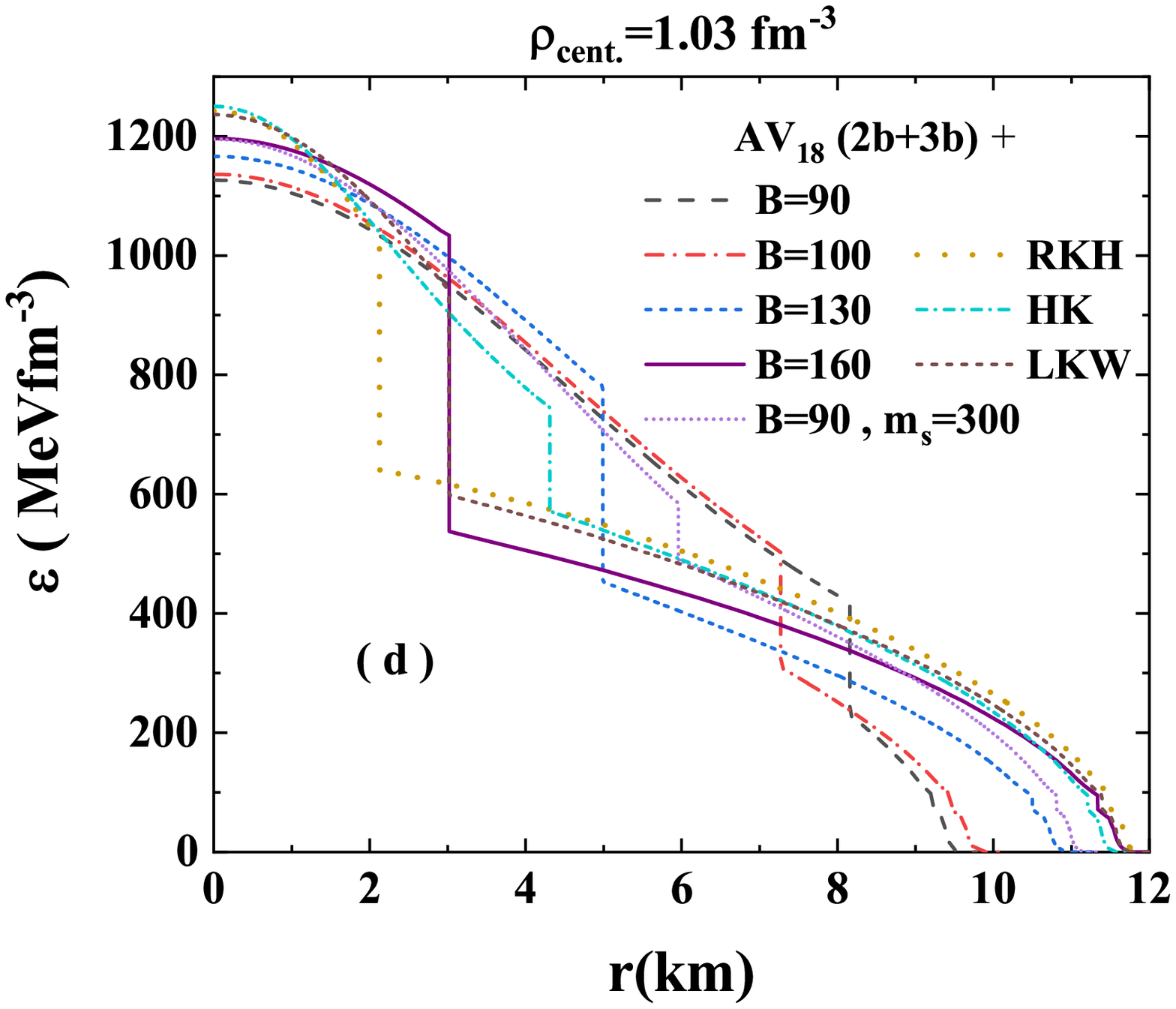}}
	\resizebox{0.305\textwidth}{!}{\includegraphics{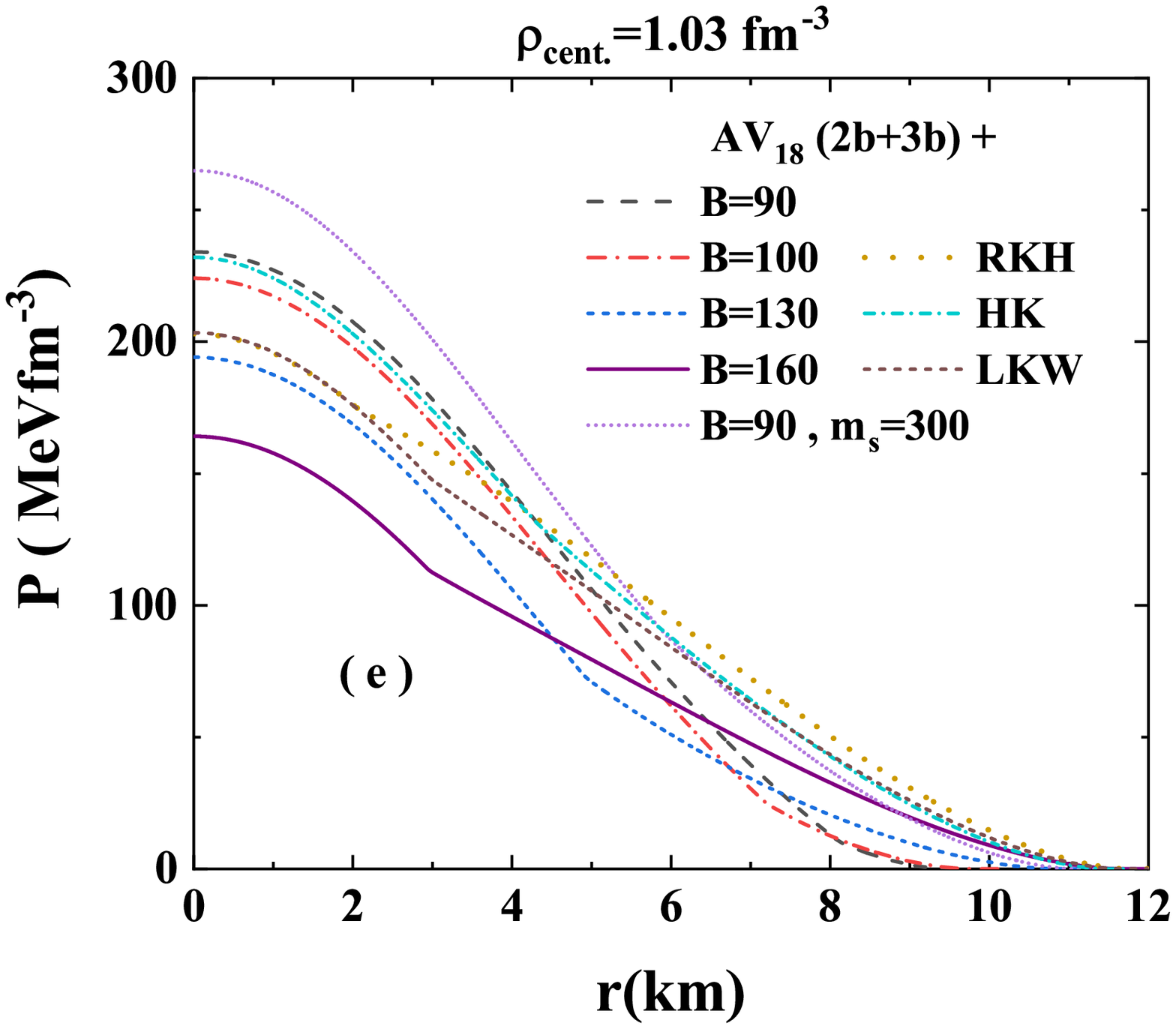}}
	\resizebox{0.305\textwidth}{!}{\includegraphics{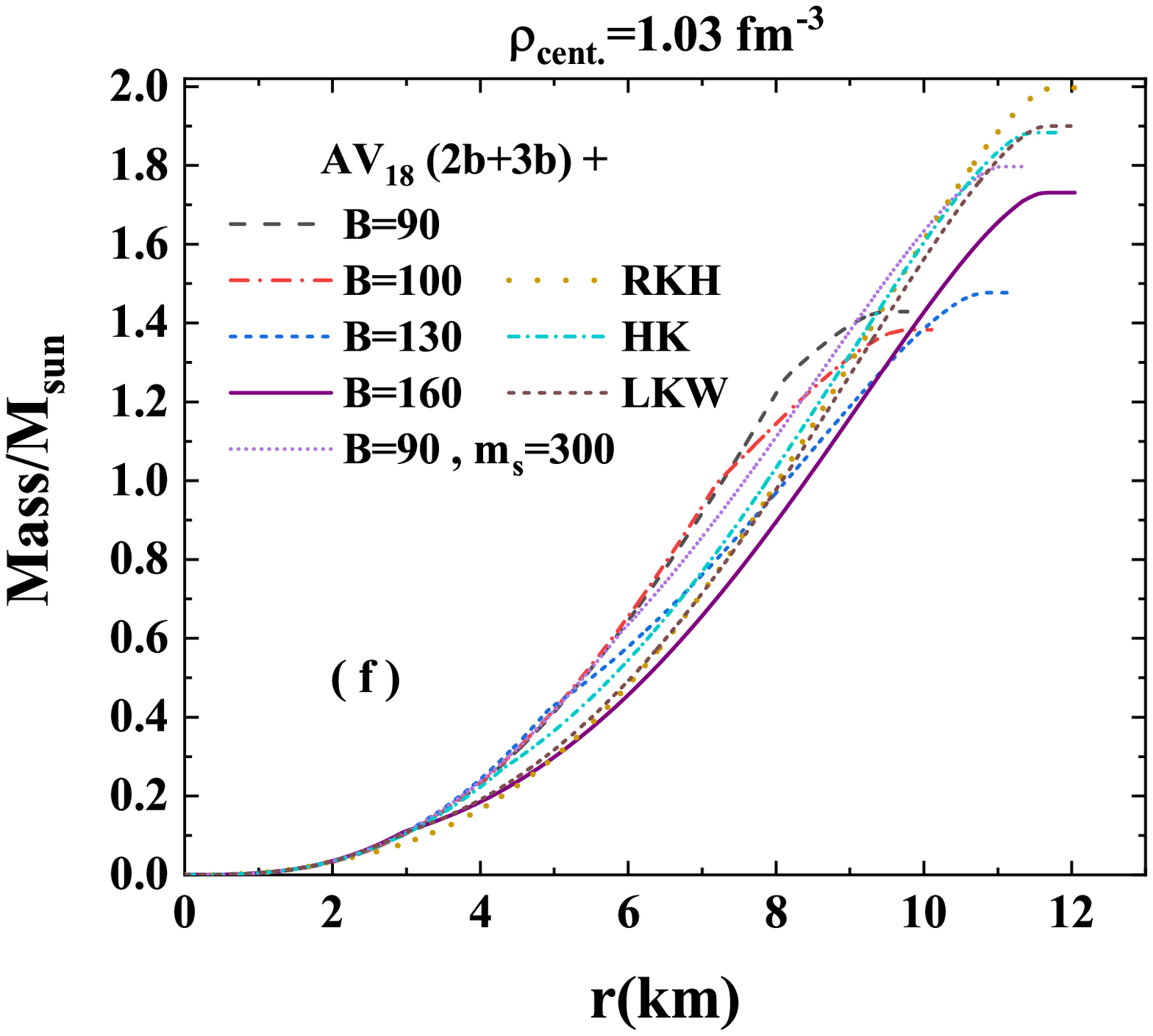}}
	\resizebox{0.305\textwidth}{!}{\includegraphics{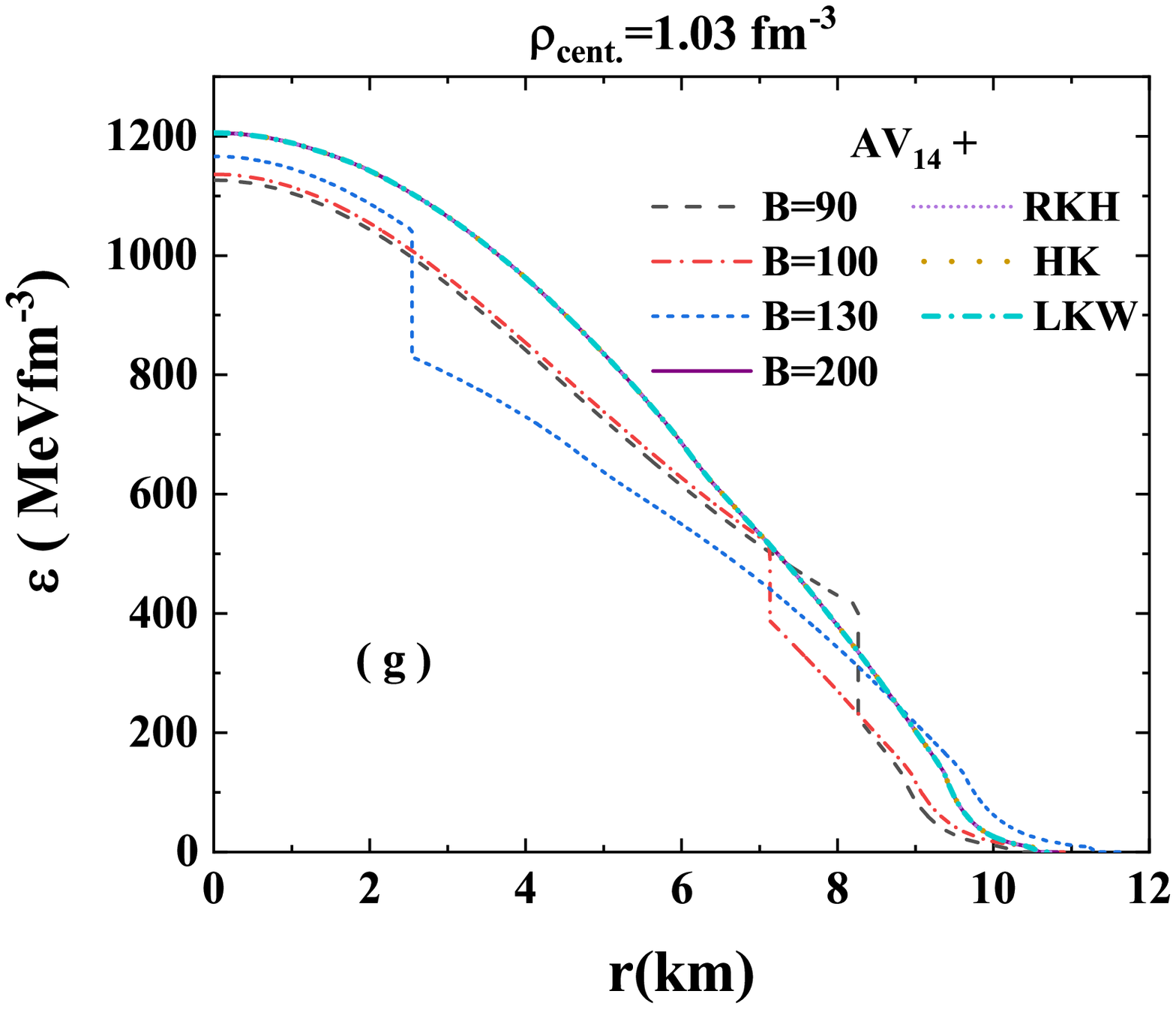}}
	\resizebox{0.305\textwidth}{!}{\includegraphics{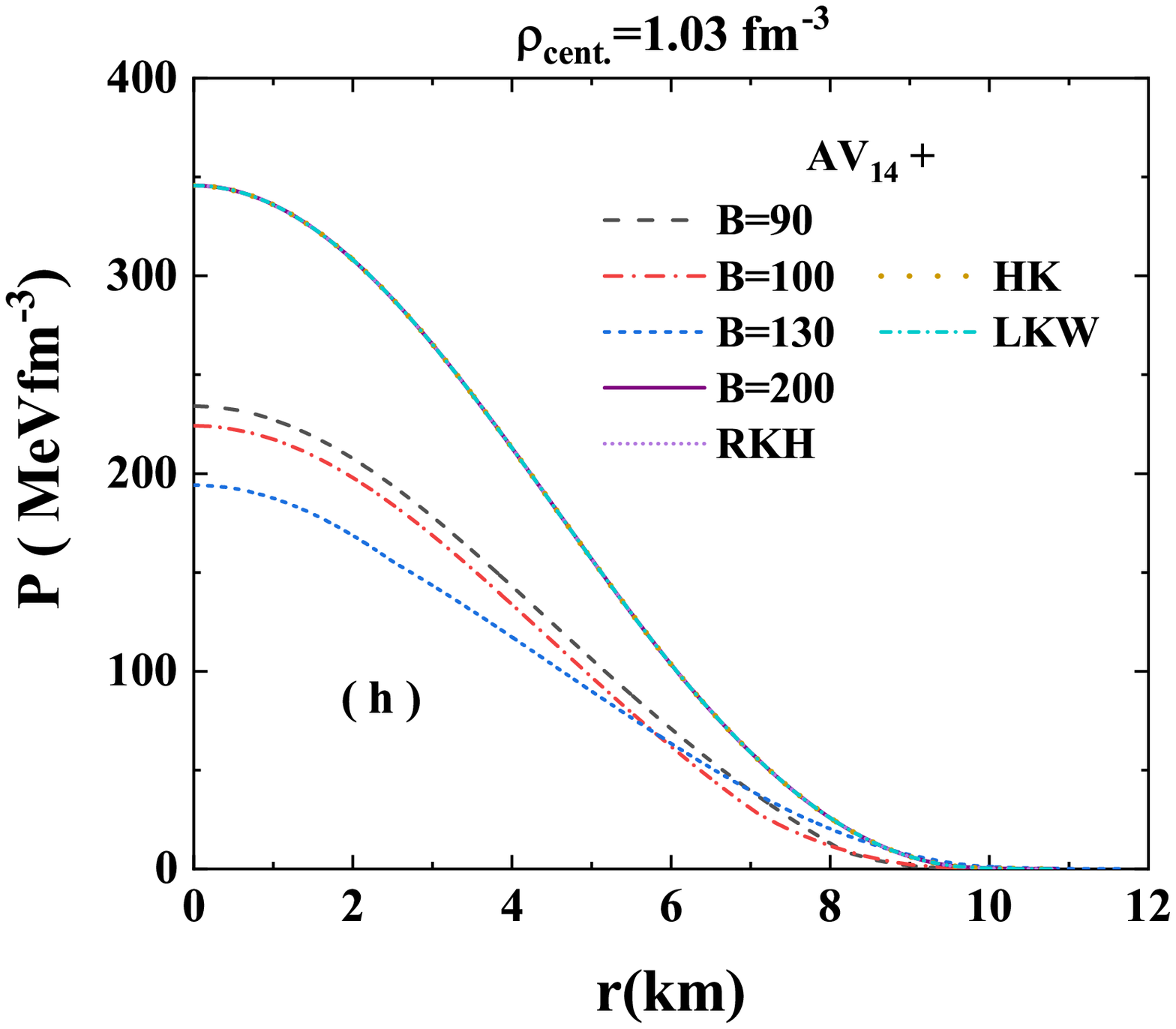}}
	\resizebox{0.305\textwidth}{!}{\includegraphics{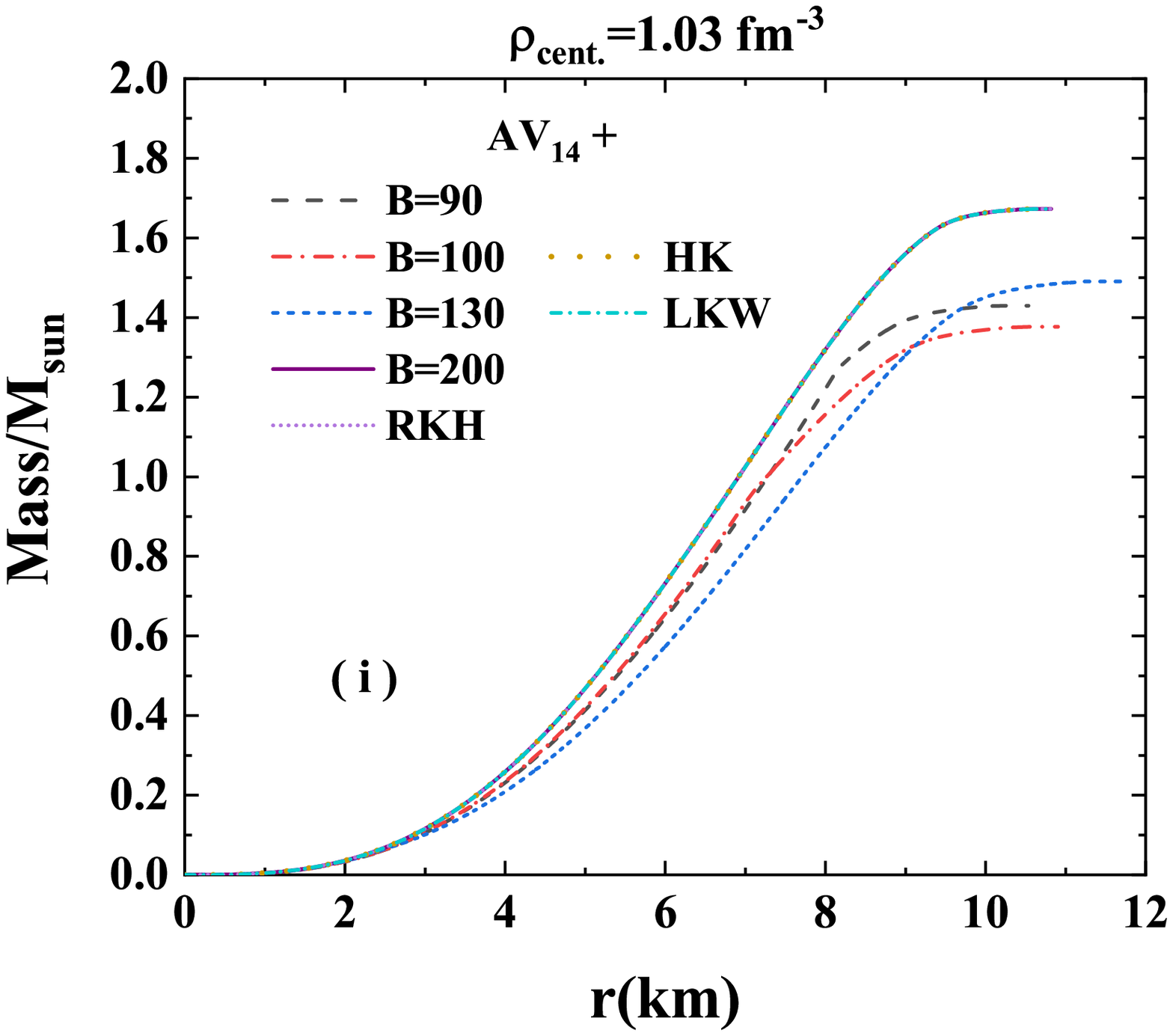}}
	\resizebox{0.305\textwidth}{!}{\includegraphics{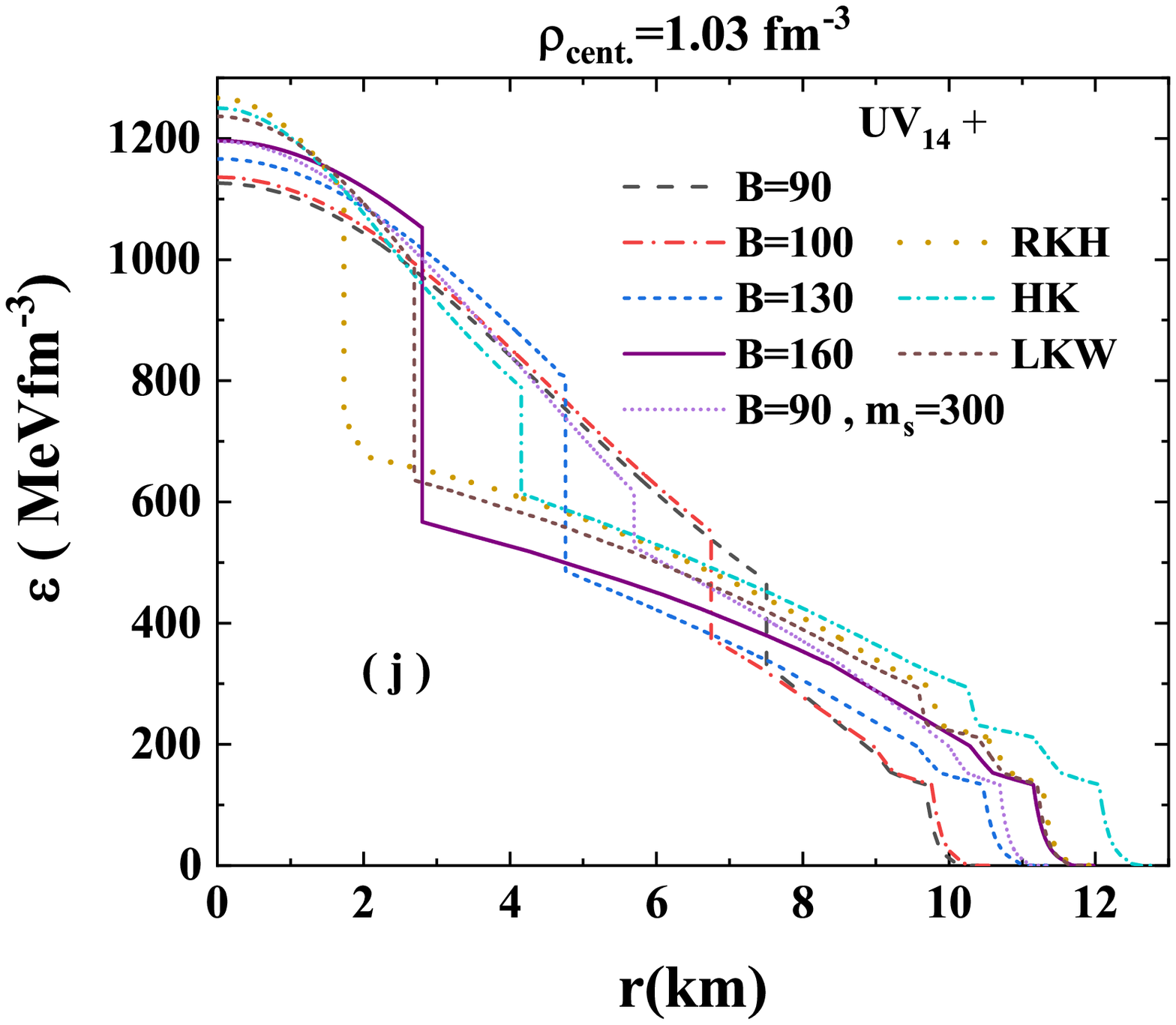}}
	\resizebox{0.305\textwidth}{!}{\includegraphics{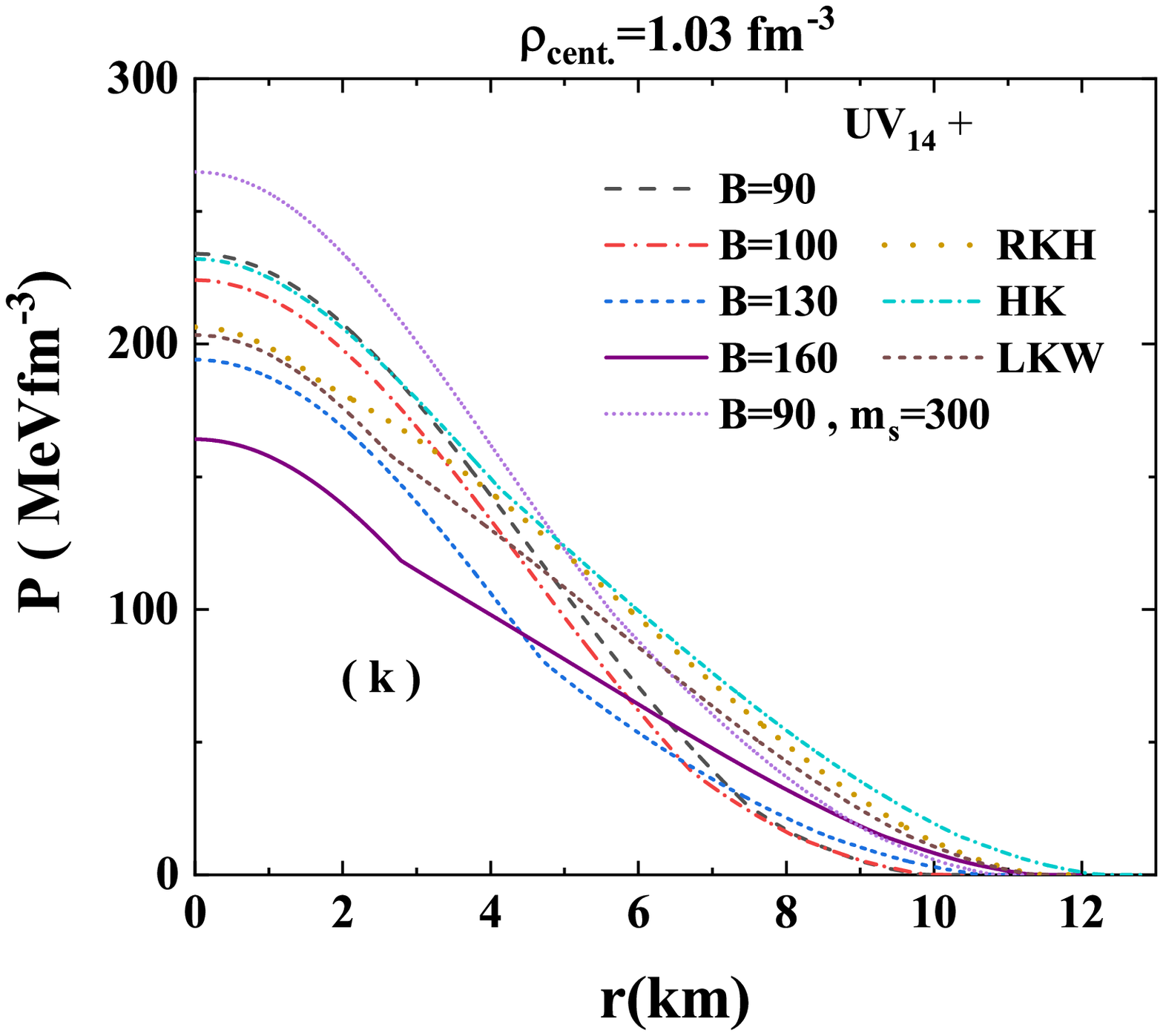}}
	\resizebox{0.305\textwidth}{!}{\includegraphics{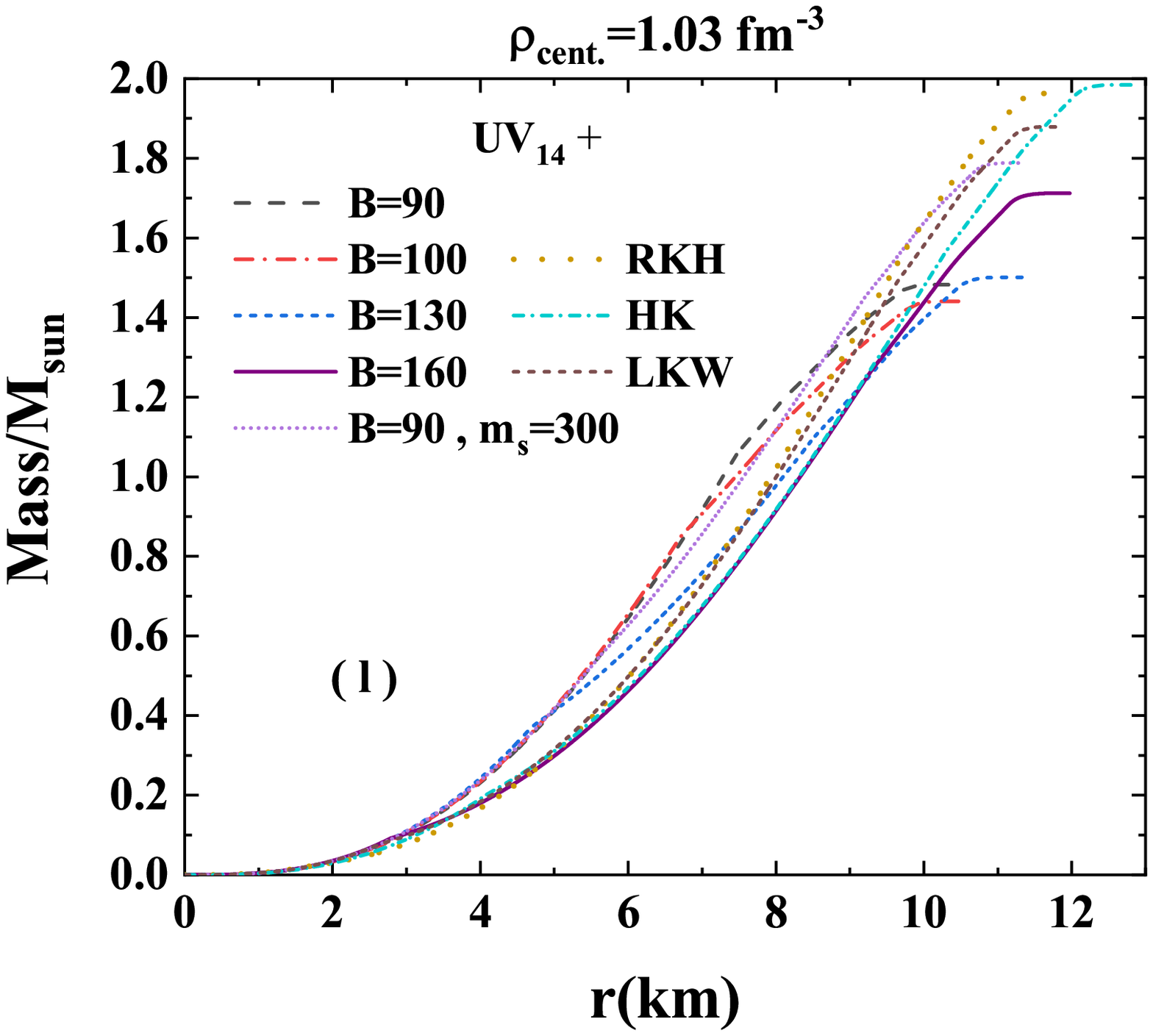}}
	\resizebox{0.305\textwidth}{!}{\includegraphics{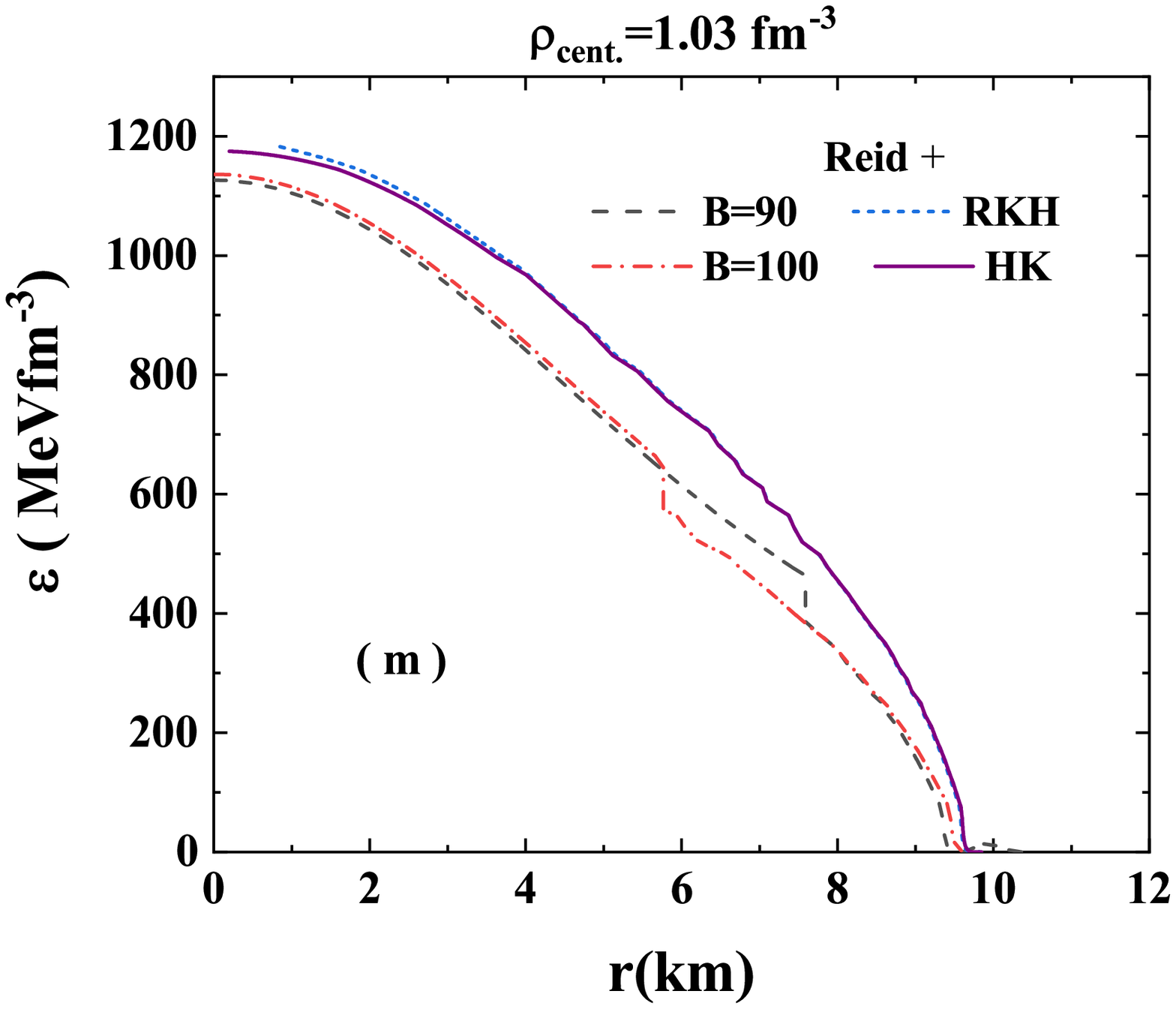}}
	\resizebox{0.305\textwidth}{!}{\includegraphics{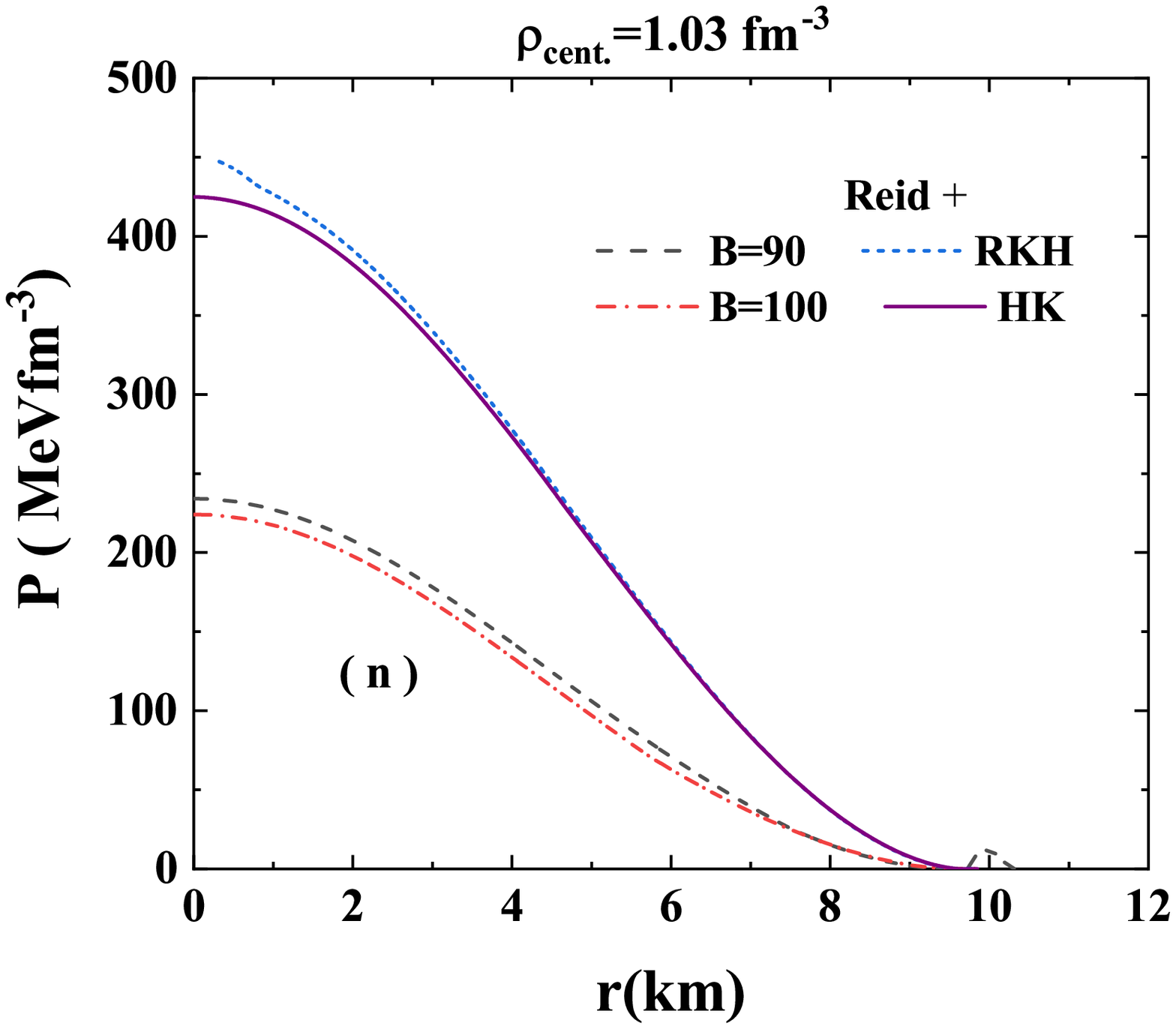}}
	\resizebox{0.305\textwidth}{!}{\includegraphics{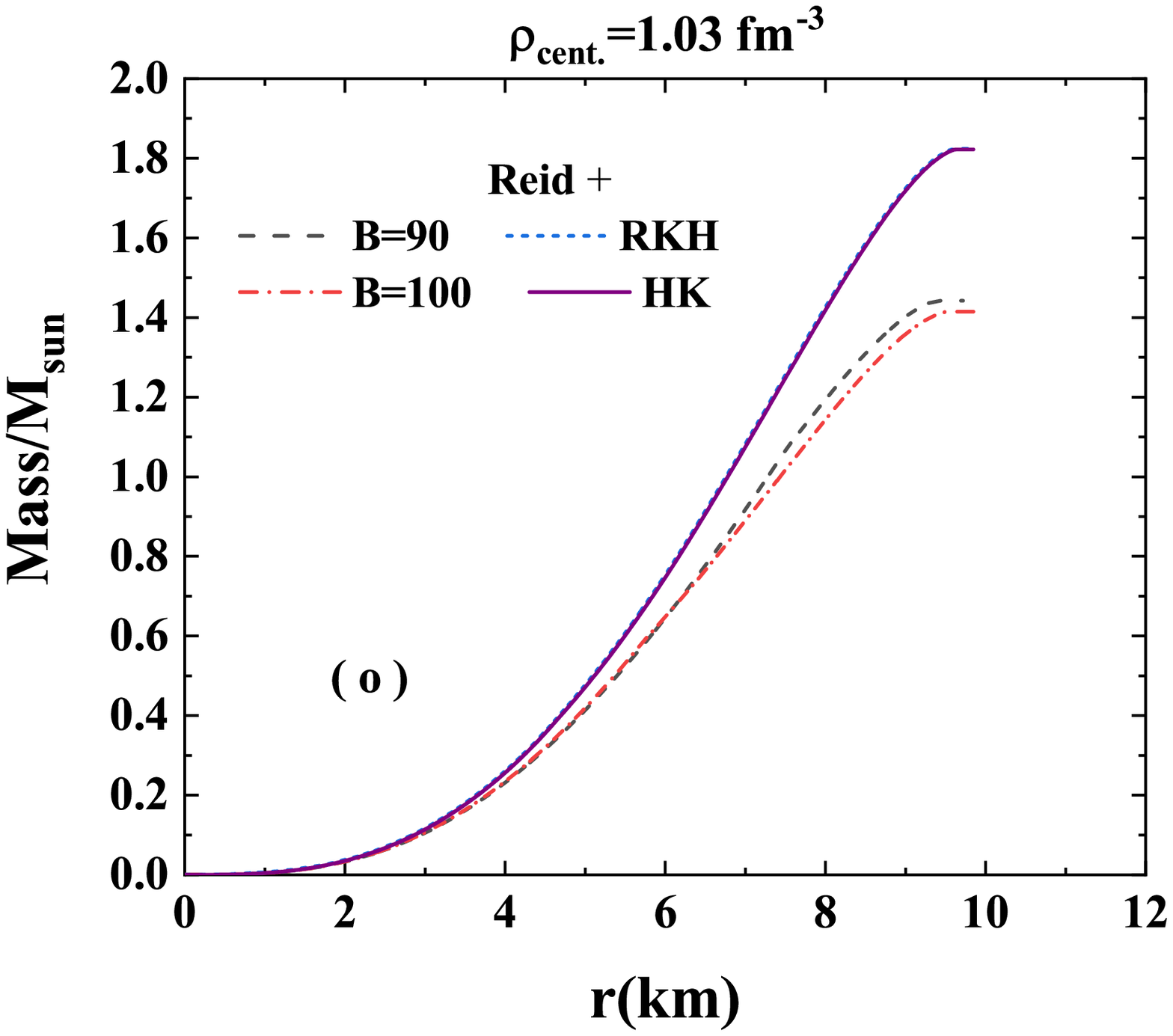}}
	\begin{center}
		\caption{{\small Panels a , b and  c: The energy density, pressure and mass profiles of HS's  with central density of $\rho  _{BC}=1.03 $ fm$ ^{-3} $for $ AV_{18} $ interaction without TBF and various quark models. The same quantities are shown in panels d , e and f for $ AV_{18} $,  g , h and i for $ AV_{14} $,  j , k and l for $ UV_{14} $ and m , n and o for Reid intractions supplemented by TBF.} \label{fig9}}
	\end{center}
\end{figure*}

We close this subsection by displaying the energy density, pressure, and mass profiles of the HSs investigated in the paper for a sample central density of $ \rho_{cent}=1.03 $ fm$ ^{-3} $ in Fig.~\ref{fig9}. For $ m_{s}=300 $ MeV, we display the profiles only for bag constant of B= $ 90 $ MeVfm$ ^{-3} $.	As  seen in Figs~\ref{fig9}(a), \ref{fig9}(d), \ref{fig9}(g), \ref{fig9}(j), and \ref{fig9}(m), the energy density decreases from the center to the surface of the star, and in phase transition density, there is a sudden decreasing in the amount of energy density that is characteristic of Maxwell construction. As is clear in the figures, this energy discontinuity is obviously large in RKH and LKW parameter sets in NJL model and also high bag constants of MIT bag model which, as we point out before, probably is the driving factor for the instability of corresponding HSs. Some of the energy density profiles do not have this discontinuity since, the sample central density of $ \rho_{cent}=1.03 $ fm$ ^{-3} $ for those HS's is lower than the transition density therefore the corresponding star is purely hadron star. As seen in seen in Figs~\ref{fig9}(b), \ref{fig9}(e), \ref{fig9}(h), \ref{fig9}(k), and \ref{fig9}(n), the pressure of star monotonically decreases from the center to the surface of the star as we expected from the TOV equations. As seen in Figs~\ref{fig9}(c), \ref{fig9}(f), \ref{fig9}(i), \ref{fig9}(l), and \ref{fig9}(o), the mass of the star increases monotonically from zero at the center to the star, mass at the surface of the star as we expected from the TOV equations.

\subsection{Tidal deformability }\label{III-C}

Gravitational waves from the final stage of inspiraling binary NSs are expected to be one of the most important sources for gravitational wave detectors. Large finite-size (tidal) effects are measurable toward the end of inspiral~\cite{Bildsten:1992my}, but the gravitational wave signal is expected to be very complex during this period. Tidal effects during the early part of the evolution will form a very small correction, but during this phase, the signal is very clean~\cite{Flanagan:2007ix}. The tidal fields induce quardupole moments on the NSs. This response of each star to the external disturbance is described by the second (tidal) Love number $ k_{2} $,   which is a dimensionless coefficient given by the ratio of the induced quardrupole moment and the applied tidal field. The dimensionless tidal Love number $ k_{2} $ depends on the structure of the NS and therefore on the mass and EOS of dense matter. The quantity $ \Lambda $ is the induced quadrupole polarizability (tidal deformability). The dimensionless tidal deformability $ \Lambda $ is defined as ~\cite{TheLIGOScientific:2017qsa}
\begin{equation}
\Lambda=\frac{2}{3}k_{2}(\frac{c^{2}R}{GM})^{5}=\frac{2}{3}k_{2}(\frac{1}{C})^{5}
\end{equation}
where $  R $ and  $ M $ are the radius and mass of the NS, $ G $ is the gravitational constant, and $ c $ is the speed of light. The quantity $ C $ is the NS compactness which is defined as $ C=\dfrac{GM}{c^{2}R} $. Clearly, $ \Lambda $ is extremely sensitive to the compactness parameter $ C $ and also proportional to the tidal Love number $ k_{2} $, which depends on both $ C $ and $ y_{R} $, the latter being a dimensionless parameter that is sensitive to the entire EOS~\cite{Hinderer:2007mb,Hinderer:2009ca}, 
\begin{eqnarray}
&&k_{2}(C,y_{R})=\dfrac{8}{5}C^{5}(1-2C)^{2}[(2-y_{R})+2C(y_{R}-1)]
\nonumber\\
&&\times\{2C(6-3y_{R}+3C(5y_{R}-8))
\nonumber\\
&&+4C^{3}[(13-11y_{R})+C(3y_{R}-2)+2C^{2}(1+y_{R})]
\nonumber\\
&&+3(1-2C)^{2}[(2-y_{R})+2C(y_{R}-1)]\log(1-2C)\}^{-1}~.\nonumber\\
\end{eqnarray}

Now we proceed to describe a few details involved in to the computation of $ y_{R} $ (for more details, see Refs.~\cite{Hinderer:2007mb,Hinderer:2009ca,Postnikov:2010yn,Piekarewicz:2018sgy} and references contained therein). As mentioned before, an external tidal field induces a mass quadrupole in the star. The external tidal field plus the induced stellar quadrupole combine to produce a nonspherical component to the gravitational potential that in the limit of axial symmetry is proportional to the spherical harmonic $ Y_{20}(\theta,\varphi) $. In turn, the coefficient of $ Y_{20}(\theta,\varphi) $, commonly referred to as $ H(r) $, is a spherically symmetric function that encodes the dynamical changes to the gravitational potential and satisfies a linear, homogeneous, second-order differential equation~\cite{Hinderer:2007mb},

\begin{eqnarray}
&&\dfrac{d^{2}H(r)}{dr^{2}}+(\dfrac{1+F(r)}{r})\dfrac{dH(r)}{dr}+Q(r)H(r)=0~,
\nonumber\\
&&\underset{r\to 0}{\mathop{\lim }}\,H(r)\simeq {{r}^{2}}
\end{eqnarray}
where $ F(r) $ and $ Q(r) $ are functions of the mass, pressure,  and energy density profiles assumed to have been obtained by solving the TOV equations and are given by the expressions
\begin{equation}
F(r)=\dfrac{1-4\pi Gr^{2}(\epsilon(r)-P(r))}{(1-\dfrac{2G M(r)}{r})}
\end{equation}

\begin{eqnarray}
&&Q(r)=
\nonumber\\
&&\dfrac{4\pi}{(1-\dfrac{2G M(r)}{r})}(5\epsilon(r)+9P(r)+\dfrac{\epsilon(r)+P(r)}{c_{s}^{2}(r)}-\dfrac{6}{4\pi r^{2}})
\nonumber\\
&&-4[\dfrac{G(M(r)+4\pi r^{3}P(r))}{r^{2}(1-\dfrac{2GM(r)}{r})}]
\end{eqnarray}
in which $c_{s}^{2}(r)=dP(r)/d\epsilon(r)  $ is the speed of sound at radius $ r $.
Once the differential equation is solved, the value of $ y_{R} $ from the logarithmic derivative of $ H(r) $ evaluated at the surface of the star is $ y_{R}=(rH'(r)/H(r))_{r=R} $. However, since all that is needed to compute the second (tidal) Love number $ k_{2} $ is the logarithmic derivative of $ H(r) $, it is more efficient to solve directly for $ y_{R} $, which, in turn, satisfies the following nonlinear, first-order differential equation~\cite{Postnikov:2010yn,Fattoyev:2012uu}:

\begin{eqnarray}
&&r\frac{dy(r)}{dr}+y^{2}(r)+F(r)y(r)+r^{2}Q(r)=0~;
\nonumber\\
&&\text{with}~y(0)=2 ~~~\text{and}~~~ y_{R}=y(r=R)\nonumber\\
\end{eqnarray}

Tidal effects will form a very small correction in which the accumulated phase shift can be characterized by a single quantity $ \tilde{\Lambda} $, which is a weighted average of the induced quadrupole polarizability (tidal deformability) for the individual stars, $ \Lambda_{1} $ and $ \Lambda_{2} $. Since both NSs have the same EOS, the weighted average $ \tilde{\Lambda}({\cal M})$, as a function of chirp mass $ {\cal M}=m_{1}^{3/5}m_{2}^{3/5}/(m_{1}+m_{2})^{1/5} $, is relatively insensitive to the mass ratio $ m_{1}/m_{2} $~\cite{Hinderer:2009ca}. We therefore focus on the behavior of the quadrupole polarizability $ \Lambda $ of the individual stars~\cite{Postnikov:2010yn}.

Since the tidal deformability hides within a higher-order coefficient in the post-Newtonian expansion of the gravitational wave form, its extraction becomes a challenging proposition. As such, GW170817 could only establish upper limits on the tidal deformability of a $ 1.4 M_{\odot} $ NS, i.e., $ {\Lambda}_{1.4}\leq800 $~\cite{Piekarewicz:2018sgy}, extracted from the original discovery paper~\cite{TheLIGOScientific:2017qsa}. More over, the authors in Ref.[68], by employing a parametrized manner on a very large range of physically plausible EOS for compact stars and by the use of the constraints on mass, upper limit of tidal deformability and the recent suggesstion on lower limit of the tidal deformability, deduced more constraints on tidal deformability and radii of neutron and hybrid stars.  In that study, for a purely hadronic star with a mass of $ 1.4 M_{\odot} $, the radius of the NS is considered to be $ 12.00<R_{1.4}$ (km) $<13.45 $ at a 2-$\sigma $ level; similarly, the smallest weighted average dimensionless tidal deformability is $ \tilde{\Lambda}_{1.4}>375 $, again at a  2-$\sigma $ level. For HSs since EOS with a phase transition allow for very compact stars on the "twin star" branch, small radii are possible~\cite{Montana:2018bkb}; therefore,  the radius varies in a much broader range, $ 8.35<R_{1.4}$  (km) $ <13.74  $ at a  2-$\sigma $  level , with $ \tilde{\Lambda}_{1.4}>35.5 $ at a 3-$\sigma $ level. As we mentioned earlier, we have computed the tidal deformabililty for individual stars with the mass of 1.4 $ M\odot $, $ \Lambda_{1.4} $ instead of $  \tilde{\Lambda}_{1.4} $~\cite{Hinderer:2009ca,Postnikov:2010yn}.

We have calculated  $ y_{R} $, the compactness of the stars $ C $, tidal Love number $ k_{2} $, and dimensionless tidal deformability $ \Lambda $ for purely NSs with the mass of 1.4 $ M_{\odot} $ with $ AV_{18} $ interaction with and without TBF and also $ AV_{14} $, $ UV_{14} $, and Reid interaction supplemented by TBF in order to compare them with the constraints set by the binary NS system GW170817. We have also done the calculations for several HSs in this study with the MIT and NJL models combined with these hadron interactions in LOCV formalism. The results are summarized in Table~\ref{t8}. As seen in this table, the radius and dimensionless tidal deformability of  purely NSs with  $ AV_{18} $ interaction without considering TBF are $ R_{1.4}=9.497 $ km and $ \Lambda_{1.4}=88.428 $ and for NS with the Reid interaction are $ R_{1.4}=10.34 $ km and $\Lambda_{1.4}=147.186   $, which are not in compatible with the constraints predicted for those quantities. It is worth noting that the maximum mass of purely NSs predicted with these models are $ 1.77M_{\odot} $ and $ 1.91 M_{\odot}$ which are  lower than the maximum mass constraint. the radius for purely NS in $ AV_{18} $, $ AV_{14} $, and $ UV_{18} $ interactions supplemented by TBF are $ R_{1.4}=12.32 $ km , $ R_{1.4}=12.18 $ km and $ R_{1.4}=12.28 $ km and  tidal deformabilities are $\Lambda_{1.4}=446.892  $, $\Lambda_{1.4}=453.142   $, and  $\Lambda_{1.4}=461.541   $, respectively, which all are in the ranges set by GW170817. The calculated maximum masses of the neutron stars with these models are $ 2.319M_{\odot} $, $ 1.76 M_{\odot}$, and $ 2.24M_{\odot} $, respectively. Therefore, the calculated properties of purely NSs within the LOCV framework with $ AV_{18} $ and  $ UV_{14} $ interactions supplemented by TBF are completely compatible with the extracted constraints from the observations. For HSs, as seen in the second and third rows of Table~\ref{t8}, with B= $ 90, 100 $ MeVfm$ ^{-3}  $ combined with all the interactions, the star becomes more compact (less radii and more compactness $ C $), and the dimensionless tidal deformability becomes much lower than the purely neutron star; however, again, they are all in the range extracted from  GW170817 for the HSs, mentioned earlier. The maximum masses of all HSs within these models are about $ 1.5M_{\odot} $ (as seen in Table~\ref{t4} ), which is much lower than the maximum mass constraints. As it is seen in the forth row of the table~\ref{t8}, HS's with  $B=  130, 160, 200 $ MeVfm$ ^{-3}  $ and $ m_{s}=150 $ and   $B=  90 $ MeVfm$ ^{-3}$ with $ m_{s}=300 $ MeV and also NJL model in any parameter sets combined with all the interactions,  are less compacted with respect to the purely NS. In these cases, the dimensionless tidal deformability increases and again are in the ranges extracted from observations for HSs. In these cases, the central density in which the star reaches $ 1.4M_{\odot} $ occurs in the hadron branch; therefore, it is the same as the central density in the pure hadron star. However, since the radii of HSs are a little higher with respect to the purely hadron star,  HSs are less compact; therefore, their tidal deformability increases. 
In the cases of $ AV_{18} $, $ AV_{14} $, and $ UV_{14} $ interactions supplemented by TBF combined with the mentioned models, the radii  are $ R_{1.4}=12.42 $ km , $ R_{1.4}=12.28 $ km, and $ R_{1.4}=12.38 $ km, and  the dimensionless tidal deformabilities are $\Lambda_{1.4}=469.980  $, $\Lambda_{1.4}=476.815   $, and $\Lambda_{1.4}=485.665   $ respectively, which are all in the ranges set by GW170817 even for purely hadron stars. Only the maximum masses of a HSs with $ AV_{18} $ ($ UV_{14} $) interaction supplemented by TBF combined with RKH parameter set are in the range of the maximum mass constraint. For these cases, the maximum mass of the HS for $ AV_{18} $ ($ UV_{14} $) is $ 2.009M_{\odot} $ ($ 1.97M_{\odot} $). The same situation happens for a HS with $ AV_{18} $ supplemented by TBF combined with the MIT bag model with B= $  200 $ MeVfm$ ^{-3}$ (B= $  130,160,200 $ MeVfm$ ^{-3}$) and $ m_{s}=150 $ MeV ($ m_{s}=300 $ MeV). In these cases, the maximum mass of HS is $ 1.962M_{\odot} $ ($ 1.96M_{\odot},2.05M_{\odot},2.13M_{\odot} $). The calculated maximum mass in other models are below the observed value.

\section{CONCLUSION}\label{IV}

We have studied the hadron-quark phase transition at high densities, which may occur in the core of massive NSs. We have adopted the LOCV formalism to describe the nuclear matter phase, while the MIT bag and NJL models have been implied for describing the quark matter phase. With $ m_{s}=150 $ MeV, the stable HSs occur in B= $ 90,100$ MeVfm$ ^{-3} $  combined with all the N-N interactions supplemented by TBF and also in B= $ 130$ MeVfm$ ^{-3} $  combined with  $ AV_{14} $ supplemented by TBF with the maximum mass of about $ 1.5M_{\odot} $.The maximum mass of the HSs with $ B=200 $ MeVfm$ ^{-3} $ combined with $ AV_{18} $ interaction supplemented by TBF is compatible with observations. This HS is unstable because of the large energy density discontinuity in transition region which  manifest itself as a cusp in mass-radius curve. 

We examined the effect of N-N forces with other bare two-body interactions, $ AV_{14} $, $UV_{14} $ and Reid 68 supplemented by TBF. The results are almost the same as the $AV_{18} $ potential combined with TBF. 
We also checked the influence of the TBF absence in the nuclear matter. We found that in this situation the phase transition of hadron to quark matter took place in much larger densities and stable HS with B= $ 90,100$ MeVfm$ ^{-3} $ was predicted. We also examined the effect of increasing the strange mass $ m_{s} $ on the results gained with the MIT model. With B= $ 160$ and $200 $ {MeV}{fm$ ^{-3} $} with $ m_{s} =300$ MeV, although the maximum masses of the HSs are  $ 2.05$ and $2.13M_{\odot} $,  which is compatible with the observations, the HSs become unstable as soon as the onset of quark phase because of the cusp in mass-radius curves. We found stable HSs with  $ m_{s}=300 $ MeV combined with $ AV_{18} $  and $UV_{14} $ potentials supplemented by TBF with the maximum masses of  $ 1.796M_{\odot} $ and $ 1.788M_{\odot} $ respectively. By increasing the bag constant, the HSs were rendered unstable.
	
Within the NJL model, a stable HS was calculated in  HK parameter set combined with  $ AV_{18} $ ($UV_{14} $ ) 
 interaction with TBF with the maximum mass of  $  1.896 M_{\odot}$  ($1.882 M_{\odot}$). However, the stable HS was not predicted with other hybrid EOSs, again with the reason of high-energy density discontinuity in the transition region, which manifests itself as a cusp in mass-radius curves. It means that the pressure of the quark matter is unable to counteract the additional downward attraction. The maximum mass of 2.01$ M_{\odot} $ was calculated within the RKH parameter set combined with $ AV_{18} $ interaction supplemented by TBF, however this HS is unstable. With the NJL model, we examined the influence of the absence of TBF, and we found  that the phase transition of hadron to quark matter took place in much higher densities (about $ 8\rho_{0} $ ), the energy discontinuity became higher and the value of the maximum mass became lower . 
	
	All of our results in the MIT model as well as the NJL model were in good concurrence with other works in hybrid stars with other hadron EOSs.
	
We also computed the properties of the purely neutron and hybrid stars  with the mass of $ 1.4M_{\odot} $ to compare them with the new constraints set by the binary neutron star system, GW170817. The radii and tidal deformability of purely NS with the mass of $ 1.4M_{\odot} $,  within the framework of  $ AV_{18} $, $ AV_{14} $ and $ UV_{14} $, supplemented by TBF,  are in compatible with the ranges extracted from the observations. For HSs within all models existing in this study, the mentioned quantities are compatible with the ranges for HSs. Moreover, the radii and tidal deformabilities of HSs with $ B=130,160,200 $ MeVfm$ ^-3 $ with $ m_{s}=150 $MeV and  $ B=90 $ MeVfm$ ^-3 $ with $ m_{s}=300 $MeV and also NJL model in all three parameter sets combined with all interactions except Reid 68 supplemented by TBF, are in the ranges set by binary GW170817 for hadron stars.

\section*{Acknowledgment}
S. K. and H. R. M.  warmly appreciate Micheal Buballa for his helpful comments in NJL model and also would like to thank Research Council, University of Tehran. S. A. T. is grateful to School of Particles and Accelerators, Institute for Research in Fundamental Sciences.


%
%

%

%

\begin{table*}
	\begin{tabular}{ccccccc}
		\hline\hline
		
		Hadron interaction & Bag const.$ ({\text{MeV}}{\text{fm}^{-3}}) $ & $\mu _{B}$(MeV) &   $ {\rho_{B}^{(1)}}/{\rho_{0}}$ & $ {\rho_{B}^{(2)}}/{\rho_{0}}$ & $\epsilon ^{(1)}({\text{MeV}}{\text{fm}^{-3}})$ & $%
		\epsilon ^{(2)}({\text{MeV}}{\text{fm}^{-3}})$

		\\ \hline
		$ AV_{18} $ (2BF) & 90 & 1358.1 & 6.06 & 7 & 1051.5 & 1253.4 \\ 
		& 100 & 1407.94 & 6.43 & 7.87 & 1130.6 & 1450.5 \\ 
		& 130 & 1500.2 & 7.12 & 9.62 & 1284.3 & 1884.7 \\ 
		& 160 & 1561.9 & 7.5 & 10.93 & 1385.4 & 2231.5 \\ 
		& 200 & 1624.6 & 7.93 & 12.31 & 1488.6 & 2634.0 \\ \hline
		$ AV_{18} $ (2BF+3BF) & 90 & 1006.0 & 1.5 & 2.68 & 229.8 & 419.4 \\ 
		& 100 & 1056.4 & 2 & 3.12 & 309.6 & 505.8 \\ 
		& 130 & 1181.7 & 2.75 & 4.5 & 453.2 & 781.5 \\ 
		& 160 & 1207.7 & 3.18 & 5.62 & 537.4 & 1033.8 \\ 
		& 200 &  1351  &  3.56 & 6.87 & 614.4 & 1338.7 \\ \hline
		$ AV_{14} $ & 90 & 1001.8 & 1.43 & 2.62 & 224.6 & 413.5 \\ 
		& 100 & 1003.2 & 2.43 & 3.18 & 387.3 & 516.92 \\ 
		& 130 & 1281.0 & 4.81 & 5.81 & 830.0 & 1042.6 \\ 
		& 160 & 1401.3 & 5.81 & 7.75 & 1048.9 & 1484.6 \\ 
		& 200 & 1448.0 & 6.5 & 8.43 & 1216.13 & 1944.9 \\ \hline
		$ UV_{14} $& 90 & 1040.7 & 2.12 & 2.93 & 332.2 & 470.7 \\ 
		& 100 & 1084.0 &2.31 & 3.37 & 375.2 & 552.7 \\ 
		& 130 & 1192.7 & 2.93 & 5.83 & 486.0 & 807.2 \\ 
		& 160 & 1274.3 & 3.37 & 5.75 & 567.3 & 1053.1 \\ 
		& 200 & 1357.2 & 3.75 & 7 & 647.1 & 1360.1 \\ \hline
		Reid 68 & 90 & 1036.3 & 2.43 & 2.93 & 316.4 & 463.9 \\ 
		& 100 & 1136.5 & 3.5 & 3.93 & 593.1 & 652.9 \\ 
		& 130 & 1337.9 & 5.06 & 6.68 & 877.7 & 1223.4 \\ 
		& 160 & 1426.9 & 5.75 & 8.18 & 1029.6 & 1587.4 \\ 
		& 200 & 1504.9 & 6.18 & 9.68 & 1126.4 & 1977.5 \\ 
		\hline\hline
	\end{tabular}
	\caption{{\small Hadron-quark phase transition properties for various N-N interactions and various bag constants with $ m_{s}=150 $ MeV where $ \mu_{B} $ is critical baryon chemical potential ,  $ {\rho_{B}}/{\rho_{0}}$ is the ratio of baryon density  to the saturation density and  $ \epsilon $ is energy density  at starting $ (1) $ and ending point $ (2) $ of phase transition. }\label{t2}}
\end{table*}

\begin{table*}
	\begin{tabular}{ccccccc}
		\hline\hline
		Hadron interaction & NJL & $\mu _{B}$(MeV) & $ {\rho_{B}^{(1)}}/{\rho_{0}}$ & $ {\rho_{B}^{(2)}}/{\rho_{0}}$ & $\epsilon^{(1)}({\text{MeV}}{\text{fm}^{-3}})$ & $%
		\epsilon^{(2)}({\text{MeV}}{\text{fm}^{-3}})$ \\ \hline
		$ AV_{18} $ (2BF) & RKH & 1678.5 & 8.25 & 13.62 & 1583.9 & 2947.6 \\ 
		& HK & 1668.1 & 8.18 & 12.75 & 1561.1 & 2780.3 \\ 
		& LKW & 1645.6 & 8 & 12.5 & 1516.5 & 2703.0 \\ \hline
		$ AV_{18} $ (2BF+TBF)& RKH & 1379.7 & 3.68 & 5.43 & 640.5 & 1039.76 \\ 
		& HK & 1302.9 & 3.31 & 4.12 & 570.10 & 742.4 \\ 
		& LKW & 1333.6 & 3.43 & 5.12 & 598.10 & 941.8 \\ \hline
		$ AV_{14} $ & RKH & 1578.83 & 7.06 & 10.56 & 1349.5 & 2355.3 \\ 
		& HK & 1572.0 & 7 & 10.18 & 1338.6 & 2255.8 \\ 
		& LKW & 1540.9 & 6.8 & 9.68 & 1288.6 & 2080.9 \\ \hline
		$ UV_{14} $& RKH & 1392.8 & 4.12 & 5.81 & 739.1 & 1112.9 \\ 
		& HK & 1323.8 & 3.56 & 4.37 & 614.5 & 789.2 \\ 
		& LKW & 1345.5 & 3.68 & 5.31 & 635.7 & 993.8 \\ \hline
		Reid 68 & RKH & 1573.5 & 6.63 & 10.5 & 1241.6 & 2320.2 \\ 
		& HK & 1567.2 & 6.62 & 9.93 & 1238.9 & 2223.3 \\ 
		& LKW & 1540.9 & 6.25 & 9.62 & 1145.3 & 2080.7 \\ 
		\hline\hline
	\end{tabular}
	\caption{{\small Same as table.~\ref{t2} but with the NJL  model in various parameter sets for quark phase.}\label{t3}}
\end{table*}

\begin{table*}
	\begin{center}
		\begin{tabular}{ccccccccc}
			\hline\hline
			Hadron interaction & Bag const.$ (\frac{\text{MeV}}{\text{fm}^{3}}) $ & $ \rho_{BC\text{max}} $ $ ({\text{fm}^{-3}}) $ & ${\text{R}}_\text{max}$(km) & ${\text{M}_{\text{max}}}({\text{M}_{\odot}})$     \\ \hline
			
			$ AV_{18} $ (2BF)& NS & 1.62 & 8.5 & 1.77 \\			
			& 90 &1.44 & 9.16 & 1.57   \\
			& 100 & 1.40 & 9.25 & 1.61  \\
			HS $ \Rrightarrow $
			& 130 &  1.57 & 9.19 & 1.67 \\ 
			& 160 &  1.76 & 9.10 & 1.70\\
			& 200 &  1.98 & 9.00 & 1.72\\ \hline
			
			$ AV_{18} $ (2BF+TBF) & NS& 0.94 & 10.95 & 2.319 \\  
			& 90 & 1.57 & 9.16 & 1.50 \\
			& 100 & 1.63 & 9.81 & 1.46\\
			HS $ \Rrightarrow $& 130  & 0.73 & 12.4 & 1.53 \\
			& 160  & 1.27 & 11.2 & 1.609 \\
			& 200   & 1.11 & 12.19  & 1.962 \\ \hline
			$ AV_{14} $ & NS & 1.53 & 9.59 & 1.76 \\
			& 90  & 1.55 & 9.75 & 1.5 \\
			& 100 & 1.62 & 9.85 & 1.46  \\
			HS $ \Rrightarrow$& 130  & 1.19 & 11.27& 1.49  \\
			& 160  & 1.27 & 11.2& 1.61 \\
			& 200   & 1.54 & 10.79& 1.67 \\ \hline
			$ UV_{14} $ & NS & 1.00 & 10.76 & 2.24\\
			& 90 &  1.49 & 9.71 & 1.53\\
			& 100 & 1.51 &  9.73 & 1.49\\
			HS$ \Rrightarrow $& 130  & 0.75 & 12.33 & 1.55\\
			& 160  & 0.92 & 12.21 & 1.76 \\
			& 200   & 1.13 & 12.04 & 1.918 \\ \hline
			Reid 68 &NS  & 1.44 & 9.15 & 1.91 \\
			& 90 &  1.55 & 9.14 & 1.513\\
			& 100  & 1.56 & 9.17 & 1.48\\
			HS $ \Rrightarrow $& 130  & 1.14 & 10.2 & 1.618\\
			& 160   & 1.33 & 10.09 & 1.719\\
			& 200   & 1.56 & 9.95 & 1.72 \\ 
			\hline\hline
		\end{tabular}
		\caption{{\small  Pure NS and HS structure properties in which $ {\text{M}_{\text{max}}}({\text{M}_{\odot}}) $ is the Maximum mass of the star in mass of the sun unit, $\rho_{BC\text{max}}   $ is the central density,  and  $ {\text{R}}_\text{max}$(Km) is radius of the star according to the maximum mass of the star, for various hadron interactions  and bag constants with $ m_\text{{s}}=150  $ MeV.}\label{t4}}
	\end{center}
\end{table*}

\begin{table*}
	\begin{tabular}{ccccccc}
		\hline\hline
		
		Hadron interaction & Bag const.$ ({\text{MeV}}{\text{fm}^{-3}}) $ & $\mu _{B}$(MeV) &   $ {\rho_{B}^{(1)}}/{\rho_{0}}$ & $ {\rho_{B}^{(2)}}/{\rho_{0}}$ & $\epsilon ^{(1)}({\text{MeV}}{\text{fm}^{-3}})$ & $%
		\epsilon ^{(2)}({\text{MeV}}{\text{fm}^{-3}})$

		\\ \hline
		$ AV_{18} $ (2BF) & 90 & 1648 & 8.05 & 11.16 & 1519.86 & 2338.4 \\
		$ AV_{18} $ (2BF+3BF) & 90 & 1218.9 & 2.96 & 3.45 & 490.4 & 584.9 \\ 
		$ AV_{14} $& 90 & 1451 & 6.2 & 7 & 1137.0 & 1331.6 \\ 
		$ UV_{14} $& 90 & 1232.3 & 3.16 & 3.62 & 525.9 & 617.5 \\ 
		Reid& 90 & 1475 & 5.82 & 7.48 & 1053.7 & 1431.5 \\ \hline 
		
		$ AV_{18} $ (2BF+3BF)& 100 & 1250.0 & 3.11 & 4.43 & 520.19 & 787.1 \\ 
		& 130 & 1349.3 & 3.55 & 5.82 & 612.7 & 1102.3 \\ 
		& 160 & 1416.7 & 3.82 & 6.5 & 672.5 & 1369.4 \\ 
		& 200 & 1484.9 & 4.09 & 8.114 & 734.09 & 1689.3 \\ \hline \hline
		
	\end{tabular}
	\caption{{\small Same as table.~\ref{t2} but with $ m_\text{{s}}=300 $ Mev. }\label{t5}}
\end{table*}

\begin{table*}
	\begin{center}
		\begin{tabular}{ccccccccc}
			\hline\hline
			Hadron interaction & Bag const.$ (\frac{\text{MeV}}{\text{fm}^{3}}) $ & $ \rho_{BC\text{max}} $ $ ({\text{fm}^{-3}}) $ & ${\text{R}}_\text{max}$(km) & ${\text{M}_{\text{max}}}({\text{M}_{\odot}})$     \\ \hline
			
			$ AV_{18} $ (2BF)	& 90 &1.79 & 8.97 & 1.73  \\			
			$ AV_{18} $ (2BF+TBF)& 90 &0.92 & 11.59 & 1.796 \\
			$ AV_{14} $& 90 & 1.27 & 10.8 & 1.65\\
			$ UV_{14} $	& 90 &  1.57 & 9.19 & 1.67 \\ 
			Reid 68 & 90 &  0.99 & 11.43 & 1.788\\
			\hline
			
			$ AV_{18} $ (2BF+TBF)   & 100 & 0.74 & 12.31 & 1.75 \\
			& 130 & 0.94& 12.19 & 1.96\\
			& 160  & 1.104 & 12.07 & 2.05\\
			& 200  & 1.298 & 11.94 & 2.13 \\
			\hline\hline
		\end{tabular}
		\caption{{\small  Same as table.~\ref{t4}  but with $ m_\text{{s}}=300  $ MeV.}\label{t6}}
	\end{center}
\end{table*}

\begin{table*}
	\begin{center}
		\begin{tabular}{cccccccc}
			\hline\hline
			Hadron interaction & NJL & $ \rho_{BC\text{max}} $ $ (\frac{\text{MeV}}{\text{fm}^{3}}) $ & ${\text{R}}_\text{max}$(Km) & ${\text{M}_{\text{max}}}({\text{M}_{\odot}})$     \\ \hline
			$ AV_{18} $ (2BF)&NS & 1.62  & 8.5 & 1.77 \\
			&RKH & 2.15 & 8.92 & 1.74\\
			Hs  $ \Rrightarrow $&HK & 2.05 & 8.94 & 1.738\\
			& LKW & 2.01 & 8.97 & 1.73 \\ \hline
			$ AV_{18} $ (2BF+TBF) &NS & 0.94 & 10.95 & 2.319 & \\
			&RKH & 0.89 & 12.13 & 2.009 \\
			HS $ \Rrightarrow $&HK & 0.82 & 12.03 & 1.896 &  \\
			&LKW & 0.83 & 12.21 & 1.932  \\ \hline
			$ AV_{14} $ &NS & 1.53 & 9.59 & 1.76  \\ 
			&RKH & 1.77 & 10.49 & 1.77  \\ 
			Hs $ \Rrightarrow$&HK & 1.71 & 10.51 & 1.71 \\
			& LKW &  1.61 & 10.60 & 1.69  \\ \hline
			$ UV_{14} $ &NS & 1.00 & 10.76 & 2.24  \\
			&RKH  & 0.94 & 11.97 & 1.97  \\
			HS $ \Rrightarrow $	& HK  & 0.84 & 11.92 & 1.88 \\
			& LKW   & 0.87 & 12.06 & 1.90 \\ \hline
			Reid 68&NS & 1.44 & 9.15 & 1.91 \\
			& RKH & 1.76 & 9.83 & 1.82\\
			HS $ \Rrightarrow $& HK  & 1.6z  9 & 9.84 & 1.822\\
			& LKW & 1.61 & 9.87 & 1.81 \\ \hline
			
		\end{tabular}
		\caption{{\small Same as table.~\ref{t4}  but with the NJL model with various parameter sets for quark model.}\label{t7}}
	\end{center}
\end{table*}

\begin{table*}
	\begin{center}
		\begin{tabular}{ccccccccc}
			\hline\hline
			Quark model&Hadron int. & $ \rho_{C} $ $ (\frac{\text{MeV}}{\text{fm}^{3}}) $ & ${\text{R}}$(Km) & ${\text{C}}$ &$y_R$&$ k_{2} $&$\Lambda$\\ \hline
			
			&$ AV_{18} $ (2BF) &0.86 & 9.497  &0.218&0.388&0.0655&88.428 \\			
			& $ AV_{18} $ (2BF+TBF) &0.42 & 12.32 & 0.169&0.356 &0.0948 &446.892& \\
			Pure NS $ \Rrightarrow$ &$ AV_{14} $ & 0.7 & 12.18 & 0.169&0.355 &0.09518 &453.142 &\\
			&$ UV_{14} $ &  0.44 & 12.28 & 0.169&0.361 &0.09517 & 461.541& \\ 
			& 	Reid 68 &  0.7 & 10.34 & 0.202& 0.377 & 0.0745   &147.186  \\
			\hline
			
			& $ AV_{18} $ (2BF)  & 0.86 & 9.54 & 0.217 & 0.387 & 0.0662 &91.697  \\
			& $ AV_{18} $ (2BF+TBF) & 0.97& 9.86 & 0.209&0.384& 0.0702 & 116.232\\
			MIT	B=90, $ m_{s}=150 $   + 	&  $ AV_{14} $ & 0.97 & 10.66 & 0.194& 0.371 & 0.0794 & 192.899 \\
			& $ UV_{14} $  & 0.86 & 10.74 & 0.192&0.372 & 0.0803 & 203.267 \\
			& 	Reid 68 &  0.94 & 9.83 & 0.210& 0.384 & 0.0699 & 114.188\\
			
			\hline
			& $ AV_{18} $ (2BF)  & 0.86 & 9.54 & 0.217& 0.387 & 0.0662 &91.697 \\
			& $ AV_{18} $ (2BF+TBF) & 1.08& 10.02 & 0.206& 0.399& 0.0712& 127.683\\
			MIT	B=100, $ m_{s}=150 $  +	&  $ AV_{14} $ & 1.10 & 10.73 & 0.192& 0.370 & 0.0802 & 201.869\\
			& $ UV_{14} $  & 0.93 & 10.78& 0.192& 0.409 & 0.0784& 200.849 \\
			& 	Reid 68 &  1.0 & 9.9 & 0.209& 0.383 & 0.0705 & 117.675\\
			
			\hline
			& $ AV_{18} $ (2BF)  & 0.86 & 9.54 & 0.217& 0.387 & 0.0662 &91.697 \\
			& $ AV_{18} $ (2BF+TBF) & 0.42& 12.42 & 0.168& 0.356 & 0.0951 & 469.980\\
			MIT	 (B=130,160,200, $ m_{s} $=150), (B=90, $ m_{s} $=300), 	&  $ AV_{14} $ & 0.7 & 12.28 & 0.168& 0.354 & 0.0961 & 476.815 \\
			NJL (RKH, HK, LKW) +& $ UV_{14} $  & 0.44 & 12.38 & 0.167& 0.360 & 0.0961 & 485.665 \\
			& 	Reid 68 &  0.7 & 10.40 & 0.201&0.376 & 0.0752 & 153.183\\
			
			\hline\hline
		\end{tabular}
		\caption{{\small  Central density $ \rho_{c} $ (fm$ ^{-3}$), radius R(km), compactness $ C $, $ y_{R} $, tidal Love number $ k_{2} $ and dimensionless tidal deformability $ \Lambda $ for several purely neutron and hybrid stars with the mass of $ 1.4M_{\odot} $ studied in the paper.}\label{t8}}
	\end{center}
\end{table*}

\end{document}